\documentclass[11pt]{article}
\usepackage{amsmath}
\usepackage{amssymb}
\usepackage[a4paper,top=3cm,bottom=3cm,left=2cm,right=2.5cm,bindingoffset=5mm]{geometry}
\usepackage{amscd}

\newtheorem{theorem}{Theorem}

\newtheorem{definition}{Definition}
\newtheorem{example}{Example}

\newtheorem{lemma}{Lemma}

\newtheorem{proposition}{Proposition}
\newtheorem{remark}{Remark}

\numberwithin{equation}{section}
\input{tcilatex}
\begin{document}

\title{Cautious Belief and Iterated Admissibility\thanks{%
A previous version of this paper has been circulated under the title
\textquotedblleft Common Assumption of Cautious Rationality and Iterated
Admissibility.\textquotedblright \ We are indebted to Pierpaolo Battigalli
and Amanda Friedenberg for important inputs and suggestions about our work.
We also thank Gabriele Beneduci, Adam Brandenburger, Martin Dufwenberg, Nicol%
\`{o} Generoso, Edward Green, Byung Soo Lee, Julien Manili, Burkhard
Schipper, Marciano Siniscalchi, Elias Tsakas and the attendants of our talks
at ESEM 2022 conference, LOFT 2014, AMES 2013 conference, Stern School of
Business, Bocconi University, Scuola Normale Superiore and Politecnico di
Milano for their valuable comments. Financial support from European Research
Council (STRATEMOTIONS--GA 324219) and from the Italian Ministry of
Education, PRIN 2017, Grant Number 2017K8ANN4, are gratefully acknowledged.}}
\author{Emiliano\ Catonini \\
{\small New York University in Shanghai, Department of Economics}\\
{\small emiliano.catonini@gmail.com} \and Nicodemo De Vito \\
{\small Department of Decision Sciences - Bocconi University}\\
{\small nicodemo.devito@unibocconi.it}}
\date{Draft of May 2023}
\maketitle

\begin{abstract}
We define notions of cautiousness and cautious belief to provide epistemic
conditions for iterated admissibility in finite games. We show that iterated
admissibility characterizes the behavioral implications of \textquotedblleft
cautious rationality and common cautious belief in cautious
rationality\textquotedblright \ in a terminal lexicographic type structure.
For arbitrary type structures, the behavioral implications of these
epistemic assumptions are characterized by the solution concept of
self-admissible set (Brandenburger, Friedenberg and Keisler 2008). We also
show that analogous conclusions hold under alternative epistemic
assumptions, in particular if cautiousness is \textquotedblleft
transparent\textquotedblright \ to the players.


\end{abstract}

\section{Introduction}

The iterated deletion of weakly dominated strategies, called \textit{%
iterated admissibility} (henceforth IA), is an important and widely applied
solution concept for games in strategic form. Shimoji (2004) shows that, in
many dynamic games with generic payoffs at terminal nodes, IA is
outcome-equivalent to Pearce's (1984) extensive-form rationalizability, a
prominent solution concept whose foundations are well understood (Battigalli
and Siniscalchi 2002). Applications of IA in games of interest are, for
instance, voting (Moulin 1984) and money-burning games (Ben-Porath and
Dekel, 1992). Yet, while IA has an independent intuitive appeal, its
theoretical foundations have proved to be elusive (see Samuelson 1992).
Thus, the decision-theoretic principles and the hypotheses about strategic
reasoning that yield IA require careful scrutiny.

A recent literature---starting with the seminal contribution of
Brandenburger, Friedenberg and Keisler (2008, henceforth BFK)---has tackled
this issue building on two key ideas. The decision-theoretic aspects of the
problem have been modeled through the \emph{lexicographic expected utility }%
theory of Blume et al. (1991a). Lexicographic expected utility\ preferences
are represented by \emph{lexicographic probability systems }(henceforth
LPS's), i.e., lists of probabilistic conjectures in a priority order, each
of which becomes relevant when the previous ones fail to identify a unique
best alternative. In games with complete information, opponents' strategies
constitute the only payoff-relevant uncertainty. In order to come up with an
educated conjecture about opponents' strategies, a player naturally starts
reasoning about opponents' beliefs and choice criteria. BFK modeled this
aspect with the tools of \emph{epistemic game theory}, the formal,
mathematical analysis of how players reason about each other in games.%
\footnote{%
See Dekel and Siniscalchi (2015)\ for a recent survey.}

Inspired by BFK, in this paper we adopt lexicographic expected utility\ and
epistemic game theory for our foundation of IA in finite games.
Specifically, we use the formalism of lexicographic type structures to model
players' interactive beliefs. However, we start from partially different
basic principles. To motivate our analysis, we briefly mention BFK's results.

BFK showed that, for every natural number $m$, the strategies that survive $%
m+1$ rounds of iterated elimination of inadmissible (i.e., weakly dominated)
strategies are those consistent with the epistemic conditions of \textit{%
rationality and }$m$\textit{th-order assumption of rationality}\ (henceforth
R$m$AR). Such epistemic conditions are represented as events in a type
structure, and the result requires a \textquotedblleft
richness\textquotedblright \ property---called completeness---of the type
structure. BFK's notion of \textquotedblleft rationality\textquotedblright \
incorporates a full-support requirement, which reflects the idea that
nothing is ruled out by the players. As we intuitively explain below, the
concept of \textit{assumption}\ can be thought of as a strong form of
\textquotedblleft virtually persistent belief\textquotedblright \ in an
event. \textit{Rationality and common assumption of rationality}\
(henceforth RCAR) is the condition that R$m$AR\ holds for every $m$. BFK
considered the natural conjecture that, in complete type structures, RCAR is
an epistemic condition for IA---that is, the latter characterizes the
behavioral implications of the former. Yet, they show that RCAR is empty in
complete and \textit{continuous} type structures, such as the canonical,
universal type structure (see Yang 2015, and Catonini and De Vito 2018),
which represents all hierarchies of lexicographic beliefs. Therefore, the
event RCAR falls short of providing a general justification of IA.

\paragraph{Our contribution}

In this paper we propose an alternative approach to the foundations of IA.
Specifically, we provide notions of \emph{rationality}, \emph{cautiousness}
and \emph{cautious belief} that justify the choice of iteratively admissible
strategies in the following way: in \textquotedblleft rich\textquotedblright
\ type structures, IA characterizes the behavioral implications of \textbf{%
cautious rationality and common cautious belief in cautious rationality}
(henceforth R$^{\text{c}}$CB$^{\text{c}}$R$^{\text{c}}$). A prominent
example of \textquotedblleft rich\textquotedblright \ type structure that
works for this purpose is precisely the canonical, universal type structure.
Thus, our first result (Theorem \ref{Theorem main result}) shows that R$^{%
\text{c}}$CB$^{\text{c}}$R$^{\text{c}}$, unlike RCAR, is possible in a
complete and continuous type structure.

We explain in more detail the main ingredients of our analysis. First, we
define \textbf{rationality} as lexicographic expected utility maximization.
The notion of \textbf{cautiousness} requires that all payoff-relevant
consequences be deemed possible by the player; so, it describes a cautious
attitude of the player towards the opponents' strategies. \textbf{Cautious
rationality} is given by the conjunction of rationality and cautiousness.

\textbf{Cautious belief} is a strengthening of the notion of \textit{weak
belief} (Catonini and De Vito 2020), which is based on the preference-based
concept of \textquotedblleft infinitely more likely\textquotedblright \ due
to Lo (1999). Intuitively, a player deems an event $E$ infinitely more
likely than $F$ if she strictly prefers to bet on $E$ rather than on $F$
regardless of the size of the winning prizes for the two bets (given the
same losing outcome). With this, we say that a player \textit{cautiously
believes} an event $E$ if she deems each payoff-relevant component of $E$
infinitely more likely than not-$E$. Put differently, cautious belief
requires that: (1) the player deems $E$ infinitely more likely than not-$E$,
and (2) before entertaining the possibility that $E$ does not occur, she
takes into account all the possible payoff-relevant consequences of $E$.
Condition (1) corresponds to weak belief; condition (2) says that the player
is cautious towards the (weakly) believed event. Thus cautious belief in $E$
is stronger than weak belief as it captures cautiousness relative to $E$.

We further show how our approach allows to provide alternative epistemic
conditions for IA when there is common certainty\ that everybody is
cautious. Precisely, we prove that IA is also justified by rationality, 
\textbf{transparency of cautiousness}, and common cautious belief thereof
(Theorem \ref{Theorem (alternative) main result}). Here, we say that an
event $E$\ is\emph{\ }transparent if it is true and there is common certain
belief in $E$.\footnote{%
The precise, formal definition of \textquotedblleft
transparency\textquotedblright \ for an event is given in Section \ref%
{Section transparency of cautiousness}. To gain intuition, the reader can
temporarily think of transparency as common knowledge, even if they are
distinct notions.} Hence, our epistemic conditions motivate IA as a suitable
solution concept for a context in which there is common certainty that
everyone is cautious. Such result departs from previous justifications of
IA---that is, those based on the concept of assumption or weak assumption
(BFK, Dekel et. al 2016, Yang 2015). In particular, these justifications
require that, for some games, players' caution (however defined) \emph{cannot%
} be certainly believed, as we elaborate in Section \ref{Section: Discussion}%
.

Theorems \ref{Theorem main result} and \ref{Theorem (alternative) main
result} identify \textit{terminality} as the relevant richness property of
type structures for the justification of IA. Such property is satisfied by
the canonical type structure, which represents all hierarchies of beliefs
satisfying an intuitive coherency condition. Differently from most of the
results in epistemic game theory (see, e.g., Dekel and Siniscalchi 2015),
completeness plays no role in the statements and proofs of our results. As
shown by Friedenberg and Keisler (2021), completeness may not be the
appropriate notion of richness to provide epistemic foundations for iterated
strict dominance. Leveraging on the ideas and proofs in Friedenberg and
Keisler (2021), we show (Theorem \ref{Theorem insufficiency
belief-completeness}) that, for any non-degenerate game, there exists a
complete type structure such that IA does not characterize the behavioral
implications of R$^{\text{c}}$CB$^{\text{c}}$R$^{\text{c}}$.

Given this fact, we address the following question: What are the
implications of our epistemic assumptions across all type structures? Note
that considering small (\textquotedblleft non-rich\textquotedblright ) type
structures is essentially equivalent to hypothesize that belief-related,
non-behavioral events are transparent to the players. In other words, there
is a \textquotedblleft context\textquotedblright \ (previous history, social
conventions etc.) in which the game is played, and such context specifies
what beliefs players do vs. do not consider possible. For arbitrary type
structures, we show (Theorems \ref{Theorem SAS} and \ref{Theorem SAS transp
cautious}) that the behavioral implications of the aforementioned epistemic
assumptions are characterized by the solution concept of \textit{%
self-admissible set} (henceforth SAS). This concept was introduced by BFK,
who note that a finite game may admit many SAS's, and the IA set is one of
them. BFK show that, for a fixed type structure associated with a given
game, the behavioral implications of RCAR constitute an SAS; the first part
of Theorem \ref{Theorem SAS} shows that an analogous conclusion holds for R$%
^{\text{c}}$CB$^{\text{c}}$R$^{\text{c}}$. It should be noted that, since
RCAR\ and R$^{\text{c}}$CB$^{\text{c}}$R$^{\text{c}}$\ are distinct
epistemic conditions, they can yield different SAS's in a given type
structure.\footnote{%
For instance, consider the canonical type structure. The behavioral
implications of R$^{\text{c}}$CB$^{\text{c}}$R$^{\text{c}}$ and RCAR\ are
characterized by, respectively, the IA set and the empty set. By definition,
the empty set is an SAS.} Nonetheless, as BFK show for the case of RCAR, the
second part of Theorem \ref{Theorem SAS} shows that every SAS represents the
behavioral implications of R$^{\text{c}}$CB$^{\text{c}}$R$^{\text{c}}$ in
some type structure. Theorem \ref{Theorem SAS transp cautious} shows
analogous results when cautiousness is transparent to the players.

\paragraph{Related literature}

Cautious belief has many similarities with BFK's notion of assumption. Yet,
the two concepts are distinct, and the main difference relies on how players
are cautious in \textquotedblleft believing\textquotedblright \ an event $E$%
. To gain some intuition, also assumption requires that $E$\ be deemed
infinitely more likely than not-$E$; yet, as for cautious belief, this is
not enough. Roughly speaking, assumption requires that event $E$ be deemed
infinitely more likely that not-$E$ conditional on \textit{every}
\textquotedblleft virtual\ observation\textquotedblright \ consistent with $%
E $.\footnote{%
We use the expression \textquotedblleft virtual\
observation\textquotedblright \ to emphasize that there is no real
observation in this static setting. It is just a suggestive language.} This
is in line with the full-support requirement in BFK's notion of rationality,
which reflects the idea that \textquotedblleft everything is
possible.\textquotedblright \ By contrast, cautious belief requires that
event $E$ be deemed infinitely more likely than not-$E$ conditional on 
\textit{payoff-relevant} \textquotedblleft virtual\
observations\textquotedblright \ consistent with $E$. Thus, it requires a
weaker form of \textquotedblleft virtual persistence of
belief.\textquotedblright

The notion of \emph{weak assumption}, introduced by Catonini (2013) and Yang
(2015), solves the impossibility of RCAR by requiring only the
payoff-relevant parts of $E$ to be deemed infinitely more likely than not-$E$%
, in the same way as cautious belief. However, weak assumption maintains the
same preference-based foundation of assumption, that is, the same notion of
infinitely more likely.

We chose Lo's notion mainly for two reasons. First, as pointed out by Blume
et al. (1991a), their notion of infinitely more likely may fail to satisfy a
natural \textit{disjunction} property. To clarify, consider pairwise
disjoint events $C$, $D$ and $E$, and suppose that both $C$\ and $D$ are
infinitely more likely than $E$. Blume et al. (1991a, p. 70) show that $%
C\cup D$ (logically: $C$ or $D$) need not be infinitely more likely than $E$%
; Section \ref{Section: Discussion weak assumption}\ provides a concrete
example.

Second, Lo's (1999) notion of \textquotedblleft infinitely more
likely\textquotedblright \ is equivalent, in terms of LPS's, to the
definition used by Stahl (1995) for the solution concept of \textit{%
lexicographic rationalizability}---a concept which coincides with IA.%
\footnote{%
Stahl (1995) did not provide a preference-based foundation of infinitely
more likely. Lo's (1999) definition applies to a wide class of preferences,
including the lexicographic expected utility model.} Stahl introduced
lexicographic rationalizability as a refinement of permissibility
(Brandenburger 1992), an iterated elimination procedure for lexicographic
beliefs. Stahl's analysis is \textquotedblleft
pre-epistemic\textquotedblright \ in the following sense: there is no
epistemic apparatus to formally express events such as rationality and some
forms of \textquotedblleft belief\textquotedblright \ in rationality. In our
view, R$^{\text{c}}$CB$^{\text{c}}$R$^{\text{c}}$ matches very closely the
logic of lexicographic rationalizability, exactly as the weaker condition of
\textquotedblleft cautious rationality and common weak belief of cautious
rationality\textquotedblright \ captures the logic of permissibility, as
shown in our previous work (Catonini and De Vito 2020). In a sense, the
contribution of this paper can be best seen as providing foundations for
lexicographic rationalizability (see Section \ref{Section: Discussion
lexicographic rationalizability}).

Before moving on to the formal analysis, it is worth stressing other
features of our approach. First, unlike BFK, we do not restrict the analysis
to type structures where players' beliefs are \textit{lexicographic
conditional probability systems} (henceforth LCPS's). Loosely speaking,
LCPS's are LPS's such that the supports of the component measures are
pairwise disjoint. We instead allow for arbitrary LPS-based type structures,
as in Dekel et al. (2016). These authors argue that, from the perspective of
providing foundations for IA, it is conceptually appropriate to consider
unrestricted LPS's. With this, they show that all BFK's results have
analogues in the setting of LPS-based type structures.

Finally, we define cautiousness as a full-support condition on the set of
strategies. This notion of caution can be found, with some minor
differences, in other works (Asheim and Dufwenberg 2003, Perea 2012, Heifetz
et al. 2019, Lee 2016, Catonini and De Vito 2020). An alternative to
cautiousness is the notion that requires the full-support condition on
strategies \textit{and} types. Catonini and De Vito (2020) discuss the
difference between them. Here, we just mention that cautiousness is a
condition expressed in terms of belief hierarchies, and it is invariant to
details of the type structure that are unrelated to hierarchies (e.g., the
topology of type spaces). By contrast, the full-support condition is not
invariant even to isomorphisms between type structures. Therefore, we
provide a justification of IA and SAS's using \textit{expressible} epistemic
assumptions about rationality and beliefs, that is, assumptions which can be
expressed in a language describing primitive terms (strategies) and terms
derived from the primitives (beliefs about strategies, beliefs about
strategies and beliefs of others, etc.)---cf. Battigalli et al. (2021,
Section 3.A).

We have mentioned some important contributions on the epistemic analysis of
IA in games. We provide detailed comments on the closest related literature
in Section \ref{Section: Discussion}.

\paragraph{Structure of the paper}

Section \ref{Section: solution concepts} introduces IA\ and self-admissible
sets. Section \ref{Section: lex beliefs and type structures} provides formal
definitions of LPS's and type structures. Section \ref{Section cautiousness
cautious belief} analyzes cautious rationality and cautious belief. Section %
\ref{Section on the main result} contains our epistemic justifications of IA
and self-admissible sets. Section \ref{Section: Discussion} discusses
certain conceptual aspects of the analysis, and it compares our notion of
cautious belief to those derived from BFK's approach. Appendix A provides
decision-theoretic foundations for cautiousness and cautious belief.
Appendix B and Appendix C collect the proofs omitted from the main text.%
\footnote{%
The Supplementary Appendix contains elaborations and discussions on some
results discussed throughout the paper.}

\section{Iterated admissibility and self-admissible sets\label{Section:
solution concepts}}

Throughout, we consider finite games. A \textbf{finite game} is a structure $%
G:=\left \langle I,(S_{i},\pi _{i})_{i\in I}\right \rangle $, where (a) $I$
is a finite set of players with cardinality $\left \vert I\right \vert \geq
2 $; (b) for each player $i\in I$, $S_{i}$ is a finite, non-empty set of
strategies; and (c) $\pi _{i}:S\rightarrow \mathbb{R}$ is the payoff
function.\footnote{%
Our notation is standard. For any profile of sets $\left( X_{i}\right)
_{i\in I}$, we let $X:=\tprod_{i\in I}X_{i}$ and $X_{-i}:=\tprod_{j\neq
i}X_{j}$ with typical elements $x:=\left( x_{i}\right) _{i\in I}\in X$\ and $%
x_{-i}:=\left( x_{j}\right) _{j\neq i}$ $\in X_{-i}$.}

Each strategy set $S_{i}$\ is given the obvious topology, i.e., the discrete
topology. We let $\mathcal{M}\left( X\right) $ denote the set of all Borel
probability measures on a topological space $X$. So, given a mixed strategy
profile $\sigma \in \tprod_{i\in I}\mathcal{M}(S_{i})$, we will denote
player $i$'s expected utility simply by $\pi _{i}(\sigma _{i},\sigma _{-i})$%
, i.e.,%
\begin{equation*}
\pi _{i}(\sigma _{i},\sigma _{-i}):=\tsum_{(s_{i},\left( s_{j}\right)
_{j\neq i})\in S_{i}\times S_{-i}}\sigma _{i}(\left \{ s_{i}\right \}
)\left( \tprod_{j\neq i}\sigma _{j}(\left \{ s_{j}\right \} )\right) \pi
_{i}(s_{i},\left( s_{j}\right) _{j\neq i})\text{.}
\end{equation*}%
Similarly, given a pure strategy $s_{i}\in S_{i}$\ and a probability measure 
$\mu _{i}\in \mathcal{M}(S_{-i})$, we will denote player $i$'s expected
utility by $\pi _{i}(s_{i},\mu _{i}):=\tsum_{s_{-i}\in S_{-i}}\pi
_{i}(s_{i},s_{-i})\mu _{i}(\left \{ s_{-i}\right \} )$. With an abuse of
notation, we will also identify the pure strategy $s_{i}\in S_{i}$\ with the
mixed strategy $\sigma _{i}\in \mathcal{M}(S_{i})$\ such that $\sigma
_{i}(\left \{ s_{i}\right \} )=1$.

In the remainder of this section, we fix a finite game $G:=\left \langle
I,(S_{i},\pi _{i})_{i\in I}\right \rangle $. Let $\mathcal{Q}$ be the
collection of all subsets of $S$ with the cross-product form $Q=\tprod_{i\in
I}Q_{i}$, where $Q_{i}\subseteq S_{i}$ for every $i$.

\begin{definition}
Fix a set $Q\in \mathcal{Q}$. A strategy $s_{i}\in Q_{i}$\ is \textbf{weakly
dominated with respect} \textbf{to} $Q$\ if there exists a mixed strategy $%
\sigma _{i}\in \mathcal{M}(S_{i})$, with $\sigma _{i}\left( Q_{i}\right) =1$%
, such that $\pi _{i}(\sigma _{i},s_{-i})\geq \pi _{i}(s_{i},s_{-i})$\ for
every $s_{-i}\in Q_{-i}$ and $\pi _{i}(\sigma _{i},s_{-i}^{\prime })>\pi
_{i}(s_{i},s_{-i}^{\prime })$\ for some $s_{-i}^{\prime }\in Q_{-i}$.
Otherwise, say $s_{i}$\ is \textbf{admissible with respect} \textbf{to} $Q$.

If $s_{i}\in S_{i}$\ is \textit{weakly dominated (resp. admissible)} with
respect to $S$, say $s_{i}$ is \textbf{weakly dominated} (resp. \textbf{%
admissible}).
\end{definition}

\begin{remark}
\label{Remark second Pearce lemma}Fix a set $Q\in \mathcal{Q}$. A standard
result (Pearce 1984, Lemma 4) states that a strategy $s_{i}\in Q_{i}$\ is
admissible\textit{\ with respect to} $Q$\ if and only if there exists $\mu
_{i}\in \mathcal{M}(S_{-i})$, with $\mu _{i}(Q_{-i})=1$, such that $\mu
_{i}(\left \{ s_{-i}\right \} )>0$ for every $s_{-i}\in Q_{-i}$, and $\pi
_{i}(s_{i},\mu _{i})\geq \pi _{i}(s_{i}^{\prime },\mu _{i})$\ for every $%
s_{i}^{\prime }\in Q_{i}$.
\end{remark}

The set of iteratively admissible strategies (henceforth IA set) is defined
inductively.

\begin{definition}
For every $i\in I$, let $S_{i}^{0}:=S_{i}$, and for every $m\in \mathbb{N}$,
let $S_{i}^{m}$\ be the set of all $s_{i}\in S_{i}^{m-1}$\ that are
admissible with respect to $S^{m-1}:=\tprod_{j\in I}S_{j}^{m-1}$. A strategy 
$s_{i}\in S_{i}^{m}$\ is called $m$-\textbf{admissible}. A strategy $%
s_{i}\in S_{i}^{\infty }:=\cap _{m=0}^{\infty }S_{i}^{m}$\ is called \textbf{%
iteratively admissible}.
\end{definition}

By finiteness of the game, it follows that $S_{i}^{m}\not=\emptyset $\ for
all $m\in \mathbb{N}$, and, since $S_{i}^{m}\supseteq S_{i}^{m+1}$\ for all $%
m\in \mathbb{N}$, there exists $M\in \mathbb{N}$\ such that $S_{i}^{\infty
}=S_{i}^{M}$. Consequently, the IA set $S^{\infty }$\ is non-empty.

To formally introduce BFK's notion of self-admissible set, we need an
additional definition. Say that a strategy $s_{i}^{\prime }\in S_{i}$ 
\textbf{supports} $s_{i}\in S_{i}$, if there exists a mixed strategy $\sigma
_{i}\in \mathcal{M}(S_{i})$ such that $\sigma _{i}(\left \{ s_{i}^{\prime
}\right \} )>0$\ and $\pi _{i}(\sigma _{i},s_{-i})=\pi _{i}(s_{i},s_{-i})$
for all $s_{-i}\in S_{-i}$.

\begin{definition}
\label{Definition SAS}A set $Q\in \mathcal{Q}$\ is a \textbf{self-admissible
set} (\textbf{SAS}) if, for every player $i$,

(a) every $s_{i}\in Q_{i}$\ is admissible,

(b) every $s_{i}\in Q_{i}$\ is admissible with respect to $S_{i}\times
Q_{-i} $,

(c) for every $s_{i}\in Q_{i}$ and $s_{i}^{\prime }\in S_{i}$, if $%
s_{i}^{\prime }$\ supports $s_{i}$\ then $s_{i}^{\prime }\in Q_{i}$.
\end{definition}

Every finite game admits a non-empty SAS---in particular, the IA set is an
SAS. But, as shown by BFK, many games possess other SAS's, which can be even
disjoint from the IA set. This is illustrated by the following example,
which is taken from Brandenburger et al. (2012, Example 5.9).

\begin{example}
\label{Example: IA SAS BoSOO}Consider the following game between two
players, Ann ($a$)\ and Bob ($b$):%
\begin{equation*}
\begin{tabular}{|c|c|c|}
\hline
$a\backslash b$ & $\ell $ & $r$ \\ \hline
$u$ & $2,2$ & $2,2$ \\ \hline
$m$ & $3,1$ & $0,0$ \\ \hline
$d$ & $0,0$ & $1,3$ \\ \hline
\end{tabular}%
\end{equation*}%
There are three non-empty SAS's: $\left \{ u\right \} \times \left \{
r\right \} $, $\left \{ u\right \} \times \left \{ \ell ,r\right \} $\ and $%
\left \{ m\right \} \times \left \{ \ell \right \} $. The SAS $\left \{
m\right \} \times \left \{ \ell \right \} $\ is the IA set.\hfill $%
\blacklozenge $
\end{example}

A comprehensive analysis of the properties of SAS's in a wide class of games
is given by Brandenburger and Friedenberg (2010).\footnote{%
Example \ref{Example: IA SAS BoSOO} is the reduced strategic form of the
Battle of the Sexes with an Outside Option, which is used by Battigalli and
Friedenberg (2012a) to illustrate how \textquotedblleft extensive-form best
response sets\textquotedblright \ (EFBR's) are related to Pearce's (1984)
notion of \textquotedblleft extensive-form
rationalizability.\textquotedblright \ In the example, the EFBR's coincide
with the SAS's, and Pearce's extensive-form rationalizability coincides with
IA.}

\section{Lexicographic beliefs and lexicographic type structures\label%
{Section: lex beliefs and type structures}}

\subsection{Lexicographic probability systems}

All the sets considered in this paper are assumed to be Polish spaces (that
is, topological spaces that are homeomorphic to complete, separable metric
spaces), and they are endowed with the Borel $\sigma $-field. We let $\Sigma
_{X}$ denote the Borel $\sigma $-field of a Polish space $X$, the elements
of which are called \textit{events}. When it is clear from the context, we
suppress reference to $\Sigma _{X}$ and simply write $X$\ to denote a
measurable space.

Given a sequence $\left( X_{n}\right) _{n\in \mathbb{N}}$ of pairwise
disjoint Polish spaces, the set $X:=\cup _{n\in \mathbb{N}}X_{n}$\ is
endowed with the \textit{direct sum topology},\footnote{%
In this topology, a set $O\subseteq X$\ is open if and only if $O\cap X_{n}$%
\ is open in $X_{n}$ for all $n\in \mathbb{N}$. The assumption that the
spaces $X_{n}$\ are pairwise disjoint is without any loss of generality,
since they can be replaced by a homeomorphic copy, if needed (see Engelking
1989, p. 75).} so that $X$\ is a Polish space. Moreover, we endow each
finite or countable product of Polish spaces with the product topology,
hence the product space is Polish as well.

Recall that $\mathcal{M}\left( X\right) $\ denotes the set of Borel
probability measures on a topological space $X$. The set $\mathcal{M}\left(
X\right) $\ is endowed with the \textit{weak*}-topology. So, if $X$ is
Polish, then $\mathcal{M}\left( X\right) $\ is also Polish. We let $\mathcal{%
N}\left( X\right) $\ (resp. $\mathcal{N}_{n}\left( X\right) $) denote the
set of all finite (resp. length-$n$) sequences of Borel probability measures
on $X$, that is,%
\begin{equation*}
\mathcal{N}\left( X\right) :=\tbigcup_{n\in \mathbb{N}}\mathcal{N}_{n}\left(
X\right) :=\tbigcup_{n\in \mathbb{N}}\left( \mathcal{M}\left( X\right)
\right) ^{n}\text{.}
\end{equation*}

Each $\bar{\mu}:=\left( \mu ^{1},...,\mu ^{n}\right) \in \mathcal{N}\left(
X\right) $\ is called \textbf{lexicographic probability system} (\textbf{LPS}%
). In view of our assumptions, the topological space $\mathcal{N}\left(
X\right) $ is Polish.

For every Borel probability measure $\mu $\ on $X$, the support of $\mu $,
denoted by $\mathrm{Supp}\mu $, is the smallest closed subset $C\subseteq X$
such that $\mu \left( C\right) =1$. The support of an LPS $\bar{\mu}:=\left(
\mu ^{1},...,\mu ^{n}\right) \in \mathcal{N}\left( X\right) $\ is defined as 
$\mathrm{Supp}\bar{\mu}:=\cup _{l\leq n}\mathrm{Supp}\mu ^{l}$. So, an LPS $%
\bar{\mu}:=\left( \mu ^{1},...,\mu ^{n}\right) \in \mathcal{N}\left(
X\right) $\ is of \textbf{full-support} if $\mathrm{Supp}\bar{\mu}=X$. We
write $\mathcal{N}^{+}\left( X\right) $ for the set of full-support LPS's.

For future reference, we also record the following definition. An LPS $\bar{%
\mu}:=\left( \mu ^{1},...,\mu ^{n}\right) \in \mathcal{N}\left( X\right) $\
is called \textbf{lexicographic conditional probability system} (\textbf{LCPS%
}) if there are events\ $E_{1},...,E_{n}$\ in $X$ such that, for every $%
l\leq n$, $\mu ^{l}\left( E_{l}\right) =1$\ and $\mu ^{l}\left( E_{m}\right)
=0$\ for $m\neq l$. If $X$ is finite, then an LPS is an LCPS if and only if
the supports of the component measures are pairwise disjoint.

Fix Polish spaces $X$\ and $Y$, and a Borel map $f:X\rightarrow Y$. The map $%
\widetilde{f}:\mathcal{M}\left( X\right) \rightarrow \mathcal{M}\left(
Y\right) $, defined by%
\begin{equation*}
\widetilde{f}\left( \mu \right) \left( E\right) :=\mu \left( f^{-1}\left(
E\right) \right) ,\text{ \  \ }E\in \Sigma _{Y}\text{, }\mu \in \mathcal{M}%
\left( X\right) \text{,}
\end{equation*}%
is called the image (or pushforward) measure map of $f$. For each $n\in 
\mathbb{N}$, the map $\widehat{f}_{\left( n\right) }:\mathcal{N}_{n}\left(
X\right) \rightarrow \mathcal{N}_{n}\left( Y\right) $ is defined by%
\begin{equation*}
\left( \mu ^{1},...,\mu ^{n}\right) \mapsto \widehat{f}_{\left( n\right)
}\left( \left( \mu ^{1},...,\mu ^{n}\right) \right) :=\left( \widetilde{f}%
\left( \mu ^{k}\right) \right) _{k\leq n}\text{.}
\end{equation*}%
With his, the map $\widehat{f}:\mathcal{N}\left( X\right) \rightarrow 
\mathcal{N}\left( Y\right) $ defined by%
\begin{equation*}
\widehat{f}\left( \bar{\mu}\right) :=\widehat{f}_{\left( n\right) }\left( 
\bar{\mu}\right) \text{, }\bar{\mu}\in \mathcal{N}_{n}\left( X\right) \text{,%
}
\end{equation*}%
is called the \textbf{image LPS map of} $f$. Alternatively put, the map $%
\widehat{f}$ is the \textit{union} of the maps $\left( \widehat{f}_{\left(
n\right) }\right) _{n\in \mathbb{N}}$, and it is Borel measurable.\footnote{%
For details and proofs related to Borel measurability and continuity of the
involved maps, the reader can consult Catonini and De Vito (2018).}

Given Polish spaces $X$\ and $Y$, we let $\mathrm{Proj}_{X}$\ denote the
canonical projection from $X\times Y$\ onto $X$; in view of our assumption,
the map $\mathrm{Proj}_{X}$\ is continuous. The marginal measure of $\mu \in 
\mathcal{M}\left( X\times Y\right) $ on $X$ is defined by $\mathrm{marg}%
_{X}\mu :=\widetilde{\mathrm{Proj}}_{X}\left( \mu \right) $. Consequently,
the marginal of $\bar{\mu}\in \mathcal{N}\left( X\times Y\right) $\ on $X$
is defined by $\overline{\mathrm{marg}}_{X}\bar{\mu}:=\widehat{\mathrm{Proj}}%
_{X}\left( \bar{\mu}\right) $, and the function $\widehat{\mathrm{Proj}}_{X}:%
\mathcal{N}\left( X\times Y\right) \rightarrow \mathcal{N}\left( X\right) $
is continuous and surjective.

\subsection{Lexicographic type structures\label{Section on LPS type
structures}}

Fix a finite game $G:=\left \langle I,(S_{i},\pi _{i})_{i\in
I}\right
\rangle $. A \emph{type structure} (associated with $G$)
formalizes an implicit approach to model hierarchies of beliefs. The
following is a generalization of the standard definition\ of epistemic type
structure with beliefs represented by probability measures, i.e., length-$1$
LPS's (cf. Heifetz and Samet 1998).

\begin{definition}
\label{Definition lexico type structure}An $\left( S_{i}\right) _{i\in I}$-%
\textbf{based lexicographic type structure} is a structure $\mathcal{T}%
:=\langle S_{i},T_{i},\beta _{i}\rangle _{i\in I}$ where

\begin{enumerate}
\item for each $i\in I$, $T_{i}$\ is a Polish space;

\item for each $i\in I$, the function $\beta _{i}:T_{i}\rightarrow \mathcal{N%
}\left( S_{-i}\times T_{-i}\right) $ is Borel measurable.
\end{enumerate}

Each space $T_{i}$\ is called \textbf{type space} and each $\beta _{i}$\ is
called \textbf{belief map}.\footnote{%
Some authors (e.g., Battigalli and Siniscalchi 1999, Heifetz and Samet 1998)
use the terminology \textquotedblleft type space\textquotedblright \ for
what is called \textquotedblleft type structure\textquotedblright \ here.}
Members of type spaces, viz. $t_{i}\in T_{i}$, are called \textbf{types}.
Each element $\left( s_{i},t_{i}\right) _{i\in I}\in \tprod_{i\in I}\left(
S_{i}\times T_{i}\right) $\ is called \textbf{state} (\textbf{of the world}).
\end{definition}

In what follows, we will omit the qualifier \textquotedblleft
lexicographic,\textquotedblright \ and simply speak of \textbf{type
structures} when the underlying strategy sets $\left( S_{i}\right) _{i\in I}$%
\ are clear from the context. Furthermore, if every type in a type structure 
$\mathcal{T}$ is associated with a probability measure, then we will say
that $\mathcal{T}$ is an \textbf{ordinary type structure}.

Type structures generate a collection of hierarchies of beliefs for each
player. For instance, type $t_{i}$'s first-order belief is an LPS on $S_{-i}$%
, and is given by $\overline{\mathrm{marg}}_{S_{-i}}\beta _{i}(t_{i})$. A
standard inductive procedure (see Catonini and De Vito 2018, for details)
shows how to provide an explicit description of a hierarchy induced by a
type.

We will be interested in type structures with one or more of the following
features, which do not make reference to hierarchies of beliefs or other
type structures.

\begin{definition}
\label{Definition complete type structure}A type structure $\mathcal{T}%
:=\langle S_{i},T_{i},\beta _{i}\rangle _{i\in I}$ is

\begin{itemize}
\item \textbf{finite} if the cardinality of each type space $T_{i}$\ is
finite;

\item \textbf{compact} if each type space $T_{i}$\ is compact;

\item \textbf{belief-complete} if each belief map $\beta _{i}$ is onto;

\item \textbf{continuous} if each belief map $\beta _{i}$ is continuous.
\end{itemize}
\end{definition}

The idea of (belief-)completeness was introduced by Brandenburger (2003) and
adapted to the present context.\footnote{%
In Dekel et al. (2016), a type structure $\mathcal{T}:=\langle
S_{i},T_{i},\beta _{i}\rangle _{i\in I}$\ is said to be complete if the
range of each belief map $\beta _{i}$ is a strict superset of $\mathcal{N}%
^{+}\left( S_{-i}\times T_{-i}\right) $. A belief-complete type structure is
complete in the sense of Dekel et al. (2016); the converse does not hold.}
Note that each type space in a belief-complete type structure\ has the
cardinality of the continuum. Finite type structures are compact and
continuous, but not belief-complete. No belief-complete lexicographic type
structure is also compact and continuous. To see this, observe that if the
type structure is compact and continuous, each $\beta _{i}(T_{i})$ is
compact but the space $\mathcal{N}\left( S_{-i}\times T_{-i}\right) $ is not
compact,\footnote{%
The space $\mathcal{M}\left( X\right) $\ is compact if and only if $X$ is
compact, and this in turn implies that the space $\mathcal{N}_{n}\left(
X\right) $\ is also compact\ for every \textit{finite} $n\in \mathbb{N}$.
But the same conclusion does not hold for the space $\mathcal{N}\left(
X\right) $. This is an instance of a well-known mathematical fact (see
Theorem 2.2.3 in Engelking 1989): If $\left( X_{\theta }\right) _{\theta \in
\Theta }$\ is an indexed family of non-empty compact spaces with $%
\left
\vert X_{\theta }\right \vert >1$\ for all $\theta \in \Theta $, then
the direct sum $\cup _{\theta \in \Theta }X_{\theta }$\ is compact if and
only if the right-directed set $\Theta $\ is finite.} hence $\beta _{i}$\ is
not onto.

We next introduce the notion of type morphism, which captures the idea that
a type structure $\mathcal{T}$\ is \textquotedblleft contained
in\textquotedblright \ another type structure $\mathcal{T}^{\ast }$. In what
follows, given a type structure $\mathcal{T}:=\langle S_{i},T_{i},\beta
_{i}\rangle _{i\in I}$, we let $T$\ denote the Cartesian product of type
spaces, that is, $T:=\tprod_{i\in I}T_{i}$. Moreover, for any set $X$, we
let $\mathrm{Id}_{X}$ denote the identity map on $X$, that is, $\mathrm{Id}%
_{X}\left( x\right) :=x$\ for all $x\in X$.

\begin{definition}
\label{Definition type morphism}Fix type structures $\mathcal{T}:=\langle
S_{i},T_{i},\beta _{i}\rangle _{i\in I}$\ and $\mathcal{T}^{\ast }:=\langle
S_{i},T_{i}^{\ast },\beta _{i}^{\ast }\rangle _{i\in I}$. For each $i\in I$,
let $\varphi _{i}:T_{i}\rightarrow T_{i}^{\ast }$\ be a measurable map such
that%
\begin{equation*}
\beta _{i}^{\ast }\circ \varphi _{i}=\widehat{\left( \mathrm{Id}%
_{S_{-i}},\varphi _{-i}\right) }\circ \beta _{i}\text{.}
\end{equation*}%
where $\varphi _{-i}:=\left( \varphi _{j}\right) _{j\neq
i}:T_{-i}\rightarrow T_{-i}^{\ast }$. The function $\left( \varphi
_{i}\right) _{i\in I}:T\rightarrow T^{\ast }$ is called \textbf{type morphism%
}\textit{\ (\textbf{from} }$\mathcal{T}$\textit{\  \textbf{to} }$\mathcal{T}%
^{\ast }$\textit{).}

The morphism\ is called \textbf{bimeasurable} if the map $\left( \varphi
_{i}\right) _{i\in I}$ is Borel bimeasurable.\footnote{%
A Borel map $f:X\rightarrow Y$ between separable metrizable spaces is
bimeasurable if $f(E)$ is Borel in $Y$ provided $E$ is Borel in $X$.} The
morphism\ is called \textbf{type isomorphism} if the map $\left( \varphi
_{i}\right) _{i\in I}$ is a Borel isomorphism. Say $\mathcal{T}$\ and $%
\mathcal{T}^{\ast }$\ are \textbf{isomorphic} if there is a type isomorphism
between them.
\end{definition}

A type morphism requires consistency between the function $\varphi
_{i}:T_{i}\rightarrow T_{i}^{\ast }$\ and the induced function $\widehat{%
\left( \mathrm{Id}_{S_{-i}},\varphi _{-i}\right) }:\mathcal{N}\left(
S_{-i}\times T_{-i}\right) \rightarrow \mathcal{N}\left( S_{-i}\times
T_{-i}^{\ast }\right) $. That is, the following diagram commutes:%
\begin{equation}
\begin{CD} T_{i} @>{\beta _{i}}>> \mathcal{N}(S_{-i}\times \ T_{-i})\\
@VV\varphi_{i}V @VV\  \widehat{(\mathrm{Id}_{S_{-i}},\varphi_{-i})}V\\
T_{i}^* @>{\beta _{i}^*}>> \mathcal{N}(S_{-i}\times \ T_{-i}^*) \end{CD}%
\text{.}  \label{commutative diagram}
\end{equation}%
Thus, a type morphism maps $\mathcal{T}$\ into $\mathcal{T}^{\ast }$ in a
way that preserves the beliefs associated with types.

The notion of type morphism does not make reference to hierarchies of
beliefs. But the important property of type morphisms is that they preserve
the explicit description of lexicographic belief hierarchies: the $\left(
S_{i}\right) _{i\in I}$-based belief hierarchy generated by a type $t_{i}\in
T_{i}$ in $\mathcal{T}$ is also generated by its image $\varphi
_{i}(t_{i})\in T_{i}^{\ast }$ in $\mathcal{T}^{\ast }$. Heifetz and Samet
(1998, Proposition 5.1) show this result for the case of ordinary type
structures; the generalization to lexicographic type structures is
straightforward (see Catonini and\ De Vito 2018).

Next, we introduce the notion of terminality for a type structure.

\begin{definition}
Fix a class $\mathbb{T}$ of type structures. A type structure $\mathcal{T}%
^{\ast }:=\langle S_{i},T_{i}^{\ast },\beta _{i}^{\ast }\rangle _{i\in I}$\
is \textbf{terminal with respect to }$\mathbb{T}$ if for every type
structure $\mathcal{T}:=\langle S_{i},T_{i},\beta _{i}\rangle _{i\in I}$ in $%
\mathbb{T}$, there is a type morphism from $\mathcal{T}$\ to $\mathcal{T}%
^{\ast }$.
\end{definition}

Whenever $\mathcal{T}^{\ast }$ is terminal with respect to the class of 
\textit{all} type structures, we simply say, as customary, that $\mathcal{T}%
^{\ast }$ is\emph{\ terminal}. In Section \ref{Section on the main result}
we will show that R$^{\text{c}}$CB$^{\text{c}}$R$^{\text{c}}$ justifies IA
in every type structure which is terminal with respect to the class of all
finite type structures, and that such a type structure exists.

\section{Cautiousness and cautious belief\label{Section cautiousness
cautious belief}}

For this section, we fix a finite game $G:=\left \langle I,(S_{i},\pi
_{i})_{i\in I}\right \rangle $, and we append to $G$ a type structure $%
\mathcal{T}:=\langle S_{i},T_{i},\beta _{i}\rangle _{i\in I}$.

\subsection{Rationality\ and cautiousness\label{Section Rationality and
Cautiousness}}

For any two vectors $x:=\left( x_{l}\right) _{l=1}^{n},y:=\left(
y_{l}\right) _{l=1}^{n}\in \mathbb{R}^{n}$, we write $x\geq _{L}y$ if either
(a) $x_{l}=y_{l}$ for every $l\leq n$, or (b) there exists $m\leq n$ such
that $x_{m}>y_{m}$ and\ $x_{l}=y_{l}$ for every $l<m$; we write $x>_{L}y$\
if condition (b) holds.

\begin{definition}
\label{Definition optimality strategy} A strategy $s_{i}\in S_{i}$\ is 
\textbf{optimal under} $\beta _{i}(t_{i}):=\left( \mu _{i}^{1},...,\mu
_{i}^{n}\right) \in \mathcal{N}(S_{-i}\times T_{-i})$\ if, for every $%
s_{i}^{\prime }\in S_{i}$,%
\begin{equation*}
\left( \pi _{i}(s_{i},\mathrm{marg}_{S_{-i}}\mu _{i}^{l})\right)
_{l=1}^{n}\geq _{L}\left( \pi _{i}(s_{i}^{\prime },\mathrm{marg}_{S_{-i}}\mu
_{i}^{l})\right) _{l=1}^{n}\text{.}
\end{equation*}%
Say that $s_{i}$\ is a \textbf{lexicographic best reply to} $\overline{%
\mathrm{marg}}_{S_{-i}}\beta _{i}(t_{i})$\ if it\ is optimal under $\beta
_{i}(t_{i})$.
\end{definition}

This is the usual definition of optimality for a strategy, but this time
optimality is taken lexicographically. We next introduce the notion of
cautiousness.

\begin{definition}
\label{Definition: Cautious type}A type $t_{i}\in T_{i}$ is \textbf{cautious}
(in $\mathcal{T}$) if $\overline{\mathrm{marg}}_{S_{-i}}\beta _{i}(t_{i})\in 
\mathcal{N}^{+}\left( S_{-i}\right) $.
\end{definition}

This notion of cautiousness\ requires that the first-order belief of a type
be a full-support LPS. That is, it requires that every payoff-relevant
component, viz. $\left \{ s_{-i}\right \} \times T_{-i}$, be assigned
strictly positive probability by at least one of the measures of LPS $\beta
_{i}(t_{i})$. For each $i\in I$, we let $C_{i}$ denote the set of all pairs $%
\left( s_{i},t_{i}\right) \in S_{i}\times T_{i}$\ such that $t_{i}$ is
cautious.

For strategy-type pairs we define the following notions.

\begin{definition}
Fix a strategy-type pair $\left( s_{i},t_{i}\right) \in S_{i}\times T_{i}$.

\begin{enumerate}
\item Say $\left( s_{i},t_{i}\right) $\ is \textbf{rational} (in $\mathcal{T}
$) if $s_{i}$\ is optimal under $\beta _{i}\left( t_{i}\right) $.

\item Say $\left( s_{i},t_{i}\right) $\ is \textbf{cautiously rational} (in $%
\mathcal{T}$) if it is rational and $t_{i}$ is cautious.
\end{enumerate}
\end{definition}

We let $R_{i}$\ denote the set of all rational strategy-type pairs. As one
should expect, cautious rationality guarantees admissibility.

\begin{proposition}
\label{Proposition CR implies admissibility}If strategy-type pair $%
(s_{i},t_{i})\in S_{i}\times T_{i}$ is cautiously rational, then $s_{i}$ is
admissible.
\end{proposition}

The proof of Proposition \ref{Proposition CR implies admissibility} is in
Appendix B.

\subsection{Infinitely more likely and cautious belief\label{Section IML and
CB}}

We say that player $i$ deems event $E$ infinitely more likely than event $F$
if she prefers to bet on $E$ rather than on $F$ no matter the prizes for the
two bets. This preference-based notion of \textquotedblleft infinitely more
likely\textquotedblright \ is due to Lo (1999, Definition 1), and it is
formalized in Appendix A, where we introduce the appropriate language. Here,
we provide the equivalent definition of \textquotedblleft infinitely more
likely\textquotedblright \ in terms of the LPS that represents player $i$'s
preferences.

Given an LPS $\bar{\mu}_{i}:=(\mu _{i}^{1},...,\mu _{i}^{n})\in \mathcal{N}%
(S_{-i}\times T_{-i})$ and an event $E\subseteq S_{-i}\times T_{-i}$, let%
\begin{equation*}
\mathcal{I}_{\bar{\mu}_{i}}\left( E\right) :=\inf \left \{ l\in \left \{
1,...,n\right \} :\mu _{i}^{l}\left( E\right) >0\right \} \text{,}
\end{equation*}%
with the convention that $\inf \emptyset :=+\infty $. The following
definition is from Catonini and De Vito (2020); see also Stahl (1995).

\begin{definition}
\label{Definition IML LPS}Fix two disjoint events $E,F\subseteq S_{-i}\times
T_{-i}$. Say that $E$ is \textbf{infinitely more likely} than $F$ under $%
\bar{\mu}_{i}$\ if $\mathcal{I}_{\bar{\mu}_{i}}\left( E\right) <\mathcal{I}_{%
\bar{\mu}_{i}}\left( F\right) $.
\end{definition}

It is straightforward to see that \textquotedblleft infinitely more
likely\textquotedblright \ is monotone. That is, if $E$ is infinitely more
likely than $F$ under $\bar{\mu}_{i}$\ and $G$ is an event such that\ $%
E\subseteq G$, then $G$ is infinitely more likely than $F$ under $\bar{\mu}%
_{i}$.

Consider now the following attitudes of player $i$ towards an event $E$.
First, player $i$ deems $E$ infinitely more likely than its complement.
Second, player $i$ has a cautious attitude towards the event: Before
considering its complement, she considers all the possible payoff-relevant
consequences of the event. The notion of cautious belief captures both
attitudes.

\begin{definition}
\label{Definition assumption under an LPS}Fix a non-empty event $E\subseteq
S_{-i}\times T_{-i}$ and a type $t_{i}\in T_{i}$ with $\beta _{i}\left(
t_{i}\right) :=(\mu _{i}^{1},...,\mu _{i}^{n})$. Event $E$ is \textbf{%
cautiously believed under} $\beta _{i}\left( t_{i}\right) $ \textbf{at level}
$m\leq n$\ if the following conditions hold:

\begin{description}
\item[(i)] $\mu _{i}^{l}\left( E\right) =1$\ for all $l\leq m$;

\item[(ii)] for every elementary cylinder $\hat{C}_{s_{-i}}:=\left \{
s_{-i}\right \} \times T_{-i}$, if $E\cap \hat{C}_{s_{-i}}\not=\emptyset $\
then $\mu _{i}^{l}\left( E\cap \hat{C}_{s_{-i}}\right) >0$\ for some $l\leq
m $.
\end{description}

Event $E$ is \textbf{cautiously believed under} $\beta _{i}\left(
t_{i}\right) $ if it is cautiously believed under $\beta _{i}\left(
t_{i}\right) $ at some level\ $m\leq n$.

Type $t_{i}\in T_{i}$\  \textbf{cautiously believes} $E$\ if $E$\ is
cautiously believed under $\beta _{i}\left( t_{i}\right) $.
\end{definition}

Condition (i) captures the first attitude. Under condition (i), condition
(ii) is equivalent to saying that player $i$ deems all \emph{payoff-relevant
parts} of $E$ (i.e., the non-empty intersections of $E$ with each
strategy-based cylinder) infinitely more likely than not-$E$, so it captures
the second attitude.

The conceptual consistency between cautiousness and cautious belief is
highlighted by the following connection.

\begin{remark}
\label{Remark Cautiousness is Cautious Belief of state space}A type $%
t_{i}\in T_{i}$ is cautious if and only if $t_{i}$ cautiously believes $%
S_{-i}\times T_{-i}$.
\end{remark}

Appendix A provides a preference-based foundation for cautious belief in
terms of \textquotedblleft infinitely more likely,\textquotedblright \ as
well as a characterization in terms of infinitesimal nonstandard numbers.
Here, we just mention some properties of cautious belief that will be useful
for the proofs of our results.

\begin{proposition}
\label{Lemma on properties of assumption one direction conjunction}Fix a
type $t_{i}\in T_{i}$ with $\beta _{i}\left( t_{i}\right) :=(\mu
_{i}^{1},...,\mu _{i}^{n})$.

\begin{enumerate}
\item Fix non-empty events $E_{1},E_{2},...$\ in $S_{-i}\times T_{-i}$. If,
for each $k$, type $t_{i}$ cautiously believes $E_{k}$, then $t_{i}$
cautiously believes $\cap _{k}E_{k}$\ and $\cup _{k}E_{k}$.

\item \textit{A non-empty event }$E\subseteq S_{-i}\times T_{-i}$\textit{\
is cautiously believed under }$\beta _{i}\left( t_{i}\right) $\textit{\ if
and only if there exists }$m\leq n$\textit{\ such that }$\beta _{i}\left(
t_{i}\right) $\textit{\ satisfies condition (i) of Definition \ref%
{Definition assumption under an LPS} plus the following condition:}

\begin{description}
\item[(ii')] $\cup _{l\leq m}\mathrm{Suppmarg}_{S_{-i}}\mu _{i}^{l}=\mathrm{%
Proj}_{S_{-i}}\left( E\right) $.
\end{description}
\end{enumerate}
\end{proposition}

The proof of Proposition \ref{Lemma on properties of assumption one
direction conjunction} is in Appendix B.

Proposition \ref{Lemma on properties of assumption one direction conjunction}%
.1\ states that cautious belief satisfies one direction of conjunction as
well as one direction of disjunction. Proposition \ref{Lemma on properties
of assumption one direction conjunction}.2 can be viewed as a
\textquotedblleft marginalization\textquotedblright \ property of cautious
belief: If $E$\textit{\ }is cautiously believed under $\beta _{i}\left(
t_{i}\right) $, then

(a) $\mathrm{Proj}_{S_{-i}}\left( E\right) $\ is infinitely more likely than 
$S_{-i}\backslash \mathrm{Proj}_{S_{-i}}\left( E\right) $ under $\overline{%
\mathrm{marg}}_{S_{-i}}\beta _{i}(t_{i})$; and

(b) \emph{every} strategy in $\mathrm{Proj}_{S_{-i}}\left( E\right) $\ is
infinitely more likely than (every strategy in) $S_{-i}\backslash \mathrm{%
Proj}_{S_{-i}}\left( E\right) $ under $\overline{\mathrm{marg}}%
_{S_{-i}}\beta _{i}(t_{i})$.

The failure of one direction of conjunction reveals that, although
\textquotedblleft infinitely more likely\textquotedblright \ is monotone,
cautious belief is not. That is, if $t_{i}$ cautiously believes $E$, then $%
t_{i}$ may not cautiously believe an event $F$\ such that $E\subseteq F$.
The reason why this can occur is that player $i$ may not have towards $F$
the same cautious attitude that she has towards $E$. That is, there may be
some payoff-relevant components of $F\backslash E$\ which are not deemed
infinitely more likely than not-$F$. This is illustrated by the following
example, which is from Catonini and De Vito (2020, Example 1).

\begin{example}
\label{Example: cautious belief not monotone}Consider a finite game with two
players, Ann ($a$)\ and Bob ($b$), where the strategy set of Bob is\textit{\ 
}$S_{b}:=\left \{ s_{b}^{1},s_{b}^{2},s_{b}^{3}\right \} $\textit{. Append
to this game a type structure }$\mathcal{T}$\textit{\ such that }$%
T_{b}:=\left \{ t_{b}^{\ast }\right \} $\textit{. Consider the LPS} $\bar{\mu%
}_{a}:=(\mu _{a}^{1},\mu _{a}^{2})\in \mathcal{N}(S_{b}\times T_{b})$ 
\textit{with }$\mu _{a}^{1}\left( \left \{ \left( s_{b}^{1},t_{b}^{\ast
}\right) \right \} \right) =1$\textit{\ and }$\mu _{a}^{2}\left( \left \{
\left( s_{b}^{2},t_{b}^{\ast }\right) \right \} \right) =\mu _{a}^{2}\left(
\left \{ \left( s_{b}^{3},t_{b}^{\ast }\right) \right \} \right) =\frac{1}{2}
$\textit{. Next, consider the events }$E:=\left \{ s_{b}^{1}\right \} \times
T_{b}$\textit{\ and }$F:=\left \{ s_{b}^{1},s_{b}^{2}\right \} \times T_{b}$%
\textit{. Clearly, }$E\subseteq F$\textit{. Yet, }$E$\textit{\ is cautiously
believed under }$\bar{\mu}_{a}$\textit{\ at level }$1$\textit{, while }$F$%
\textit{\ is not cautiously believed: indeed, }$\mu _{a}^{1}\left( F\right)
=1$\textit{\ and }$\mu _{a}^{2}\left( F\right) =\frac{1}{2}$\textit{, and,
with }$l=1$\textit{, condition (ii) of Definition \ref{Definition assumption
under an LPS}\ is not satisfied for }$\hat{C}_{s_{b}^{2}}:=\left \{
s_{b}^{2}\right \} \times T_{b}$\textit{.}\hfill $\blacklozenge $
\end{example}

However, it is easy to observe that cautious belief is monotone with respect
to events with the same behavioral implications.

\begin{remark}
\label{Remark: quasi-monotonocity of assumption}Let $E_{-i},F_{-i}\subseteq
S_{-i}\times T_{-i}\ $be events such that $E_{-i}\subseteq F_{-i}$ and $%
\mathrm{Proj}_{S_{-i}}\left( E_{-i}\right) =\mathrm{Proj}_{S_{-i}}\left(
F_{-i}\right) $. If a type $t_{i}$ cautiously believes $E_{-i}$, then $t_{i}$
cautiously believes $F_{-i}$.
\end{remark}

This \textquotedblleft quasi-monotonicity\textquotedblright \ property will
play a crucial role in the proof of our main result.

For future reference, it is useful to mention the following notion of
belief, called certain belief (Halpern 2010). Fix a non-empty event $%
E\subseteq S_{-i}\times T_{-i}$ and a type $t_{i}\in T_{i}$ with $\beta
_{i}\left( t_{i}\right) :=(\mu _{i}^{1},...,\mu _{i}^{n})$. We say that $E$
is \textbf{certainly believed under} $\beta _{i}\left( t_{i}\right) $\ if $%
\mu _{i}^{l}\left( E\right) =1$ for all $l\leq n$. In other words, $E$\ is
certainly believed under $\beta _{i}\left( t_{i}\right) $ if its complement
is deemed subjectively impossible by the player (see Appendix A for a
preference-based foundation).

Certain belief satisfies monotonicity: If $E$ is certainly believed under $%
\beta _{i}\left( t_{i}\right) $\ and $F$ is an event such that\ $E\subseteq
F $, then $F$ is certainly believed under $\beta _{i}\left( t_{i}\right) $.
Furthermore, certain belief is \textit{not} a stronger concept than cautious
belief. To see this, consider a type $t_{i}\in T_{i}$ such that $\overline{%
\mathrm{marg}}_{S_{-i}}\beta _{i}(t_{i})\notin \mathcal{N}^{+}\left(
S_{-i}\right) $. Event $S_{-i}\times T_{-i}$ is certainly believed under $%
\beta _{i}\left( t_{i}\right) $, but it is not cautiously believed because $%
t_{i}$\ is not cautious (see Remark \ref{Remark Cautiousness is Cautious
Belief of state space}). Yet, it is immediate to check that, for cautious
types, certain belief implies cautious belief.

\section{Epistemic analysis\label{Section on the main result}}

\subsection{Epistemic analysis of IA\label{Section epistemic analysis IA}}

In what follows, we fix a finite game $G:=\left \langle I,(S_{i},\pi
_{i})_{i\in I}\right \rangle $. Given an associated type structure $\mathcal{%
T}:=\langle S_{i},T_{i},\beta _{i}\rangle _{i\in I}$, for each player $i\in
I $, we let $R_{i}^{1}:=R_{i}\cap C_{i}$ denote the set of cautiously
rational strategy-type pairs. Let $\mathbf{B}_{i}^{c}:\Sigma _{S_{-i}\times
T_{-i}}\rightarrow \Sigma _{S_{i}\times T_{i}}$\ be the operator\ defined by%
\begin{equation*}
\mathbf{B}_{i}^{c}\left( E_{-i}\right) :=\left \{ \left( s_{i},t_{i}\right)
\in S_{i}\times T_{i}:t_{i}\text{ cautiously believes }E_{-i}\right \} \text{%
, }E_{-i}\in \Sigma _{S_{-i}\times T_{-i}}\text{.}
\end{equation*}%
Corollary C.1 in Appendix C shows that the set $\mathbf{B}_{i}^{c}\left(
E_{-i}\right) $\ is Borel in $S_{i}\times T_{i}$ if $E_{-i}\subseteq
S_{-i}\times T_{-i}$\ is an event; so the operator $\mathbf{B}_{i}^{c}$ is
well-defined.

For each $m\geq 1$, define $R_{i}^{m+1}$\ recursively by%
\begin{equation*}
R_{i}^{m+1}:=R_{i}^{m}\cap \mathbf{B}_{i}^{c}\left( R_{-i}^{m}\right) \text{,%
}
\end{equation*}%
where $R_{-i}^{m}:=\prod_{j\neq i}R_{j}^{m}$. Note that%
\begin{equation*}
R_{i}^{m+1}=R_{i}^{1}\cap \left( \tbigcap_{l\leq m}\mathbf{B}_{i}^{c}\left(
R_{-i}^{l}\right) \right) \text{,}
\end{equation*}%
and each $R_{i}^{m}$\ is Borel in $S_{i}\times T_{i}$ (see Lemma C.2 in
Appendix C).

We write $R_{i}^{\infty }:=\cap _{m\in \mathbb{N}}R_{i}^{m}$ for each $i\in
I $. If $\left( s_{i},t_{i}\right) _{i\in I}\in \tprod_{i\in I}R_{i}^{m+1}$,
we say that there is \textbf{cautious rationality and} $m$\textbf{th-order
cautious belief in cautious rationality} (\textbf{R}$^{\text{c}}m$\textbf{B}$%
^{\text{c}}$\textbf{R}$^{\text{c}}$) at this state. If $\left(
s_{i},t_{i}\right) _{i\in I}\in \tprod_{i\in I}R_{i}^{\infty }$, we say that
there is \textbf{cautious rationality and} \textbf{common cautious belief in
cautious rationality} (\textbf{R}$^{\text{c}}$\textbf{CB}$^{\text{c}}$%
\textbf{R}$^{\text{c}}$) at this state.

With this, we state the first main result of this paper.

\begin{theorem}
\label{Theorem main result}Fix a type structure $\mathcal{T}^{\ast
}:=\langle S_{i},T_{i}^{\ast },\beta _{i}^{\ast }\rangle _{i\in I}$ which is
terminal with respect to the class of all finite type structures. Then:

\begin{description}
\item[(i)] for each $m\geq 1$, $\prod_{i\in I}\mathrm{Proj}_{S_{i}}\left(
R_{i}^{\ast ,m}\right) =\prod_{i\in I}S_{i}^{m}$;

\item[(ii)] $\prod_{i\in I}\mathrm{Proj}_{S_{i}}\left( R_{i}^{\ast ,\infty
}\right) =\prod_{i\in I}S_{i}^{\infty }$.
\end{description}
\end{theorem}

We point out that type structure $\mathcal{T}^{\ast }$ in Theorem \ref%
{Theorem main result} exists. In particular, there exists a \emph{universal}
type structure for LPS's, that is, a type structure which is terminal and
for which the type morphism from every other type structure is unique. Lee
(2016b) shows the existence of a universal type structure for a wide class
of preferences, which includes those represented by LPS's. Yang (2015) and
Catonini and De Vito (2018) construct the canonical type structure for
hierarchies of lexicographic beliefs;\ Catonini and De Vito also show that
this type structure is universal. Since the canonical type structure is
continuous and belief-complete, it follows from Theorem \ref{Theorem main
result} that R$^{\text{c}}$CB$^{\text{c}}$R$^{\text{c}}$ is possible in a
continuous, belief-complete type structure.

The idea of Theorem \ref{Theorem main result}\ stems from Theorem 3 in
Friedenberg and Keisler (2020), a result concerning epistemic foundations
for iterated strict dominance in the context of ordinary type structures. In
the spirit of such result, Theorem \ref{Theorem main result} identifies a
\textquotedblleft richness\textquotedblright \ property of the type
structure that depends on its ability to capture sufficiently many
hierarchies of beliefs (specifically, all those induced by finite type
structures).

The proof of Theorem \ref{Theorem main result}, like the proof of Theorem 3
in Friedenberg and Keisler (2020), is based on the following
\textquotedblleft embedding\textquotedblright \ argument. We first construct
a finite type structure $\mathcal{T}$\ such that, for every $m\geq 1$, the
behavioral implications of R$^{\text{c}}m$B$^{\text{c}}$R$^{\text{c}}$ are
characterized by the set of $m$-admissible strategy profiles. Then, by the
terminality property of $\mathcal{T}^{\ast }$, we map $\mathcal{T}$ in $%
\mathcal{T}^{\ast }$ via type morphism. While doing so, we show that
cautious rationality and all orders of belief in cautious rationality are
preserved. For this, we need the next three preparatory results. First, we
need to show the existence of $\mathcal{T}$.

\begin{lemma}
\label{Main Lemma finite type structure}There exists a finite type structure 
$\mathcal{T}:=\left \langle S_{i},T_{i},\beta _{i}\right \rangle _{i\in I}$\
such that, for each $i\in I$\ and each $m\geq 1$, $\mathrm{Proj}%
_{S_{i}}\left( R_{i}^{m}\right) =S_{i}^{m}$.
\end{lemma}

The proof of Lemma \ref{Main Lemma finite type structure}\ is in Appendix C.
Here we just mention that in the finite type structure we construct for
Lemma \ref{Main Lemma finite type structure} all types are cautious. This
fact will be used below (Section \ref{Section transparency of cautiousness}).

Second, we need to claim the invariance of cautious rationality under type
morphisms.

\begin{lemma}
\label{Lemma on invariance of cautiousness under type morphisms}Fix type
structures $\mathcal{T}:=\langle S_{i},T_{i},\beta _{i}\rangle _{i\in I}$\
and $\mathcal{T}^{\ast }:=\langle S_{i},T_{i}^{\ast },\beta _{i}^{\ast
}\rangle _{i\in I}$. Suppose that there exists a type morphism $\left(
\varphi _{i}\right) _{i\in I}:T\rightarrow T^{\ast }$ from $\mathcal{T}$\ to 
$\mathcal{T}^{\ast }$, and fix a strategy-type pair $\left(
s_{i},t_{i}\right) \in S_{i}\times T_{i}$. Then:

(i) $t_{i}$\ is cautious in $\mathcal{T}$ if and only if $\varphi _{i}\left(
t_{i}\right) $ is cautious in $\mathcal{T}^{\ast }$;

(ii) $\left( s_{i},t_{i}\right) $ is rational in $\mathcal{T}$ if and only
if $\left( s_{i},\varphi _{i}\left( t_{i}\right) \right) $ is rational in $%
\mathcal{T}^{\ast }$.
\end{lemma}

The proof of Lemma \ref{Lemma on invariance of cautiousness under type
morphisms}\ can be found in Catonini and De Vito (2020, Fact C.1).

Third, we need an analogous invariance property for cautious belief.

\begin{lemma}
\label{Lemma conjecture referee}Fix type structures $\mathcal{T}:=\langle
S_{i},T_{i},\beta _{i}\rangle _{i\in I}$\ and $\mathcal{T}^{\ast }:=\langle
S_{i},T_{i}^{\ast },\beta _{i}^{\ast }\rangle _{i\in I}$. Suppose that there
exists a bimeasurable type morphism $\left( \varphi _{i}\right) _{i\in
I}:T\rightarrow T^{\ast }$ from $\mathcal{T}$\ to $\mathcal{T}^{\ast }$. If
a type $t_{i}\in T_{i}$ cautiously believes event $E_{-i}\subseteq
S_{-i}\times T_{-i}$, then $\varphi _{i}(t_{i})$ cautiously believes $\left( 
\mathrm{Id}_{S_{-i}},\varphi _{-i}\right) \left( E_{-i}\right) $.
\end{lemma}

The proof of Lemma \ref{Lemma conjecture referee}\ is in Appendix C. For our
purpose, it is crucial to observe that, by Remark \ref{Remark:
quasi-monotonocity of assumption}, if $\varphi _{i}(t_{i})$ cautiously
believes $\left( \mathrm{Id}_{S_{-i}},\varphi _{-i}\right) \left(
E_{-i}\right) $, then $\varphi _{i}(t_{i})$ cautiously believes also every
Borel superset $E_{-i}^{\ast }$ such that $\mathrm{Proj}_{S_{-i}}\left(
E_{-i}^{\ast }\right) =\mathrm{Proj}_{S_{-i}}\left( \left( \mathrm{Id}%
_{S_{-i}},\varphi _{-i}\right) \left( E_{-i}\right) \right) $.

Finally, for the proof of Theorem \ref{Theorem main result}, we find it
convenient to single out the following fact, whose proof is immediate.

\begin{remark}
\label{Remark:same projection of a set under type morphism}Fix type
structures $\mathcal{T}:=\langle S_{i},T_{i},\beta _{i}\rangle _{i\in I}$\
and $\mathcal{T}^{\ast }:=\langle S_{i},T_{i}^{\ast },\beta _{i}^{\ast
}\rangle _{i\in I}$. Suppose that there exists a type morphism $\left(
\varphi _{i}\right) _{i\in I}:T\rightarrow T^{\ast }$ from $\mathcal{T}$\ to 
$\mathcal{T}^{\ast }$. Then, for every $E_{i}\subseteq S_{i}\times T_{i}$,%
\begin{equation*}
\mathrm{Proj}_{S_{i}}\left( \left( \mathrm{Id}_{S_{i}},\varphi _{i}\right)
(E_{i})\right) =\mathrm{Proj}_{S_{i}}(E_{i})\text{.}
\end{equation*}
\end{remark}

With this, we are ready to prove Theorem \ref{Theorem main result}.

\bigskip

\noindent \textbf{Proof of Theorem \ref{Theorem main result}}. By Lemma \ref%
{Main Lemma finite type structure}, there is a finite type structure $%
\mathcal{T}:=\langle S_{i},T_{i},\beta _{i}\rangle _{i\in I}$ such that $%
\mathrm{Proj}_{S_{i}}\left( R_{i}^{m}\right) =S_{i}^{m}$ for each $m\geq 1$
and for each $i\in I$.

\textbf{Part (i)}: Fix a type morphism $(\varphi _{i})_{i\in I}:T\rightarrow
T^{\ast }$ from $\mathcal{T}$ to $\mathcal{T}^{\ast }$. Structure $\mathcal{T%
}$\ is finite, so $(\varphi _{i})_{i\in I}$\ is bimeasurable. We show by
induction on $m\geq 1$ that $\left( \mathrm{Id}_{S_{i}},\varphi _{i}\right)
(R_{i}^{m})\subseteq R_{i}^{\ast ,m}$ and $\mathrm{Proj}_{S_{i}}\left(
R_{i}^{\ast ,m}\right) =S_{i}^{m}$ for each $i\in I$.

($m=1$) Fix $i\in I$. It is immediate from Lemma \ref{Lemma on invariance of
cautiousness under type morphisms} that $\left( \mathrm{Id}_{S_{i}},\varphi
_{i}\right) (R_{i}^{1})\subseteq R_{i}^{\ast ,1}$. By Remark \ref%
{Remark:same projection of a set under type morphism}, $\mathrm{Proj}%
_{S_{i}}\left( \left( \mathrm{Id}_{S_{i}},\varphi _{i}\right)
(R_{i}^{1})\right) =\mathrm{Proj}_{S_{i}}\left( R_{i}^{1}\right) $, and
since $\mathrm{Proj}_{S_{i}}\left( R_{i}^{1}\right) =S_{i}^{1}$, we obtain $%
S_{i}^{1}\subseteq \mathrm{Proj}_{S_{i}}\left( R_{i}^{\ast ,1}\right) $.
Conversely, Proposition \ref{Proposition CR implies admissibility} entails $%
\mathrm{Proj}_{S_{i}}\left( R_{i}^{\ast ,1}\right) \subseteq S_{i}^{1}$.
Therefore, $\mathrm{Proj}_{S_{i}}\left( R_{i}^{\ast ,1}\right) =S_{i}^{1}$.

($m>1$) Fix $i\in I$ and $(s_{i},t_{i})\in R_{i}^{m}$. We want to show that $%
(s_{i},\varphi _{i}(t_{i}))\in R_{i}^{\ast ,m}$. Since $R_{i}^{m}\subseteq
R_{i}^{m-1}$, the induction hypothesis yields $(s_{i},\varphi
_{i}(t_{i}))\in R_{i}^{\ast ,m-1}$. Hence, it suffices to show that $\varphi
_{i}(t_{i})$ cautiously believes $R_{-i}^{\ast ,m-1}$. Since $t_{i}$
cautiously believes $R_{-i}^{m-1}$ and the type morphism $(\varphi
_{i})_{i\in I}$ is bimeasurable, it follows from Lemma \ref{Lemma conjecture
referee} that $\varphi _{i}(t_{i})$ cautiously believes $(\mathrm{Id}%
_{S_{-i}},\varphi _{-i})(R_{-i}^{m-1})$. Note that%
\begin{eqnarray*}
\mathrm{Proj}_{S_{-i}}\left( R_{-i}^{\ast ,m-1}\right) &=&S_{-i}^{m-1} \\
&=&\mathrm{Proj}_{S_{-i}}\left( R_{-i}^{m-1}\right) \\
&=&\mathrm{Proj}_{S_{-i}}\left( (\mathrm{Id}_{S_{-i}},\varphi
_{-i})(R_{-i}^{m-1})\right) \text{,}
\end{eqnarray*}%
where the first equality is the induction hypothesis, the second equality
follows from the property of $\mathcal{T}$, and the third equality follows
from Remark \ref{Remark:same projection of a set under type morphism}. We
also know from the induction hypothesis that $\left( \mathrm{Id}%
_{S_{-i}},\varphi _{-i}\right) (R_{-i}^{m-1})\subseteq R_{-i}^{\ast ,m-1}$;
thus, Remark \ref{Remark: quasi-monotonocity of assumption} allows us to
conclude that $\varphi _{i}(t_{i})$ cautiously believes $R_{-i}^{\ast ,m-1}$.

So, we have shown that $\left( \mathrm{Id}_{S_{i}},\varphi _{i}\right)
(R_{i}^{m})\subseteq R_{i}^{\ast ,m}$. By the property of $\mathcal{T}$\ and
Remark \ref{Remark:same projection of a set under type morphism}, we obtain%
\begin{eqnarray*}
S_{i}^{m} &=&\mathrm{Proj}_{S_{i}}\left( R_{i}^{m}\right) \\
&=&\mathrm{Proj}_{S_{i}}\left( \left( \mathrm{Id}_{S_{i}},\varphi
_{i}\right) (R_{i}^{m})\right) \\
&\subseteq &\mathrm{Proj}_{S_{i}}\left( R_{i}^{\ast ,m}\right) \text{.}
\end{eqnarray*}%
To show the opposite inclusion, fix $(s_{i},t_{i}^{\ast })\in R_{i}^{\ast
,m} $.\ Since $R_{i}^{\ast ,m}\subseteq R_{i}^{\ast ,m-1}$, it follows from
the induction hypothesis that $s_{i}\in S_{i}^{m-1}$. Let $\beta _{i}^{\ast
}(t_{i}^{\ast }):=\left( \mu _{i}^{1},...,\mu _{i}^{n}\right) $. Since $%
t_{i}^{\ast }$ cautiously believes $R_{-i}^{\ast ,m-1}$ at some level $l$,
it follows from Proposition \ref{Lemma on properties of assumption one
direction conjunction}.2 and the induction hypothesis that%
\begin{equation*}
\tbigcup_{k\leq l}\mathrm{Suppmarg}_{S_{-i}}\mu _{i}^{k}=S_{-i}^{m-1}\text{.}
\end{equation*}%
So, by Proposition 1 in Blume et al. (1991b), we can form a nested convex
combination of the measures $\mathrm{marg}_{S_{-i}}\mu _{i}^{k}$, for $%
k=1,...,l$, to get a probability measure $\nu _{i}\in \mathcal{M}\left(
S_{-i}\right) $, with $\mathrm{Supp}\nu _{i}=S_{-i}^{m-1}$, under which $%
s_{i}$ is optimal. Thus, by Remark \ref{Remark second Pearce lemma}, $s_{i}$%
\ is admissible with respect to $S_{i}\times S_{-i}^{m-1}$, and a fortiori
with respect to $S_{i}^{m-1}\times S_{-i}^{m-1}$. Hence, $s_{i}\in S_{i}^{m}$%
.

\textbf{Part (ii)}: Fix $i\in I$. Since $(R_{i}^{m})_{m\in \mathbb{N}}$ and $%
(S_{i}^{m})_{m\in \mathbb{N}}$ are weakly decreasing sequences of finite,
non-empty sets, there exists $N\in \mathbb{N}$ such that $%
R_{i}^{N}=R_{i}^{\infty }$ and $S_{i}^{N}=S_{i}^{\infty }$. Lemma \ref{Main
Lemma finite type structure} implies $\mathrm{Proj}_{S_{i}}\left(
R_{i}^{\infty }\right) =S_{i}^{\infty }$. Hence, for every $s_{i}\in
S_{i}^{\infty }$, there exists $t_{i}\in T_{i}$ such that $(s_{i},t_{i})\in
R_{i}^{m}$ for every $m\in \mathbb{N}$. We have shown in the proof of Part
(i) that, for every $m\in \mathbb{N}$, $\left( \mathrm{Id}_{S_{i}},\varphi
_{i}\right) (R_{i}^{m})\subseteq R_{i}^{\ast ,m}$. So $\left( \mathrm{Id}%
_{S_{i}},\varphi _{i}\right) ((s_{i},t_{i}))\in R_{i}^{\ast ,m}$ for every $%
m\in \mathbb{N}$, which implies that $\left( \mathrm{Id}_{S_{i}},\varphi
_{i}\right) ((s_{i},t_{i}))\in R_{i}^{\ast ,\infty }$. Therefore, $%
S_{i}^{\infty }\subseteq \mathrm{Proj}_{S_{i}}\left( R_{i}^{\ast ,\infty
}\right) $. Conversely, Part (i) entails $\mathrm{Proj}_{S_{i}}\left(
R_{i}^{\ast ,N}\right) =S_{i}^{N}=S_{i}^{\infty }$.$\ $Hence $\mathrm{Proj}%
_{S_{i}}\left( R_{i}^{\ast ,\infty }\right) \subseteq S_{i}^{\infty }$. We
conclude that $S_{i}^{\infty }=\mathrm{Proj}_{S_{i}}\left( R_{i}^{\ast
,\infty }\right) $.\hfill $\blacksquare $

\subsection{Epistemic analysis of SAS's\label{Section SAS}}

Fix a finite game $G:=\left \langle I,(S_{i},\pi _{i})_{i\in
I}\right
\rangle $. The following result states that, for every type
structure associated with game $G$, the behavioral implications of R$^{\text{%
c}}$CB$^{\text{c}}$R$^{\text{c}}$ constitute an SAS. Conversely, every SAS
corresponds to the behavioral implications of R$^{\text{c}}$CB$^{\text{c}}$R$%
^{\text{c}}$ in some type structure.

\begin{theorem}
\label{Theorem SAS}

\begin{description}
\item[(i)] Fix a type structure $\mathcal{T}:=\langle S_{i},T_{i},\beta
_{i}\rangle _{i\in I}$. Then $\tprod_{i\in I}\mathrm{Proj}_{S_{i}}\left(
R_{i}^{\infty }\right) $\ is an SAS.

\item[(ii)] Fix an SAS $Q\in \mathcal{Q}$. There exists a finite type
structure $\mathcal{T}:=\langle S_{i},T_{i},\beta _{i}\rangle _{i\in I}$\
such that, for each $i\in I$,%
\begin{equation*}
\mathrm{Proj}_{S_{i}}\left( R_{i}^{\infty }\right) =Q_{i}\text{.}
\end{equation*}
\end{description}
\end{theorem}

The idea of Theorem \ref{Theorem SAS}\ stems from Theorem 8.1 in BFK, a
result concerning the epistemic justification of SAS's with rationality and
common assumption of rationality. The proof of Theorem \ref{Theorem SAS},
which is similar to that in BFK, is in Appendix C. Here we point out that in
the finite type structure we construct for Theorem \ref{Theorem SAS}.(ii) 
\textit{all} types are cautious. We have already mentioned (Section \ref%
{Section epistemic analysis IA}) that the same property holds for the finite
type structure we construct for Lemma \ref{Main Lemma finite type structure}%
. This raises the question whether one could incorporate the cautiousness
assumption in the definition of type structures for the epistemic analysis
of IA and SAS's. That is, could we restrict attention to the class $\mathbb{T%
}$\ of type structures where each type's belief over the opponents'
strategies have full support? Are there analogues of Theorems \ref{Theorem
main result}\ and \ref{Theorem SAS}\ for such a class? We explore this issue
in the next section.

\subsection{Alternative epistemic conditions for IA and SAS's \label{Section
transparency of cautiousness}}

Fix a finite game $G:=\left \langle I,(S_{i},\pi _{i})_{i\in
I}\right
\rangle $ and an associated type structure $\mathcal{T}:=\langle
S_{i},T_{i},\beta _{i}\rangle _{i\in I}$. Say that \textbf{type} $t_{i}$\ 
\textbf{certainly believes} a non-empty event $E_{-i}\subseteq S_{-i}\times
T_{-i}$\ if $E_{-i} $\ is certainly believed under $\beta _{i}(t_{i})$. For
each player $i\in I$, let $\mathbf{B}_{i}:\Sigma _{S_{-i}\times
T_{-i}}\rightarrow \Sigma _{S_{i}\times T_{i}}$\ be the operator\ defined by%
\begin{equation*}
\mathbf{B}_{i}\left( E_{-i}\right) :=\left \{ \left( s_{i},t_{i}\right) \in
S_{i}\times T_{i}:t_{i}\text{ certainly believes }E_{-i}\right \} \text{, }%
E_{-i}\in \Sigma _{S_{-i}\times T_{-i}}\text{.}
\end{equation*}%
As shown in Catonini and De Vito (2020), the set $\mathbf{B}_{i}\left(
E_{-i}\right) $\ is Borel in $S_{i}\times T_{i}$ if $E_{-i}\subseteq
S_{-i}\times T_{-i}$\ is an event; thus, the operator $\mathbf{B}_{i}$ is
well-defined.

Next, fix a non-empty event $E_{i}\subseteq S_{i}\times T_{i}$\ for every $%
i\in I$. Event $E:=\tprod_{i\in I}E_{i}$ is \textbf{self-evident} (in $%
\mathcal{T}$) if it satisfies $E\subseteq \tprod_{i\in I}\mathbf{B}%
_{i}\left( E_{-i}\right) $; standard results (e.g., Catonini and De Vito
2018, Appendix 5.2) show that $E$\ is self-evident if and only if at every
state $\left( s_{i},t_{i}\right) _{i\in I}\in E$ there is common certain
belief in $E$. With this, we say that there is \textbf{transparency of }$E$
at state $\left( s_{i},t_{i}\right) _{i\in I}$ if $\left( s_{i},t_{i}\right)
_{i\in I}\in E$ and $E$ is self-evident.

We let $C^{\infty }\subseteq \tprod \nolimits_{i\in I}S_{i}\times T_{i}$
denote the event corresponding to transparency of cautiousness in $\mathcal{T%
}$, and we let $C_{i}^{\infty }$\ denote the corresponding projection on $%
S_{i}\times T_{i}$.\footnote{%
Since the set $C_{i}$\ is a Borel subset of $S_{i}\times T_{i}$ (Catonini
and De Vito 2020, Corollary D.1), an argument by induction on the iteration
of the operator $\mathbf{B}_{i}$\ shows that $C_{i}^{\infty }$\ is also a
Borel subset (event) of $S_{i}\times T_{i}$. In particular, if $\mathcal{T}$%
\ is the canonical type structure, then it is possible to show that $%
C_{i}^{\infty }$\ is a Polish subset of $S_{i}\times T_{i}$.} With this, for
each $i\in I$, let $\hat{R}_{i}^{1}:=R_{i}\cap C_{i}^{\infty }$\textbf{.}
Then, for each $i\in I$\ and $m\geq 1$, define $\hat{R}_{i}^{m+1}$\
recursively by%
\begin{equation*}
\hat{R}_{i}^{m+1}:=\hat{R}_{i}^{m}\cap \mathbf{B}_{i}^{c}\left( \hat{R}%
_{-i}^{m}\right) \text{,}
\end{equation*}%
where $\hat{R}_{-i}^{m}:=\tprod_{j\neq i}\hat{R}_{j}^{m}$. Write $\hat{R}%
_{i}^{\infty }:=\cap _{m\in \mathbb{N}}\hat{R}_{i}^{m}$ for each $i\in I$.
Therefore, if $\left( s_{i},t_{i}\right) _{i\in I}\in \tprod_{i\in I}\hat{R}%
_{i}^{\infty }$, we say that at state $\left( s_{i},t_{i}\right) _{i\in I}$
there is (a) \textbf{rationality and} \textbf{transparency of cautiousness},
and (b) \textbf{common cautious belief in (a)}.

Say that $\mathcal{T}:=\langle S_{i},T_{i},\beta _{i}\rangle _{i\in I}$\ is
a \textbf{cautious type structure} if $C_{i}=S_{i}\times T_{i}$\ for every $%
i\in I$. Since event $S\times T:=\tprod \nolimits_{i\in I}\left( S_{i}\times
T_{i}\right) $\ is self-evident, it is easily seen that $C^{\infty }=S\times
T$ in a cautious type structure. Therefore, if $\mathcal{T}$ is a cautious
type structure, then, for each $i\in I$,%
\begin{equation*}
\hat{R}_{i}^{1}=R_{i}^{1}=R_{i}\text{;}
\end{equation*}%
that is, event $\tprod_{i\in I}\hat{R}_{i}^{1}$ is the set of states where
there is rationality. It follows by induction on $m\geq 1$ that $\hat{R}%
_{i}^{m}=R_{i}^{m}$ for each $i\in I$. So, if $\left( s_{i},t_{i}\right)
_{i\in I}\in \tprod_{i\in I}\hat{R}_{i}^{m+1}$, we can say that there is 
\textbf{rationality and} $m$\textbf{th-order cautious belief in rationality}
at this state.

We are now in a position to state a characterization result, which is an
analogue of Theorem \ref{Theorem main result}.

\begin{theorem}
\label{Theorem (alternative) main result}

\begin{description}
\item[(i)] Fix a type structure $\mathcal{T}^{\ast }:=\langle
S_{i},T_{i}^{\ast },\beta _{i}^{\ast }\rangle _{i\in I}$ which is terminal
with respect to the class of all finite type structures. Then:

\begin{description}
\item[(i.1)] for each $m\geq 1$, $\prod_{i\in I}\mathrm{Proj}_{S_{i}}\left( 
\hat{R}_{i}^{\ast ,m}\right) =\tprod_{i\in I}S_{i}^{m}$;

\item[(i.2)] $\prod_{i\in I}\mathrm{Proj}_{S_{i}}\left( \hat{R}_{i}^{\ast
,\infty }\right) =\tprod_{i\in I}S_{i}^{\infty }$.
\end{description}

\item[(ii)] The same conclusions as in (i.1) and (i.2) hold if $\mathcal{T}%
^{\ast }$\ is a cautious type structure which is terminal with respect to
the class of all finite, cautious type structures.
\end{description}
\end{theorem}

To see why Part (i) of Theorem \ref{Theorem (alternative) main result}
holds, let $\mathcal{T}:=\langle S_{i},T_{i},\beta _{i}\rangle _{i\in I}$\
be the finite type structure we construct for the proof of Lemma \ref{Main
Lemma finite type structure}. Since $\mathcal{T}$\ is a cautious type
structure, $\hat{R}_{i}^{m}=R_{i}^{m}$ for each $i\in I$\ and $m\geq 1$.
Finiteness of $\mathcal{T}$ guarantees that the type morphism $\varphi $
from $\mathcal{T}$\ to $\mathcal{T}^{\ast }$ is bimeasurable. Lemma \ref%
{Lemma on invariance of cautiousness under type morphisms}.(i) entails that $%
S\times \varphi (T)\subseteq C^{\ast }$; that is, the image of self-evident
event $C^{\infty }$ in $\mathcal{T}$ under type morphism $\varphi $ is a
subset of $C^{\ast }$. Proposition 12 in Catonini and De Vito (2018) shows
that event $S\times \varphi (T)$\ is self-evident in $\mathcal{T}^{\ast }$.%
\footnote{%
Lemma A3 in Battigalli and Friedenberg (2012b) shows an analogous result for
type structures with beliefs represented by conditional probability systems.}
But then, by the monotonicity property of certain belief, it is immediate to
see that at every state in $S\times \varphi (T)$\ there is common certain
belief in $C^{\ast }$. We therefore conclude that $S\times \varphi
(T)\subseteq C^{\ast ,\infty }$. With this, the proof of Theorem \ref%
{Theorem (alternative) main result}\ is the same as that of Theorem \ref%
{Theorem main result}---just replace sets such as $R_{i}^{\ast ,m}$ and $%
R_{-i}^{\ast ,m}$ with the corresponding sets $\hat{R}_{i}^{\ast ,m}$ and $%
\hat{R}_{-i}^{\ast ,m}$.

Theorem \ref{Theorem (alternative) main result}.(ii) is a characterization
result for $m$-admissible (resp. iteratively admissible) strategies in terms
of rationality and $m$th-order (resp. common) cautious belief in rationality.%
\textit{\ }To see why Theorem \ref{Theorem (alternative) main result}.(ii)
holds, we first point out that a terminal, cautious type structure exists.
Let $\mathcal{T}^{U}$\ be the canonical, universal type structure. By
Proposition 12 in Catonini and De Vito (2018), the self-evident event $%
C^{U,\infty }$ identifies a \textquotedblleft smaller,\textquotedblright \
cautious\ type structure $\mathcal{T}^{\ast }$. In such a structure, $\hat{R}%
_{i}^{\ast ,m}=R_{i}^{\ast ,m}$ for each $i\in I$\ and $m\geq 1$. By the
above argument, $\mathcal{T}^{\ast }$ is terminal with respect to the class
of finite, cautious type structures. With this, the proof of Theorem \ref%
{Theorem main result} yields the result.

The following characterization result, pertaining to SAS's, is an analogue
of Theorem \ref{Theorem SAS}.

\begin{theorem}
\label{Theorem SAS transp cautious}

\begin{description}
\item[(i)] Fix a type structure $\mathcal{T}:=\langle S_{i},T_{i},\beta
_{i}\rangle _{i\in I}$. Then $\tprod_{i\in I}\mathrm{Proj}_{S_{i}}\left( 
\hat{R}_{i}^{\infty }\right) $\ is an SAS.

\item[(ii)] Fix an SAS $Q\in \mathcal{Q}$. There exists a finite, cautious
type structure $\mathcal{T}:=\langle S_{i},T_{i},\beta _{i}\rangle _{i\in I}$%
\ such that, for each $i\in I$,%
\begin{equation*}
\mathrm{Proj}_{S_{i}}\left( \hat{R}_{i}^{\infty }\right) =Q_{i}\text{.}
\end{equation*}
\end{description}
\end{theorem}

The proof of Theorem \ref{Theorem SAS transp cautious}.(i) is, with some
minor modifications, identical to the proof of Theorem \ref{Theorem SAS}%
.(i). Part (ii) of Theorem \ref{Theorem SAS transp cautious} is essentially
Theorem \ref{Theorem SAS}.(ii). As noted, in the proof of Theorem \ref%
{Theorem SAS}.(ii) we construct a finite, cautious type structure $\mathcal{T%
}$\ such that $\prod_{i\in I}\mathrm{Proj}_{S_{i}}\left( R_{i}^{\infty
}\right) =Q$. In such a type structure, $\hat{R}_{i}^{\infty }=R_{i}^{\infty
}$ for each $i\in I$.

Theorem \ref{Theorem SAS transp cautious} allows us to obtain an alternative
characterization of SAS's. Specifically, if we restrict attention to the
class of cautious type structures, we can restate the results of Theorem \ref%
{Theorem SAS transp cautious} as follows: \textit{SAS's characterize the
behavioral implications of rationality and common cautious belief in
rationality across all cautious type structures}.

Some final remarks on the epistemic assumptions considered so far are in
order. First, events $\prod_{i\in I}R_{i}^{\infty }$\ and $\prod_{i\in I}%
\hat{R}_{i}^{\infty }$ are equivalent in cautious type structures, but they
are typically different in non-cautious type structures. In the
Supplementary Appendix we exhibit an example where the game in Example \ref%
{Example: IA SAS BoSOO}\ is associated with a finite, \textit{non-cautious}
type structure $\mathcal{T}$\ such that $\prod_{i\in I}R_{i}^{\infty }$\ and 
$\prod_{i\in I}\hat{R}_{i}^{\infty }$ are disjoint. The example shows that
the behavioral implications of such epistemic assumptions are characterized
by two \textit{disjoint} SAS's---specifically, the sets $\left \{ m\right \}
\times \left \{ \ell \right \} $\ and $\left \{ u\right \} \times \left \{
r\right \} $, respectively.

To understand this point, fix an arbitrary type structure $\mathcal{T}$\
associated with a\ finite game. Let us consider the first two steps in the
definitions of $\prod_{i\in I}R_{i}^{\infty }$\ and $\prod_{i\in I}\hat{R}%
_{i}^{\infty }$. At the first step, we have $\hat{R}_{i}^{1}\subseteq
R_{i}^{1}$\ for each $i\in I$, because transparency of cautiousness implies
cautiousness. Does an analogous conclusion hold for the sets $\hat{R}%
_{i}^{2} $\ and $R_{i}^{2}$ ($i\in I$)? The answer is no. Recall that, for
each $i\in I$,%
\begin{eqnarray*}
\hat{R}_{i}^{2} &:&=\hat{R}_{i}^{1}\cap \mathbf{B}_{i}^{c}\left( \hat{R}%
_{-i}^{1}\right) \text{,} \\
R_{i}^{2} &:&=R_{i}^{1}\cap \mathbf{B}_{i}^{c}\left( R_{-i}^{1}\right) \text{%
.}
\end{eqnarray*}%
Since cautious belief is \textit{not} monotonic (see Section \ref{Section
IML and CB}), for some player $i\in I$ we can have $\mathbf{B}_{i}^{c}\left( 
\hat{R}_{-i}^{1}\right) \nsubseteq \mathbf{B}_{i}^{c}\left(
R_{-i}^{1}\right) $\ even if $\hat{R}_{-i}^{1}\subseteq R_{-i}^{1}$; this is
illustrated by the example in the Supplementary Appendix. Yet, it should be
noted that $\prod_{i\in I}\hat{R}_{i}^{2}\subseteq \prod_{i\in I}R_{i}^{2}$
holds whenever $\prod_{i\in I}\mathrm{Proj}_{S_{i}}\left( \hat{R}%
_{i}^{1}\right) =\prod_{i\in I}\mathrm{Proj}_{S_{i}}\left( R_{i}^{1}\right) $%
. Indeed, under such condition, the \textquotedblleft
quasi-monotonicity\textquotedblright \ property of cautious belief (Remark %
\ref{Remark: quasi-monotonocity of assumption}) entails that $\mathbf{B}%
_{i}^{c}\left( \hat{R}_{-i}^{1}\right) \subseteq \mathbf{B}_{i}^{c}\left(
R_{-i}^{1}\right) $\ for each $i\in I$.

In the Supplementary Appendix we show that, if $\mathcal{T}$\ is a
\textquotedblleft rich\textquotedblright \ type structure, then%
\begin{equation*}
\tprod_{i\in I}\hat{R}_{i}^{\infty }\subseteq \tprod_{i\in I}R_{i}^{\infty }%
\text{ \  \ and \  \ }\tprod_{i\in I}\mathrm{Proj}_{S_{i}}\left( \hat{R}%
_{i}^{\infty }\right) =\tprod_{i\in I}\mathrm{Proj}_{S_{i}}\left(
R_{i}^{\infty }\right) \text{.}
\end{equation*}%
In words, if structure $\mathcal{T}$ is \textquotedblleft
rich,\textquotedblright \ then the epistemic assumption of \textquotedblleft
rationality, transparency of cautiousness, and common cautious belief in
both\textquotedblright \ is stronger than R$^{\text{c}}$CB$^{\text{c}}$R$^{%
\text{c}}$; nonetheless, they are equivalent in terms of behavioral
implications.

As one should expect, examples of \textquotedblleft rich\textquotedblright \
type structures satisfying the aforementioned property are terminal
structures such as $\mathcal{T}^{\ast }$\ in the statement of Theorem \ref%
{Theorem main result}. Other examples are belief-complete type structures
(Definition \ref{Definition complete type structure}): this is formally
shown in the Supplementary Appendix, whose proof makes explicit use of the
\textquotedblleft quasi-monotonicity\textquotedblright \ property of
cautious belief. Thus, if $\mathcal{T}$\ is belief-complete, we can
conclude, by Theorem \ref{Theorem SAS}, that the behavioral implications of
both epistemic assumptions are characterized by one specific SAS. But this
SAS could be different from the IA set, as the following section illustrates.

\subsection{Belief-completeness vs terminality\label{Discussion subsection
completeness terminality}}

Theorem \ref{Theorem main result} identifies a \textquotedblleft
richness\textquotedblright \ condition on type structures for the epistemic
justification of IA. A related \textquotedblleft richness\textquotedblright
\ condition is belief-completeness (Definition \ref{Definition complete type
structure}): a belief-complete type structure induces all possible beliefs
about types. In light of this, one might conjecture that the conclusions of
Theorem \ref{Theorem main result} continue to hold for belief-complete type
structures. However, this is not the case.

Call a finite game $G:=\left \langle I,(S_{i},\pi _{i})_{i\in
I}\right
\rangle $\  \textbf{non-degenerate} if $\left \vert
S_{i}\right
\vert \geq 2$\ for each $i\in I$. The following result states
that, for each non-degenerate, finite game, there exists a continuous,
belief-complete type structure where R$^{\text{c}}$CB$^{\text{c}}$R$^{\text{c%
}}$ is not possible at any state. It follows that, in such a structure, the
behavioral implications of R$^{\text{c}}$CB$^{\text{c}}$R$^{\text{c}}$
constitute the empty SAS (cf. Theorem \ref{Theorem SAS}.(i)).

\begin{theorem}
\label{Theorem insufficiency belief-completeness}\textit{Fix a
non-degenerate finite game }$G:=\left \langle I,(S_{i},\pi _{i})_{i\in
I}\right \rangle $. \textit{There exists a continuous, belief-complete type
structure }$\mathcal{T}:=\langle S_{i},T_{i},\beta _{i}\rangle _{i\in I}$%
\textit{\ such that}%
\begin{equation*}
\tprod \nolimits_{i\in I}R_{i}^{\infty }=\emptyset \text{.}
\end{equation*}
\end{theorem}

Theorem \ref{Theorem insufficiency belief-completeness} is inspired by
Friedenberg and Keisler (2021, Theorem 1). Specifically, Friedenberg and
Keisler consider finite games satisfying a non-triviality condition that is
stronger than non-degeneracy. With this, they show the existence of a
belief-complete, ordinary type structure in which there is no state
consistent with rationality and common belief in rationality.

The proof of Theorem \ref{Theorem insufficiency belief-completeness} adapts
the arguments in Friedenberg and Keisler (2021) to the lexicographic
framework, and it can be found in Appendix C. Here, we briefly explain 
\textit{why} Theorem \ref{Theorem insufficiency belief-completeness}\ holds.

Fix a belief-complete type structure $\mathcal{T}$. In such a structure, $%
\prod_{i\in I}R_{i}^{m}\neq \emptyset $ for each $m\geq 1$. In particular,
it can be shown that, for each $m\geq 1$, the behavioral implications of R$^{%
\text{c}}m$B$^{\text{c}}$R$^{\text{c}}$ are characterized by the set of $m$%
-admissible strategy profiles. The reason why the set of states $\prod_{i\in
I}R_{i}^{\infty }$\ can be empty is---conceptually---the same as in
Friedenberg and Keisler (2021): While a belief-complete (lexicographic) type
structure induces all beliefs about types, it need not induce all possible
hierarchies of beliefs. Specifically, Theorems \ref{Theorem main result} and %
\ref{Theorem insufficiency belief-completeness} imply that a belief-complete
type structure may not induce all hierarchies of beliefs that can arise in
finite type structures.

In the context of ordinary type structures, Friedenberg (2010, Theorem 3.1)\
shows that a belief-complete type structure is terminal if each type space
is compact and each belief map is continuous. An analogue of Friedenberg's
result\ does not exist in the lexicographic framework: as already remarked
(see Section \ref{Section on LPS type structures}), a belief-complete,
lexicographic type structure cannot be compact and continuous. This explains
why the terminality property of the type structure is made explicit in the
statements (and proofs) of our results (Theorems \ref{Theorem main result}
and \ref{Theorem (alternative) main result}).

\section{Discussion\label{Section: Discussion}}

Given the previous analysis, it is now possible to discuss in detail a set
of conceptual issues, some of which were informally addressed in the
Introduction.

\subsection{IA and lexicographic rationalizability\label{Section: Discussion
lexicographic rationalizability}}

In the Introduction, we have informally claimed that R$^{\text{c}}$CB$^{%
\text{c}}$R$^{\text{c}}$ matches very closely the logic of lexicographic
rationalizability (Stahl 1995), an iterated elimination procedure for
lexicographic beliefs. Here we make this informal claim precise.

Fix a player $i\in I$ and a non-empty set $Q_{-i}\subseteq S_{-i}$. We let $%
r_{i}(\bar{\mu}_{i})$ denote the set of player $i$'s strategies which are
optimal under $\bar{\mu}_{i}\in \mathcal{N}(S_{-i})$, and%
\begin{equation*}
\mathcal{B}_{c}^{+}(Q_{-i}):=\left \{ \left( \mu _{i}^{1},...,\mu
_{i}^{n}\right) \in \mathcal{N}^{+}(S_{-i}):\exists m\leq
n,\tbigcup_{l=1}^{m}\mathrm{Supp}\mu _{i}^{l}=Q_{-i}\right \}
\end{equation*}%
is the set of all full-support LPS's $\bar{\mu}_{i}$\ such that $Q_{-i}$ is
\textquotedblleft cautiously believed\textquotedblright \ under $\bar{\mu}%
_{i} $ (cf. Proposition \ref{Lemma on properties of assumption one direction
conjunction}.2). In words, $\mathcal{B}_{c}^{+}(Q_{-i})$ is the set of all
full-support first-order beliefs under which player $i$\ deems \emph{every}
co-players' strategy profile in $Q_{-i}$\ infinitely more likely---in the
sense of Lo (1999) or Stahl (1995)---than every profile in $S_{-i}\backslash
Q_{-i}$. Note: if $Q_{-i}^{\prime }\subseteq S_{-i}$\ is such that $%
Q_{-i}\subseteq Q_{-i}^{\prime }$, then it is not necessarily true that $%
\mathcal{B}_{c}^{+}(Q_{-i})\subseteq \mathcal{B}_{c}^{+}(Q_{-i}^{\prime })$.
Specifically, $\mathcal{B}_{c}^{+}(Q_{-i})\nsubseteq \mathcal{B}%
_{c}^{+}(Q_{-i}^{\prime })$ whenever there exists $s_{-i}\in Q_{-i}^{\prime
}\backslash Q_{-i}$ which is not deemed---under some $\bar{\mu}_{i}\in 
\mathcal{B}_{c}^{+}(Q_{-i})$---infinitely more likely than every profile in $%
S_{-i}\backslash Q_{-i}^{\prime }$ (cf. Section \ref{Section IML and CB} and
Example \ref{Example: cautious belief not monotone}). Furthermore, note that 
$\mathcal{B}_{c}^{+}(S_{-i})=\mathcal{N}^{+}(S_{-i})$.

With this, we can define a sequence $(\hat{S}^{m})_{m\geq 0}$\ of subsets of 
$S$ as follows. For each $i\in I$, let $\hat{S}_{i}^{0}:=S_{i}$. Also, let $%
\hat{S}^{0}:=\tprod_{i\in I}\hat{S}_{i}^{0}$ and $\hat{S}_{-i}^{0}:=%
\tprod_{j\neq i}\hat{S}_{j}^{0}$.\ Recursively define, for $m\geq 1$,%
\begin{eqnarray*}
\hat{S}_{i}^{m} &:&=\left \{ s_{i}\in S_{i}:\exists \bar{\mu}_{i}\in
\tbigcap_{l=0}^{m-1}\mathcal{B}_{c}^{+}(\hat{S}_{-i}^{l}),s_{i}\in r_{i}(%
\bar{\mu}_{i})\right \} \text{,} \\
\hat{S}^{m} &:&=\tprod_{i\in I}\hat{S}_{i}^{m}\text{.}
\end{eqnarray*}%
Profiles in $\hat{S}^{\infty }:=\cap _{m=0}^{\infty }\hat{S}^{m}$\ are
called lexicographic rationalizable. Stahl (1995) shows that $S^{m}=\hat{S}%
^{m}$\ for every $m\in \mathbb{N}$, that is, the set of $m$-admissible
strategies coincides with the set of strategies surviving the first $m$
steps of the procedure.

Similarly, SAS's can be given a characterization in terms of justifiability
(\textquotedblleft best reply to some belief\textquotedblright ). As shown
by De Vito (2023), a set $Q\in \mathcal{Q}$ is an SAS if and only if, for
each $i\in I$ and each $s_{i}\in Q_{i}$, there exists $\bar{\mu}_{i}\in 
\mathcal{B}_{c}^{+}(Q_{-i})$\ such that $s_{i}\in r_{i}(\bar{\mu}_{i})$ and $%
r_{i}(\bar{\mu}_{i})\subseteq Q_{i}$. Such characterization of SAS is
similar to the definition of \textquotedblleft extensive-form best response
set\textquotedblright \ (EFBRS, Battigalli and Friedenberg 2012a). Of
course, SAS's and EFBRS's are distinct concepts. In particular, EFBRS's are
defined in terms of conditional probability systems and strong belief
(Battigalli and Siniscalchi 2002).\footnote{%
The recent paper by Brandenburger et al. (2023) establishes a relationship
between strong belief and BFK's notion of assumption.} Examples of the
difference between SAS's and EFBRS's can be found in Battigalli and
Friedenberg (2012b, Section 8.c).

\subsection{Comparison to (weak) assumption\label{Section: Discussion weak
assumption}}

In Catonini (2013) and Yang (2015), an event $E\subseteq S_{-i}\times T_{-i}$
is \textquotedblleft weakly assumed\textquotedblright \ under an LCPS $\bar{%
\mu}:=(\mu ^{1},...,\mu ^{n})$ if there exists $m\leq n$ such that

\begin{description}
\item[(i)] $\mu ^{l}\left( E\right) =1$\ for all $l\leq m$,

\item[(ii)] $\mu ^{l}\left( E\right) =0$\ for all $l>m$,

\item[(iii)] for every elementary cylinder $\hat{C}_{s_{-i}}:=\left \{
s_{-i}\right \} \times T_{-i}$, if $E\cap \hat{C}_{s_{-i}}\not=\emptyset $\
then $\mu ^{l}\left( E\cap \hat{C}_{s_{-i}}\right) >0$\ for some $l\leq m$.
\end{description}

The difference between weak assumption and BFK's assumption relies on
condition (iii): BFK require that, for every open set $O\subseteq
S_{-i}\times T_{-i}$, if $E\cap O\not=\emptyset $ then $\mu ^{l}\left( E\cap
O\right) >0$\ for some $l\leq m$. BFK's assumption is stronger than weak
assumption because, technically, every elementary cylinder is an open set.
The definition of weak assumption can be extended to all LPS's while
preserving its preference-based foundation---in the same way Dekel et al.
(2016) extend BFK's assumption---as follows.\footnote{%
Such extension can be formally shown by slightly adapting the proofs in
Dekel et al. (2016). The preference-based definition of BFK's assumption is
based on two axioms: \textit{Strict Determination} and \textit{Nontriviality}%
. Similarly, weak assumption requires Strict Determination and a weaker
axiom than Nontriviality.} Say that event $E\subseteq S_{-i}\times T_{-i}$
is \textbf{weakly assumed}\ under LPS $\bar{\mu}:=(\mu ^{1},...,\mu ^{n})$
if there exists $m\leq n$ such that conditions (i) and (iii)\ above hold, and

\begin{description}
\item[(ii)'] for each $l>m$, there exists $\left( \alpha _{1}^{l},...,\alpha
_{m}^{l}\right) \in \mathbb{R}^{m}$\ such that $\mu ^{l}\left( F\right)
=\tsum_{k=1}^{m}\alpha _{k}^{l}\mu ^{k}\left( F\right) $ for each Borel set $%
F\subseteq E$.
\end{description}

Cautious belief requires only conditions (i) and (iii). Thus, weak
assumption implies cautious belief, but the converse does not hold. We show
this by means of an example, which is taken from BFK (cf. Dekel and
Siniscalchi 2015, Example 12.10).

\begin{example}
\label{Example Boss game}Consider the following game with two players, Ann ($%
a$)\ and Bob ($b$):%
\begin{equation*}
\begin{tabular}{|c|c|c|c|}
\hline
$a\backslash b$ & $\ell $ & $c$ & $r$ \\ \hline
$u$ & $4,0$ & $4,1$ & $0,1$ \\ \hline
$m$ & $0,0$ & $0,1$ & $4,1$ \\ \hline
$d$ & $3,0$ & $2,1$ & $2,1$ \\ \hline
\end{tabular}%
\end{equation*}%
The IA set is $S^{\infty }=\left \{ u,m,d\right \} \times \left \{
c,r\right
\} $. By Theorem \ref{Theorem SAS transp cautious}.(ii), we can
append to this game a \emph{cautious}, finite type structure $\mathcal{T}%
:=\langle S_{i},T_{i},\beta _{i}\rangle _{i\in \left \{ a,b\right \} }$ such
that $S^{\infty }=\mathrm{Proj}_{S_{a}}\left( R_{a}^{\infty }\right) \times 
\mathrm{Proj}_{S_{b}}\left( R_{b}^{\infty }\right) $. Consider any type $%
t_{a}$ of Ann such that

(1) strategy $d$ is optimal under $\beta _{a}(t_{a}):=(\mu _{a}^{1},...,\mu
_{a}^{n})$, and

(2) $t_{a}$ cautiously believes Bob's (cautious) rationality.

We show that the event corresponding to Bob's rationality, viz. $R_{b}^{1}$,
cannot be weakly assumed under $\beta _{a}(t_{a})$. To ease notation, let $%
E_{s_{b}}:=\left \{ s_{b}\right \} \times T_{b}$ for each $s_{b}\in \left \{
\ell ,c,r\right \} $. There is no rational strategy-type pair of Bob in $%
E_{\ell }$; hence, $R_{b}^{1}=E_{c}\cup E_{r}$. As $t_{a}$ is cautious,
there exists $k\leq n$\ such that $\mathrm{marg}_{S_{b}}\mu _{a}^{k}\left(
\left \{ \ell \right \} \right) =\mu _{a}^{k}(E_{\ell })>0$. Let $k^{\ast
}:=\inf \left \{ k\leq n:\mu _{a}^{k}(E_{\ell })>0\right \} $. Since $%
R_{b}^{1} $\ is cautiously believed under $\beta _{a}(t_{a})$, it is the
case that $\mu _{a}^{1}\left( R_{b}^{1}\right) =1$, which\ implies $\mu
_{a}^{1}\left( E_{\ell }\right) =0$. This in turn yields $k^{\ast }\geq 2$.

Next note that, for every $\nu \in \mathcal{M}(S_{b})$ such that $\mathrm{%
Supp}\nu \subseteq \left \{ c,r\right \} $, strategy $d$ is optimal under $%
\nu $ if and only if $\nu (\left \{ c\right \} )=\nu (\left \{ r\right \}
)=1/2$; this entails that also $u$ and $m$ are optimal under $\nu $. Hence
we must have $\mu _{a}^{l}(E_{c})=\mu _{a}^{l}(E_{r})=1/2$ for every $%
l=1,...,k^{\ast }-1$. But $d$\ must be optimal also under $\mu _{a}^{k^{\ast
}}$, the first component measure of $\beta _{a}(t_{a})$ which assigns
strictly positive probability to $E_{\ell }$. It follows that $\mu ^{k^{\ast
}}(E_{\ell })>0$ and $\mu _{a}^{k^{\ast }}(E_{r})>\mu _{a}^{k^{\ast
}}(E_{c}) $.\footnote{%
Let $\nu \in \mathcal{M}(S_{b})$\ be the marginal of $\mu _{a}^{k^{\ast }}$\
on $S_{b}$, so that $\nu (\left \{ \ell \right \} )>0$. We need $\nu
(\left
\{ r\right \} )>\nu (\left \{ c\right \} )$ for strategy $d$ to be
optimal under $\nu $. For, if $\nu (\left \{ r\right \} )\leq \nu (\left \{
c\right
\} )$, then $u$ would be the unique best reply to $\nu $.}
Furthermore, $\mu _{a}^{k^{\ast }}\left( R_{b}^{1}\right) <1$ (for, if $\mu
_{a}^{k^{\ast }}\left( R_{b}^{1}\right) =1$, we would have $\mu ^{k^{\ast
}}(E_{\ell })=0$). With this, we conclude that conditions (i) and (iii) of
weak assumption are satisfied for event $R_{b}^{1}$\ at level $m:=k^{\ast
}-1 $\ of LPS $\beta _{a}(t_{a})$.

Yet, condition (ii)' of weak assumption does not hold. To see this, note
that $E_{c}$ and $E_{r}$ are Borel subsets of $R_{b}^{1}$, and $\mu
_{a}^{m+1}\left( E_{r}\right) >\mu _{a}^{m+1}\left( E_{c}\right) $. Suppose,
per contra, that condition (ii)' is satisfied. Then, there exists $\left(
\alpha _{1},...,\alpha _{m}\right) \in \mathbb{R}^{m}$ such that 
\begin{equation*}
\mu _{a}^{m+1}\left( E_{r}\right) =\tsum_{l=1}^{m}\alpha _{l}\mu
_{a}^{l}\left( E_{r}\right) =\tfrac{1}{2}\tsum_{l=1}^{m}\alpha
_{l}=\tsum_{l=1}^{m}\alpha _{l}\mu _{a}^{l}\left( E_{c}\right) =\mu
_{a}^{m+1}\left( E_{c}\right) \text{,}
\end{equation*}%
which contradicts $\mu _{a}^{m+1}\left( E_{r}\right) >\mu _{a}^{m+1}\left(
E_{c}\right) $. We therefore conclude that every rational strategy-type pair 
$\left( d,t_{a}\right) $ is not consistent with weak assumption of $%
R_{b}^{1} $.\hfill $\blacklozenge $
\end{example}

Example \ref{Example Boss game}\ shows that weak assumption can be strictly
stronger than cautious belief. Furthermore, the example illustrates the
difference between our approach to IA and the one based on weak assumption.
To clarify, suppose that R$^{\text{c}}$CB$^{\text{c}}$R$^{\text{c}}$\ is
replaced by the epistemic notion of \textquotedblleft cautious rationality
and common weak assumption of cautious rationality\textquotedblright \ (R$^{%
\text{c}}$CA$^{\text{w}}$R$^{\text{c}}$). Are there analogues of Theorems %
\ref{Theorem main result}\ and \ref{Theorem SAS} under R$^{\text{c}}$CA$^{%
\text{w}}$R$^{\text{c}}$? Not surprisingly, the answer is Yes.\footnote{%
A proof is available upon request.}

Yet, under weak assumption, it is not possible to provide epistemic
foundations for IA if we restrict attention to cautious type structures.\
Refer back to the type structure of Example \ref{Example Boss game}. In such
a structure, events $E_{c}$ and $E_{r}$\ are proper subsets of $R_{b}^{1}$,
and condition (ii)' of weak assumption fails. This is not necessarily true 
\emph{if the type structure contains both cautious and non-cautious types}.
Indeed, it is possible to construct a non-cautious type structure whereby
event $R_{b}^{1}:=R_{b}\cap C_{b}$ (\textquotedblleft Bob's cautious
rationality\textquotedblright ) is such that $R_{b}^{1}\nsubseteq E_{c}\cup
E_{r}$, and any cautiously rational pair $\left( d,t_{a}\right) $ is
consistent with weak assumption of $R_{b}^{1}$.\footnote{%
For instance, consider LPS $\beta _{a}(t_{a}):=(\mu _{a}^{1},...,\mu
_{a}^{n})$\ such that $\mu _{a}^{l}\left( R_{b}\right) =1$\ for each $l\leq
m $, and $\mu _{a}^{l}\left( C_{b}\right) =0$\ for each $l>m$. In this case,
condition (ii)' of weak assumption is satisfied, because $\mu
_{a}^{m+1},...,\mu _{a}^{n}$ assign zero probability to Bob's cautious
rationality (in particular, we have $\mu _{a}^{m+1}\left( E_{r}\right) >0$
and $\mu _{a}^{m+1}\left( R_{b}^{1}\right) =0$).}

We spend a few words on the origin of this difference. Cautious belief is
based on the notion of \textquotedblleft infinitely more
likely\textquotedblright \ of Lo (1999), which can be expressed in words as
follows: event $E$ is deemed infinitely more likely than event $F$ if the
agent prefers to bet on $E$ rather than on $F$ no matter the (different)
winning prizes for the two bets. According to this notion of
\textquotedblleft infinitely more likely,\textquotedblright \ $\beta
_{a}(t_{a}^{d})$ deems every element of $R_{b}^{1}$ infinitely more likely
than the complement of $R_{b}^{1}$, that is, $\left \{ (\ell ,t_{b}^{\ast
})\right \} $, which is sufficient for cautious belief (and implies that $%
R_{b}^{1}$ is deemed infinitely more likely than $\left \{ (\ell
,t_{b}^{\ast })\right \} $---this notion of \textquotedblleft infinitely
more likely\textquotedblright \ satisfies disjunction).

(Weak) Assumption is instead based on the notion of \textquotedblleft
infinitely more likely\textquotedblright \ of Blume et al. (1991a).
According to this notion, $\beta _{a}(t_{a}^{d})$ deems every element of $%
R_{b}^{1}$, i.e., $\left \{ (c,t_{b}^{\ast })\right \} $\ and $\left \{
(r,t_{b}^{\ast })\right \} $, infinitely more likely than $\left \{ (\ell
,t_{b}^{\ast })\right \} $.\footnote{%
The notions of \textquotedblleft infinitely more likely\textquotedblright \
due to Lo (1999) and Blume et al. (1991a) coincide when the events under
consideration are singletons; see Blume et al. (1991a, Footnote 8).} However
(and this is the key point!), such notion of \textquotedblleft infinitely
more likely\textquotedblright \  \textit{does not satisfy disjunction}: in
Example \ref{Example Boss game}, $\left \{ (c,t_{b}^{\ast })\right \} $\ and 
$\left \{ (r,t_{b}^{\ast })\right \} $\ are both deemed under $\beta
_{a}(t_{a}^{d})$ infinitely more likely than $\left \{ (\ell ,t_{b}^{\ast
})\right \} $, but their union ($R_{b}^{1}$) is \textit{not }deemed
infinitely more likely than $\left \{ (\ell ,t_{b}^{\ast })\right \} $.
Blume et al. (1991a, p. 70) provide a similar example which shows how their
notion of \textquotedblleft infinitely more likely\textquotedblright \ fails
disjunction.

We conclude this section with an example of a game and an associated,
\textquotedblleft rich\textquotedblright \ type structure such that: (1) the
set of states consistent with R$^{\text{c}}$CA$^{\text{w}}$R$^{\text{c}}$ is
distinct from the set of states consistent with our epistemic assumptions,
but (2) the behavioral implications are the same, i.e., the IA set. Thus,
the example shows that our epistemic conditions for IA are different from
those previously studied.

\begin{example}
Refer back the the game of Example \ref{Example Boss game}. Let $\mathcal{T}%
:=\langle S_{i},T_{i},\beta _{i}\rangle _{i\in \left \{ a,b\right \} }$ be
the canonical structure associated with this game. First note that, in such
a structure, the set of states consistent with RCAR is empty. Next recall
that, according to our notation, $\hat{R}_{a}^{\infty }\times \hat{R}%
_{b}^{\infty }$ is the set of states consistent with (a) rationality, (b)
transparency of cautiousness, and (c) common cautious belief in (a) and (b).
Similarly, let $\bar{R}_{a}^{\infty }\times \bar{R}_{b}^{\infty }$\ be the
set of states consistent with R$^{\text{c}}$CA$^{\text{w}}$R$^{\text{c}}$.
As argued above, and by Theorems \ref{Theorem main result} and \ref{Theorem
(alternative) main result}, it is the case that, for every $i\in \left \{
a,b\right \} $,%
\begin{equation*}
S_{i}^{\infty }=\mathrm{Proj}_{S_{i}}\left( R_{i}^{\infty }\right) =\mathrm{%
Proj}_{S_{i}}\left( \hat{R}_{i}^{\infty }\right) =\mathrm{Proj}%
_{S_{i}}\left( \bar{R}_{i}^{\infty }\right) \text{.}
\end{equation*}%
Yet, we claim that

(1) $\hat{R}_{a}^{\infty }\times \hat{R}_{b}^{\infty }\subseteq
R_{a}^{\infty }\times R_{b}^{\infty }$, and

(2) any strategy-type pair $\left( d,t_{a}\right) \in \hat{R}_{a}^{\infty }$%
\ of Ann is such that $\left( d,t_{a}\right) \notin \bar{R}_{a}^{\infty }$.

\noindent With this, (1) and (2) entail that both $\hat{R}_{a}^{\infty
}\times \hat{R}_{b}^{\infty }$\ and $R_{a}^{\infty }\times R_{b}^{\infty }$\
are distinct from $\bar{R}_{a}^{\infty }\times \bar{R}_{b}^{\infty }$.

The set inclusion (which is strict) in (1) follows from belief-completeness
of $\mathcal{T}$ and from the \textquotedblleft
quasi-monotonicity\textquotedblright \ property of cautious belief---see
Remark \ref{Remark: quasi-monotonocity of assumption} and the discussion
below Theorem \ref{Theorem SAS transp cautious}.

To prove claim (2), pick any $\left( d,t_{a}\right) \in \hat{R}_{a}^{\infty
} $, so that strategy $d$ is optimal under $\beta _{a}(t_{a}):=(\mu
_{a}^{1},...,\mu _{a}^{n})$. In view of claim (1), Ann's type $t_{a}$
cautiously believes the event corresponding to Bob's cautious rationality,
viz. $R_{b}^{1}$. We now show that $R_{b}^{1}$ cannot be weakly assumed
under $\beta _{a}(t_{a})$. To this end, let $E_{b}:=S_{b}\times \hat{T}_{b}$%
\ denote the set of strategy-type pairs of Bob consistent with transparency
of cautiousness (i.e., according to the notation in Section \ref{Section
transparency of cautiousness}, $E_{b}$\ is the projection onto $S_{b}\times
T_{b}$ of the event $C^{\infty }$ in $\mathcal{T}$), so that%
\begin{equation*}
\mu _{a}^{1}\left( E_{b}\right) =...=\mu _{a}^{n}\left( E_{b}\right) =1\text{%
.}
\end{equation*}%
Proceeding as in Example \ref{Example Boss game}, define $%
E_{s_{b}}:=\left
\{ s_{b}\right \} \times \hat{T}_{b}$ for each $s_{b}\in
\left \{ \ell ,C,R\right \} $. There is no (cautiously) rational
strategy-type pair of Bob in $E_{\ell }$: we have $\hat{R}_{b}^{1}=E_{c}\cup
E_{r}\subseteq R_{b}^{1}$. Then, the same argument as in Example \ref%
{Example Boss game}\ shows that $R_{b}^{1}$ cannot be weakly assumed under $%
\beta _{a}(t_{a})$. Hence, every strategy-type pair $\left( D,t_{a}\right)
\in \hat{R}_{a}^{\infty }$ is not consistent with weak assumption of $%
R_{b}^{1}$.\hfill $\blacklozenge $
\end{example}

\subsection{Further comments on the related literature}

Other articles with epistemic conditions for IA include Keisler and Lee
(2023), Lee (2016a), Heifetz et al. (2019), Halpern and Pass (2019), and, in
non-lexicographic frameworks, Barelli and Galanis (2013) and Ziegler and
Zuazo-Garin (2020).

Keisler and Lee (2023) construct a \emph{discontinuous} and complete type
structure where RCAR is possible. Furthermore, they show that such type
structure generates the same set of belief hierarchies as a continuous one.
An immediate implication of this findings is that BFK's results hinge on
topological details of the type structure that cannot be expressed in terms
of belief hierarchies. Keisler and Lee conclude that BFK's negative result
stems from the fact that, in a continuous type structure, players are
\textquotedblleft too cautious\textquotedblright \ towards the assumed
events. Lee (2016a) relaxes the traditional coherency condition on belief
hierarchies while maintaining coherency of the represented preferences. With
this, he identifies hierarchies of lexicographic beliefs without an upper
bound on the length of the LPS's that cannot be represented by any type
structure but capture \textquotedblleft rationality\textquotedblright \ and
common assumption of \textquotedblleft rationality\textquotedblright .\ The
notion of \textquotedblleft rationality\textquotedblright \ used by Lee is
essentially equivalent to cautious rationality.\footnote{%
In a companion paper (Catonini and\ De Vito 2017), we show that our results
can be replicated in the hierarchical space studied by Lee (2016a).} Heifetz
et al. (2019) put forward the solution concept of comprehensive
rationalizability, and they give it an epistemic foundation in a universal
type structure for LCPS's. Comprehensive rationalizability neither refines
nor is refined by IA, but it coincides with IA in many applications. Halpern
and Pass (2019) use instead a modal-logic framework to provide an epistemic
characterization of IA. They put forward a notion of \textquotedblleft
generalized belief\textquotedblright \ which is given semantics in terms of
LPS's. Their characterization of IA relies on a corresponding\ operator
(\textquotedblleft All I know\textquotedblright ) which is taken with
respect to an appropriate language.

Within a standard Bayesian decision model, Barelli and Galanis (2013) use
the idea that each player has a list of preferences which allows her to
break ties. With this, they provide an epistemic foundation for IA in an
appropriate framework for interactive beliefs. Ziegler and Zuazo-Garin
(2020) use instead a decision model of incomplete, but continuous
preferences where each player's uncertainty is represented by a set of
beliefs. They provide foundations for IA and SAS's in terms of interactive,
ambiguous beliefs, rather than LPS's. Both Barelli and Galanis (2013) and
Ziegler and Zuazo-Garin (2020) can be regarded as complementary to the
LPS-based approach.

\section*{Appendix A. Preference basis}

We develop preference foundations for cautiousness, certain belief and
cautious belief. In so doing, we adopt the following decision-theoretic
setup. A \textit{game form} is a structure $\left \langle I,Z,\left(
S_{i}\right) _{i\in I},z\right \rangle $ where (a) $I$\ is the finite set of
players, (b) each $S_{i}$\ is the finite set of strategies, and (c) $%
z:S\rightarrow Z$ is a surjective outcome function (hence, the set $Z$\ is
finite). Each player is viewed as a Decision Maker (DM) facing a problem
where his co-players' strategies are part of the description of the states,
and mixed strategies are the feasible acts. So, fix an $\left( S_{i}\right)
_{i\in I}$-based lexicographic type structure $\mathcal{T}:=\langle
S_{i},T_{i},\beta _{i}\rangle _{i\in I}$. We fix a player $i\in I$ (the DM),
and, to ease notation, we set $\Omega :=S_{-i}\times T_{-i}$. With this, the
DM is uncertain about what \textquotedblleft state\textquotedblright \
(strategy-type profiles of the co-players)\ in $\Omega $ will be realized,
and he is endowed with a preference relation over all (Borel) measurable
functions that assign to each element of $\Omega $ an objective
randomization on $Z$. In Blume et al. (1991a), an \textit{act} is a function
from an abstract, finite domain of uncertainty $\Omega $ to $\mathcal{M}%
\left( Z\right) $.

A \textit{game} is obtained by adding to the game form a profile of von
Neumann-Morgenstern utility functions $\left( v_{i}\right) _{i\in I}$, which
represent players' preferences over lotteries of consequences, according to
expected utility calculations. In what follows, we assume that the codomain
of any act on $\Omega $ is in utils, i.e., randomizations on material
consequences are replaced by their von Neumann-Morgenstern utilities, which
take value in the interval $\left[ 0,1\right] $. Thus, an \textbf{act} on $%
\Omega $\ is a Borel measurable function $f:\Omega \rightarrow \left[ 0,1%
\right] $.\footnote{%
We omit the formalism for randomizations as all preferences considered below
agree on constant acts on $Z$, hence the utilities are uniquely defined.
Moreover, the definition of act used here is also used in BFK and Dekel et
al. (2016).} We let \textrm{ACT}$\left( \Omega \right) $\ denote the set of
all acts on $\Omega $.

The DM has preferences over elements of \textrm{ACT}$\left( \Omega \right) $%
. For $x\in \left[ 0,1\right] $, we write $\overrightarrow{x}$\ for the
constant act associated with $x$, i.e., $\overrightarrow{x}(\omega ):=x$\
for all $\omega \in \Omega $.\textbf{\ }Given a Borel set $E\subseteq \Omega 
$ and acts $f,g\in $\textrm{ACT}$\left( \Omega \right) $, we define $%
(f_{E},g_{\Omega \backslash E})\in $\textrm{ACT}$\left( \Omega \right) $ as
follows:%
\begin{equation*}
(f_{E},g_{\Omega \backslash E})(\omega ):=\left \{ 
\begin{tabular}{ll}
$f(\omega )$, & if $\omega \in E\text{,}$ \\ 
$g(\omega )$, & if $\omega \in \Omega \backslash E\text{.}$%
\end{tabular}%
\right.
\end{equation*}

Let $\succsim $ be a preference relation on \textrm{ACT}$\left( \Omega
\right) $ and write $\succ $ (resp. $\sim $) for strict preference (resp.
indifference). We assume that preference relation $\succsim $\ satisfies the
standard axioms of Order and Independence (see BFK for a formal definition).
Thus, we let $\succsim _{E}$\ denote the \emph{conditional preference} given 
$E$, that is, $f\succsim _{E}g$ if and only if $(f_{E},h_{\Omega \backslash
E})\succsim (g_{E},h_{\Omega \backslash E})$ for some $h\in $\textrm{ACT}$%
\left( \Omega \right) $. Standard results (see Blume et al., 1991a, for a
proof) show that, under the axioms of Order and Independence, $%
(f_{E},h_{\Omega \backslash E})\succsim (g_{E},h_{\Omega \backslash E})$
holds for all $h\in $\textrm{ACT}$\left( \Omega \right) $ if it holds for
some $h$.

An event $E\subseteq \Omega $\ is \textbf{Savage-null} under $\succsim $ if $%
f\sim _{E}g$\ for all $f,g\in $\textrm{ACT}$\left( \Omega \right) $. Say
that $E$\ is \textbf{non-null} under $\succsim $ if it is not Savage-null
under $\succsim $. With this, we can introduce the notion of certain belief
in terms of the preference relation $\succsim $.

\bigskip

\noindent \textbf{Definition A.1 }\textit{Event }$E\subseteq \Omega $\textit{%
\ is \textbf{certainly believed} under }$\succsim $\textit{\ if }$f\sim
_{\Omega \backslash E}g$\textit{\ for all }$f,g\in $\textrm{ACT}$\left(
\Omega \right) $\textit{.}

\bigskip

Throughout, we maintain the assumption that $\bar{\mu}\in \mathcal{N}\left(
\Omega \right) $\ is a lexicographic expected utility representation of $%
\succsim $, i.e., $\succsim =\succsim ^{\bar{\mu}}$.\footnote{%
To ease notation, we drop player $i$'s subscript from LPS $\bar{\mu}_{i}$ on 
$\Omega $.} Savage-null events and certain belief can be characterized in
terms of LPS's as follows.

\bigskip

\noindent \textbf{Proposition A.1} \textit{Fix }$\bar{\mu}:=(\mu
^{1},...,\mu ^{n})\in \mathcal{N}(\Omega )$\textit{\ and event }$E\subseteq
\Omega $\textit{. Then:}

\textit{(i)} $E$ \textit{is Savage-null under }$\succsim ^{\bar{\mu}}$%
\textit{\ if and only if }$\mu ^{m}\left( E\right) =0$\textit{\ for all }$%
m\leq n$\textit{;}

\textit{(ii)} $E$\textit{\ is certainly believed under }$\succsim ^{\bar{\mu}%
}$\textit{\ if and only if it is certainly believed under }$\bar{\mu}$%
\textit{.}

\bigskip

The proof of Part (i) of Proposition A.1 is quite immediate, and it can be
found in Dekel et al. (2016, Remark 2.1). Part (ii) follows from Part (i).

The following definition is due to Catonini and De Vito (2020).

\bigskip

\noindent \textbf{Definition A.2 }\textit{Fix }$\bar{\mu}\in \mathcal{N}%
(\Omega )$\textit{\ and a set of acts }\textrm{ACT}$^{\ast }\left( \Omega
\right) \subseteq $\textrm{ACT}$\left( \Omega \right) $\textit{. Say that }$%
\bar{\mu}$\textit{\ exhibits \textbf{cautiousness} with respect to }\textrm{%
ACT}$^{\ast }\left( \Omega \right) $\textit{\ if, for all }$f,g\in $\textrm{%
ACT}$^{\ast }\left( \Omega \right) $\textit{, the following condition holds:}

\textit{(*) if }$f(\omega )\geq g(\omega )$\textit{\ for each }$\omega \in
\Omega $\textit{\ and }$f(\omega ^{\prime })>g(\omega ^{\prime })$\textit{\
for some }$\omega ^{\prime }\in \Omega $\textit{, then }$f\succ ^{\bar{\mu}%
}g $\textit{.}

\textit{\bigskip }

Cautiousness is defined with respect to a set of acts that are conceivable
given the potential ability of the states to influence utilities. Since the
DM is a player $i$ in a game, and the domain of uncertainty is $\Omega
:=S_{-i}\times T_{-i}$, we find it appropriate to consider \textrm{ACT}$%
^{\ast }\left( \Omega \right) $ as the set of acts $f\in $\textrm{ACT}$%
\left( \Omega \right) $ such that, for all $s_{-i}\in S_{-i}$ , the map $%
f\left( s_{-i},\cdot \right) :T_{-i}\rightarrow \left[ 0,1\right] $ is
constant. We let \textrm{ACT}$^{S_{-i}}\left( \Omega \right) $ denote this
set of acts. In words, \textrm{ACT}$^{S_{-i}}\left( \Omega \right) $\ is the
set of all acts which are independent of \textit{payoff-irrelevant} \textit{%
components} of states $(s_{-i},t_{-i})\in \Omega $, i.e., the types of $i$'s
co-players. Note: every mixed strategy $\sigma _{i}\in \mathcal{M}\left(
S_{i}\right) $\ in a game can be identified with the (feasible) act $%
f_{\sigma _{i}}:S_{-i}\times T_{-i}\rightarrow \left[ 0,1\right] $ such that 
$f(s_{-i},t_{-i})=v_{i}\left( \sigma _{i},s_{-i}\right) $ for all $%
(s_{-i},t_{-i})\in \Omega $; hence, $f_{\sigma _{i}}\in $\textrm{ACT}$%
^{S_{-i}}\left( \Omega \right) $.

The following result, which is proved in Catonini and De Vito (2020),
provides the preference-based foundation for the type-based definition of
cautiousness (Definition \ref{Definition: Cautious type}).

\bigskip

\noindent \textbf{Proposition A.2} \textit{Fix a type structure }$\mathcal{T}%
:=\langle S_{i},T_{i},\beta _{i}\rangle _{i\in I}$\textit{\ and a type }$%
t_{i}\in T_{i}$\textit{. Then }$t_{i}$\textit{\ is cautious in }$\mathcal{T}$%
\textit{\ if and only if }$\beta _{i}(t_{i})$\textit{\ exhibits cautiousness
with respect to }\textrm{ACT}$^{S_{-i}}\left( S_{-i}\times T_{-i}\right) $%
\textit{.}

\bigskip

A \textbf{bet} (or \textbf{binary act}) on $\Omega $ is an act of the form $(%
\overrightarrow{x}_{E},\overrightarrow{y}_{\Omega \backslash E})$, where $%
x,y\in \left[ 0,1\right] $\ and $E\subseteq \Omega $ is an event.

\bigskip

\noindent \textbf{Definition A.3 }\textit{Fix events }$E,F\subseteq \Omega $%
\textit{. Event }$E$\textit{\ is \textbf{more likely} than }$F$\textit{\
under }$\succsim ^{\bar{\mu}}$\textit{\ if for all }$x,y\in \left[ 0,1\right]
$\textit{\ with }$x>y$\textit{,}%
\begin{equation*}
(\overrightarrow{x}_{E},\overrightarrow{y}_{\Omega \backslash E})\succsim ^{%
\bar{\mu}}(\overrightarrow{x}_{F},\overrightarrow{y}_{\Omega \backslash F})%
\text{.}
\end{equation*}

\textit{Event }$E$\textit{\ is deemed \textbf{infinitely more likely} than }$%
F$\textit{\ under }$\succsim ^{\bar{\mu}}$\textit{, and write }$E\gg ^{\bar{%
\mu}}F$\textit{,\ if for all }$x,y,z\in \left[ 0,1\right] $\textit{\ with }$%
x>y$\textit{,}%
\begin{equation*}
(\overrightarrow{x}_{E},\overrightarrow{y}_{\Omega \backslash E})\succ ^{%
\bar{\mu}}(\overrightarrow{z}_{F},\overrightarrow{y}_{\Omega \backslash F})%
\text{.}
\end{equation*}

\bigskip

In words, $E$\ is more likely than $F$\ if the DM prefers to bet on $E$
rather than on $F$ given the same prizes for the two bets. Event $E$\ is
infinitely more likely than $F$\ if betting on $E$ is \textit{strictly}
preferable to betting on $F$, and strict preference persists no matter how
bigger the prize $z$ for winning the $F$ bet is. This notion of
\textquotedblleft infinitely more likely\textquotedblright \ is due to Lo
(1999, Definition 1). Note that, if $E\gg ^{\bar{\mu}}F$, then $E$\ is
non-null under $\succsim ^{\bar{\mu}}$, while $F$ may, but \textit{need not}%
, be Savage-null under $\succsim ^{\bar{\mu}}$. When $\succsim ^{\bar{\mu}}$
has a subjective expected utility representation, $E\gg ^{\bar{\mu}}F$
implies that $F$ is Savage-null.

As pointed out by Lo (1999), the likelihood relation $\gg ^{\bar{\mu}}$
possesses some natural properties, such as irreflexivity, asymmetry and
transitivity. Furthermore, if $E\gg ^{\bar{\mu}}F$, then

\begin{description}
\item[(P1)] $E$ is infinitely more likely than every Borel subset of $F$; and

\item[(P2)] every Borel superset of $E$ is infinitely more likely than $F$.
\end{description}

The next step is to characterize the likelihood order $\gg ^{\bar{\mu}}$
between pairwise \textit{disjoint} events in terms of LPS's representing $%
\succsim ^{\bar{\mu}}$ (see Definition \ref{Definition IML LPS} of Section %
\ref{Section IML and CB}). Recall that, given $\bar{\mu}:=(\mu ^{1},...,\mu
^{n})\in \mathcal{N}(\Omega )$ and non-empty event $E\subseteq \Omega $,%
\begin{equation*}
\mathcal{I}_{\bar{\mu}}\left( E\right) :=\inf \left \{ l\in \left \{
1,...,n\right \} :\mu ^{l}\left( E\right) >0\right \} \text{,}
\end{equation*}%
with the convention that $\inf \emptyset :=+\infty $. The proof of the
following result can be found in Catonini and De Vito (2020).

\bigskip

\noindent \textbf{Proposition A.3} \textit{Fix} $\bar{\mu}:=(\mu
^{1},...,\mu ^{n})\in \mathcal{N}(\Omega )$ \textit{and disjoint events }$%
E,F\subseteq \Omega $\textit{. Then, }$E\gg ^{\bar{\mu}}F$\textit{\ if and
only if }$\mathcal{I}_{\bar{\mu}}\left( E\right) <\mathcal{I}_{\bar{\mu}%
}\left( F\right) $\textit{.}

\bigskip

We now introduce the notion of cautious belief in terms of the likelihood
order $\gg ^{\bar{\mu}}$. Recall that $\hat{C}_{s_{-i}}\subseteq \Omega $\
is called \textbf{elementary cylinder} if $\hat{C}_{s_{-i}}:=\left \{
s_{-i}\right \} \times T_{-i}$ for some $s_{-i}\in S_{-i}$. Given $s_{-i}$
and event $E$, we say that $E_{s_{-i}}$ is a \textbf{relevant part} of the
event $E$\ if $E_{s_{-i}}:=E\cap \hat{C}_{s_{-i}}\not=\emptyset $ for some $%
\hat{C}_{s_{-i}}$. Clearly, every non-empty event $E$\ can be written as a
finite, disjoint union of all its relevant parts.

\bigskip

\noindent \textbf{Definition A.4}\textit{\ Fix }$\bar{\mu}\in \mathcal{N}%
(\Omega )$. \textit{A non-empty event} $E\subseteq \Omega $ \textit{is%
\textbf{\ cautiously believed }under} $\succsim ^{\bar{\mu}}$ \textit{if it
satisfies the following condition:}

\textit{(*) for every relevant part }$E_{s_{-i}}$\textit{\ of }$E$\textit{, }%
$E_{s_{-i}}\gg ^{\bar{\mu}}\Omega \backslash E$\textit{.}

\bigskip

In words, event $E$ is cautiously believed under $\succsim ^{\bar{\mu}}$ if
every relevant part of $E$ is deemed infinitely more likely than $\Omega
\backslash E$. Since $E$ can be written as a finite, disjoint union of all
its relevant parts, it follows from (P2) that $E$ is deemed infinitely more
likely than $\Omega \backslash E$, i.e., $E\gg ^{\bar{\mu}}\Omega \backslash
E$.

However, the converse need not hold. That is, if $E\gg ^{\bar{\mu}}\Omega
\backslash E$, then $E$\ is non-null under $\succsim ^{\bar{\mu}}$, and
there exists at least one relevant part $E_{s_{-i}}$\ of $E$\ such that $%
E_{s_{-i}}\gg ^{\bar{\mu}}\Omega \backslash E$. But this does not rule out
the existence of different relevant parts of $E$ that \textit{do not}
satisfy this property.

\bigskip

\noindent \textbf{Example A.1}\textit{\ Refer back to Example \ref{Example:
cautious belief not monotone}\ in Section \ref{Section IML and CB}. Event }$%
F:=\left \{ s_{b}^{1},s_{b}^{2}\right \} \times T_{b}$\textit{\ is
infinitely more likely under }$\bar{\mu}_{a}$\textit{\ than its complement }$%
\left \{ s_{b}^{3}\right \} \times T_{b}$\textit{. Yet }$F$\textit{\ is not
cautiously believed under }$\succsim ^{\bar{\mu}_{a}}$\textit{: the relevant
part }$F_{s_{b}^{2}}:=\left \{ s_{b}^{2}\right \} \times T_{b}$\textit{\ is
more likely than }$\left \{ s_{b}^{3}\right \} \times T_{b}$\textit{, but }$%
F_{s_{b}^{2}}$\textit{\ is not infinitely more likely than }$\left \{
s_{b}^{3}\right \} \times T_{b}$\textit{.}\hfill $\blacklozenge $

\bigskip

We say that event $E\subseteq \Omega $\ is \textbf{weakly believed}\textit{\ 
}under $\succsim ^{\bar{\mu}}$ if $E\gg ^{\bar{\mu}}\Omega \backslash E$.
Event $F$ in Example A.1 is weakly, but not cautiously believed under $%
\succsim ^{\bar{\mu}_{a}}$.

We next state and prove the characterization result for cautious belief. For
the reader's convenience, we restate the LPS-based definition of cautious
belief given in the main text, but in terms of relevant parts.

\bigskip

\noindent \textbf{Definition A.5 }\textit{Fix }$\bar{\mu}:=(\mu ^{1},...,\mu
^{n})\in \mathcal{N}(\Omega )$. \textit{A non-empty event} $E\subseteq
\Omega $ \textit{is }\textbf{cautiously believed under} $\bar{\mu}$ \textbf{%
at level} $m\leq n$ \textit{if:}

\textit{(i) }$\mu ^{l}\left( E\right) =1$\textit{\ for all }$l\leq m$\textit{%
;}

\textit{(ii) for every relevant part }$E_{s_{-i}}$ \textit{of }$E$, $\mu
^{l}\left( E_{s_{-i}}\right) >0$\textit{\ for some }$l\leq m$\textit{.}

\textit{Event }$E$\textit{\ is \textbf{cautiously believed under} }$\bar{\mu}
$\textit{\ if it is cautiously believed under }$\bar{\mu}$\textit{\ at some
level\ }$m\leq n$\textit{.}

\bigskip

\noindent \textbf{Theorem A.1} \textit{Fix} $\bar{\mu}:=(\mu ^{1},...,\mu
^{n})\in \mathcal{N}(\Omega )$ \textit{and a non-empty event }$E\subseteq
\Omega $\textit{. Then }$E$\textit{\ is cautiously believed under }$\succsim
^{\bar{\mu}}$\textit{\ if and only if }$E$\textit{\ is cautiously believed
under }$\bar{\mu}$\textit{.}

\bigskip

\noindent \textbf{Proof}. The proof is immediate if $\Omega \backslash E$\
is Savage-null under $\succsim ^{\bar{\mu}}$, so, in what follows, let $%
\Omega \backslash E$ be non-null under $\succsim ^{\bar{\mu}}$.

Suppose first that $E$\ is cautiously believed under $\succsim ^{\bar{\mu}}$%
. Since every relevant part $E_{s_{-i}}$ of $E$ satisfies $E_{s_{-i}}\gg ^{%
\bar{\mu}}\Omega \backslash E$, Proposition A.3 yields $\mathcal{I}_{\bar{\mu%
}}\left( E_{s_{-i}}\right) <\mathcal{I}_{\bar{\mu}}\left( \Omega \backslash
E\right) $. Hence, $\mathcal{I}_{\bar{\mu}}\left( \Omega \backslash E\right)
\geq 2$. Let $m:=\mathcal{I}_{\bar{\mu}}\left( \Omega \backslash E\right) -1$%
. Then $\mathcal{I}_{\bar{\mu}}\left( E_{s_{-i}}\right) \leq m$. Moreover,
for every $k\leq m$, we have $\mu ^{k}(\Omega \backslash E)=0$, hence $\mu
^{k}(E)=1$. Therefore conditions (i)-(ii) of Definition A.5 are satisfied.

Conversely, if $E$\ is cautiously believed under $\bar{\mu}$ at level $m$,
then condition (i) of Definition A.5 implies $\mathcal{I}_{\bar{\mu}}\left(
\Omega \backslash E\right) >m$. With this, condition (ii) yields that each $%
E_{s_{-i}}$\ satisfies $\mathcal{I}_{\bar{\mu}}\left( E_{s_{-i}}\right) $ $<%
\mathcal{I}_{\bar{\mu}}\left( \Omega \backslash E\right) $. Hence, by
Proposition A.3, $E_{s_{-i}}\gg ^{\bar{\mu}}\Omega \backslash E$.\hfill $%
\blacksquare $

\bigskip

Cautious belief in $E$ can be given an alternative axiomatic treatment. The
Supplementary Appendix proposes two axioms: \textit{Relevance} says that,
conditional on every relevant part of $E$, the DM can have strict
preferences. \textit{Weak Dominance Determination} says that, for any pair
of acts $f,g\in $\textrm{ACT}$^{S_{-i}}\left( \Omega \right) $, whenever $f$%
\  \textquotedblleft weakly dominates\textquotedblright \ $g$\ on $E$, the DM
prefers $f$\ to $g$\ unconditionally. This notion of weak dominance is
preference-based, as it corresponds to the notion of \textit{P-weak dominance%
} of Dekel et al. (2016) for acts in \textrm{ACT}$^{S_{-i}}\left( \Omega
\right) $.\footnote{%
P-weak dominance is defined by Dekel et al. (2016) for any pair of acts
belonging to the set \textrm{ACT}$\left( \Omega \right) $ (the set of 
\textit{all }acts).}

We conclude this section by providing a characterization of cautious belief
in terms of infinitesimal nonstandard numbers. A preference relation $%
\succsim $ on $\Omega $ that admits a lexicographic expected utility
representation can be equivalently described by an $\mathbb{F}$-valued
probability measure on $\Omega $. Here, $\mathbb{F}$\ is a non-Archimedean
ordered field which is a strict extension of the set of real numbers $%
\mathbb{R}$ (see Blume et al. 1991a, Section 6). For instance, the LPS $\bar{%
\mu}:=(\mu ^{1},\mu ^{2})$ can be represented by a nonstandard real valued
probability $\nu :=(1-\varepsilon )\mu ^{1}+\varepsilon \mu ^{2}$, where $%
\varepsilon >0$ is an infinitesimal nonstandard real such that $%
x>n\varepsilon $ for each real number $x>0$ and each $n\in \mathbb{N}$.

Given nonstandard reals $x$ and $y$, we say that $x$ is \textbf{infinitely
greater} than $y$ if $x>ny$ for each $n\in \mathbb{N}$. As discussed in
Catonini and De Vito (2020), the notion of infinitely more likely in
Definition \ref{Definition IML LPS} corresponds exactly to the
\textquotedblleft infinitely greater\textquotedblright \ relation between
the nonstandard probability values that provide an equivalent representation
of preferences. With this in mind, we show that cautious belief can be given
an easy, nonstandard characterization.

To this end, we first recall the notion of \textquotedblleft standard
part\textquotedblright \ of a nonstandard real number. Fix a nonstandard
real $x$ such that $-r<x<r$\ for some real number $r>0$. The standard part
of $x$, which is denoted by $\mathrm{st}\left( x\right) $, is the unique
real number $y$\ such that $\left \vert y-x\right \vert $ is an
infinitesimal. It is easy to check that, given positive nonstandard numbers $%
x$ and $y$, if $x$\ is infinitely greater than $y$, then $\mathrm{st}\left( 
\frac{y}{x}\right) =0$; the reverse implication is also true---see Halpern
(2010, p. 159). Next, fix a non-empty event $E\subseteq \Omega $ and an $%
\mathbb{F}$-valued probability measure $\nu $ representing $\succsim $.
Event $E$\ is cautiously believed under $\nu $ if, for every relevant part $%
E_{s_{-i}}$ of $E$, it is the case that $\mathrm{st}\left( \frac{\nu \left(
\Omega \backslash E\right) }{\nu \left( E_{s_{-i}}\right) }\right) =0$.
Finally note that, as each $\nu \left( E_{s_{-i}}\right) $\ is infinitely
greater than $\nu \left( \Omega \backslash E\right) $, so is $\nu \left(
E\right) $. This in turn implies $\mathrm{st}\left( \nu \left( E\right)
\right) =1$, i.e., event $E$\ is weakly believed under $\nu $\ (see Halpern
2010, and Catonini and De Vito 2020).

\section*{Appendix B. Proofs for Section \protect \ref{Section cautiousness
cautious belief}}

We begin with the proof of Proposition \ref{Proposition CR implies
admissibility}.

\bigskip

\noindent \textbf{Proof of Proposition \ref{Proposition CR implies
admissibility}}.\textbf{\ }By definition, if $(s_{i},t_{i})\in R_{i}\cap
C_{i}$, then $s_{i}$ is a lexicographic best reply to $\overline{\mathrm{marg%
}}_{S_{-i}}\beta _{i}(t_{i})\in \mathcal{N}^{+}(S_{-i})$. Proposition 1 in
Blume et al. (1991b) says that for every $\bar{\mu}_{i}\in \mathcal{N}%
^{+}\left( S_{-i}\right) $\ and for every lexicographic best reply $%
s_{i}^{\prime }$ to $\bar{\mu}_{i}$, there exists a probability measure $\nu
_{i}\in \mathcal{M}\left( S_{-i}\right) $ such that $\mathrm{Supp}\nu
_{i}=S_{-i}$ and $\pi _{i}(s_{i}^{\prime },\nu _{i})\geq \pi
_{i}(s_{i}^{\prime \prime },\nu _{i})$\ for every $s_{i}^{\prime \prime }\in
S_{i}$. Thus, by Remark \ref{Remark second Pearce lemma}, $s_{i}$ is
admissible.\hfill $\blacksquare $

\bigskip

Next, we prove Proposition \ref{Lemma on properties of assumption one
direction conjunction}. To this end, we find it convenient to state and
prove an auxiliary result, which is the analogue of Lemma B.1 in BFK.

\bigskip

\noindent \textbf{Lemma B.1} \textit{Fix a type structure }$\mathcal{T}%
:=\langle S_{i},T_{i},\beta _{i}\rangle _{i\in I}$\textit{. Fix also a type }%
$t_{i}\in T_{i}$\textit{\ with }$\beta _{i}\left( t_{i}\right) :=(\mu
_{i}^{1},...,\mu _{i}^{n})$ \textit{and a non-empty event }$E\subseteq
S_{-i}\times T_{-i}$\textit{. Then, }$E$\textit{\ is cautiously believed
under }$\beta _{i}\left( t_{i}\right) $\textit{\ if and only if there exists 
}$m\leq n$\textit{\ such that }$\beta _{i}\left( t_{i}\right) $\textit{\
satisfies condition (i) of Definition \ref{Definition assumption under an
LPS} plus the following condition:}

\textit{(ii\textquotedblright ) }$E\subseteq \left( \cup _{l\leq m}\mathrm{%
Suppmarg}_{S_{-i}}\mu _{i}^{l}\right) \times T_{-i}$.

\bigskip

\noindent \textbf{Proof}. Suppose that $E$\ is cautiously believed under $%
\beta _{i}\left( t_{i}\right) :=(\mu _{i}^{1},...,\mu _{i}^{n})$\ at level $%
m $. We show that $\beta _{i}\left( t_{i}\right) $\ satisfies condition
(ii\textquotedblright ). For every $s_{-i}\in $\textrm{Proj}$_{S_{-i}}\left(
E\right) $, we have%
\begin{equation*}
\left( \left \{ s_{-i}\right \} \times T_{-i}\right) \cap E\neq \emptyset 
\text{.}
\end{equation*}%
By condition (ii) of Definition \ref{Definition assumption under an LPS},
there exists $k\leq m$ such that $\mu _{i}^{k}\left( \left \{
s_{-i}\right
\} \times T_{-i}\right) >0$. Thus, $s_{-i}\in \mathrm{Suppmarg}%
_{S_{-i}}\mu _{i}^{k}$. Hence,%
\begin{equation*}
E\subseteq \mathrm{Proj}_{S_{-i}}\left( E\right) \times T_{-i}\subseteq
\left( \tbigcup_{l\leq m}\mathrm{Supp}\text{\textrm{marg}}_{S_{-i}}\mu
_{i}^{l}\right) \times T_{-i}\text{.}
\end{equation*}%
Conversely, suppose that conditions (i) and (ii\textquotedblright ) hold. We
show that condition (ii) of Definition \ref{Definition assumption under an
LPS} holds. Fix $s_{-i}\in S_{-i}$ such that $E_{s_{-i}}:=\left( \left \{
s_{-i}\right \} \times T_{-i}\right) \cap E\neq \emptyset $. By condition
(ii\textquotedblright ), $E_{s_{-i}}\subseteq \left( \cup _{l\leq m}\mathrm{%
Supp}\text{\textrm{marg}}_{S_{-i}}\mu _{i}^{l}\right) \times T_{-i}$. Hence,
there exists $k\leq m$ such that $s_{-i}\in \mathrm{Supp}$\textrm{marg}$%
_{S_{-i}}\mu _{i}^{k}$. Thus, $\mu _{i}^{k}\left( \left \{ s_{-i}\right \}
\times T_{-i}\right) >0$. Moreover, by condition (i), $\mu _{i}^{k}\left(
E\right) =1$. Therefore, $\mu _{i}^{k}\left( E_{s_{-i}}\right) >0$, as
desired.\hfill $\blacksquare $

\bigskip

\noindent \textbf{Proof of Proposition \ref{Lemma on properties of
assumption one direction conjunction}}. \textbf{Part 1}: Let $\bar{\mu}%
_{i}:=\left( \mu _{i}^{1},...,\mu _{i}^{n}\right) $\ and suppose that, for
each $k$, $E_{k}$ is cautiously believed under $\bar{\mu}_{i}$ at some level 
$m_{k}$. Let $m_{K}:=\min \left \{ m_{k}:k=1,2,...\right \} $. We show that $%
E:=\cap _{k}E_{k}$ is cautiously believed at level $m_{K}$. For each $k$, it
holds that $\mu _{i}^{l}\left( E_{k}\right) =1$\ for all $l\leq m_{K}$. By
the $\sigma $-additivity property of probability measures, it follows that $%
\mu _{i}^{l}\left( E\right) =1$\ for all $l\leq m_{K}$. Fix an elementary
cylinder $\hat{C}_{s_{-i}}:=\left \{ s_{-i}\right \} \times T_{-i}$ such
that $E\cap \hat{C}_{s_{-i}}\not=\emptyset $. Let $E_{m_{K}}$ be an event in 
$\left( E_{k}\right) _{k\geq 1}$\ which is cautiously believed at level $%
m_{K} $. Obviously, $E_{m_{K}}\cap \hat{C}_{s_{-i}}\not=\emptyset $. Since $%
E_{m_{K}}$ is cautiously believed, by condition (ii) of Definition \ref%
{Definition assumption under an LPS}, we have $\mu _{i}^{l}\left(
E_{m_{K}}\cap \hat{C}_{s_{-i}}\right) >0$\ for some $l\leq m_{K}$. Since $%
\mu _{i}^{l}\left( E\right) =1$, we obtain 
\begin{equation*}
0<\mu _{i}^{l}\left( E_{m_{K}}\cap \hat{C}_{s_{-i}}\right) =\mu
_{i}^{l}\left( E_{m_{K}}\cap \hat{C}_{s_{-i}}\cap E\right) \leq \mu
_{i}^{l}\left( E\cap \hat{C}_{s_{-i}}\right) \text{.}
\end{equation*}

Next, let $m_{K}:=\max \left \{ m_{k}:k=1,2,...\right \} $. We show that $%
E=\cup _{k}E_{k}$ is cautiously believed at level $m_{K}$. Let $E_{m_{K}}$
be an event in $\left( E_{k}\right) _{k\geq 1}$\ which is cautiously
believed at level $m_{K}$. For each $l\leq m_{K}$, we have $1=\mu
_{i}^{l}\left( E_{m_{K}}\right) \leq \mu _{i}^{l}\left( E\right) $. For each
elementary cylinder $\hat{C}_{s_{-i}}:=\left \{ s_{-i}\right \} \times
T_{-i} $ with $E\cap \hat{C}_{s_{-i}}\not=\emptyset $, there is $k$ such
that $E_{k}\cap \hat{C}_{s_{-i}}\not=\emptyset $. By condition (ii) of
Definition \ref{Definition assumption under an LPS}, it follows that $0<\mu
_{i}^{l}\left( E_{k}\cap \hat{C}_{s_{-i}}\right) \leq \mu _{i}^{l}\left(
E\cap \hat{C}_{s_{-i}}\right) $\ for some $l\leq m_{k}\leq m_{K}$.

\textbf{Part 2}: Suppose that condition (i) of Definition \ref{Definition
assumption under an LPS}\ and condition (ii') are satisfied. Then condition
(ii') implies%
\begin{equation*}
E\subseteq \mathrm{Proj}_{S_{-i}}^{-1}\left( \text{\textrm{Proj}}%
_{S_{-i}}\left( E\right) \right) =\mathrm{Proj}_{S_{-i}}^{-1}\left( \left(
\tbigcup_{l\leq m}\mathrm{Supp}\text{\textrm{marg}}_{S_{-i}}\mu
_{i}^{l}\right) \right) =\left( \tbigcup_{l\leq m}\mathrm{Supp}\text{\textrm{%
marg}}_{S_{-i}}\mu _{i}^{l}\right) \times T_{-i}\text{,}
\end{equation*}%
i.e., condition (ii\textquotedblright ) in Lemma B.1 holds. Hence $E$\ is
cautiously believed under $\beta _{i}\left( t_{i}\right) $.

For the converse, suppose that $E$\ is cautiously believed under $\beta
_{i}\left( t_{i}\right) :=(\mu _{i}^{1},...,\mu _{i}^{n})$ at level $m$. By
Lemma B.1, it follows that%
\begin{equation*}
\text{\textrm{Proj}}_{S_{-i}}\left( E\right) \subseteq \text{\textrm{Proj}}%
_{S_{-i}}\left( \left( \tbigcup_{l\leq m}\mathrm{Supp}\text{\textrm{marg}}%
_{S_{-i}}\mu _{i}^{l}\right) \times T_{-i}\right) =\tbigcup_{l\leq m}\mathrm{%
Supp}\text{\textrm{marg}}_{S_{-i}}\mu _{i}^{l}\text{.}
\end{equation*}%
To show that this set inclusion holds with equality, let $s_{-i}\notin $%
\textrm{Proj}$_{S_{-i}}\left( E\right) $. Then $\left( \left \{
s_{-i}\right
\} \times T_{-i}\right) \cap E=\emptyset $. By condition (i)
of Definition \ref{Definition assumption under an LPS}, $\mu _{i}^{l}\left(
E\right) =1$ for each $l\leq m$, so%
\begin{equation*}
\mu _{i}^{l}\left( \left \{ s_{-i}\right \} \times T_{-i}\right) =\mathrm{%
marg}_{S_{-i}}\mu _{i}^{l}\left( \left \{ s_{-i}\right \} \right) =0\text{.}
\end{equation*}%
This implies $s_{-i}\notin \mathrm{Supp}$\textrm{marg}$_{S_{-i}}\mu _{i}^{l}$%
.\hfill $\blacksquare $

\section*{Appendix C. Proofs for Section \protect \ref{Section on the main
result}}

In this section we first show that, for a given type structure $\mathcal{T}%
:=\langle S_{i},T_{i},\beta _{i}\rangle _{i\in I}$, the sets $R_{i}^{m}$, $%
m>1$, are Borel subsets of $S_{i}\times T_{i}$. Then we provide the proofs
of Lemmas \ref{Main Lemma finite type structure}-\ref{Lemma conjecture
referee}, as well as the proofs of Theorem \ref{Theorem SAS} and Theorem \ref%
{Theorem insufficiency belief-completeness}.

We begin by showing that, for a given type structure $\mathcal{T}:=\langle
S_{i},T_{i},\beta _{i}\rangle _{i\in I}$, the set $\mathbf{B}%
_{i}^{c}(E)\subseteq S_{i}\times T_{i}$\ is Borel for every event $%
E\subseteq S_{-i}\times T_{-i}$.

\bigskip

\noindent \textbf{Lemma C.1} \textit{Fix a type structure }$\mathcal{T}%
:=\langle S_{i},T_{i},\beta _{i}\rangle _{i\in I}$\textit{\ and non-empty
event }$E\subseteq S_{-i}\times T_{-i}$\textit{. Then the set of all }$\bar{%
\mu}\in \mathcal{N}(S_{-i}\times T_{-i})$\textit{\ under which }$E$\textit{\
is cautiously believed is Borel in }$\mathcal{N}(S_{-i}\times T_{-i})$%
\textit{.}

\bigskip

\noindent \textbf{Proof}. Recall that, for any event $E\subseteq
S_{-i}\times T_{-i}$, the set of probability measures $\mu $\ satisfying $%
\mu \left( E\right) =p$\ for $p\in \mathbb{Q}\cap \left[ 0,1\right] $\ is
measurable in $\mathcal{M}(S_{-i}\times T_{-i})$. So the sets of all $\mu
\in \mathcal{M}(S_{-i}\times T_{-i})$\ satisfying $\mu \left( E\right) =1$
or $\mu \left( E\right) =0$\ are Borel in $\mathcal{M}(S_{-i}\times T_{-i})$%
. Now, fix $n$\ and $m\leq n$. By the above argument and by definition of $%
\mathcal{N}_{n}(S_{-i}\times T_{-i})$, it turns out that the set%
\begin{equation*}
C_{n,m}^{1}:=\tbigcap_{l\leq m}\left \{ \bar{\mu}\in \mathcal{N}%
_{n}(S_{-i}\times T_{-i}):\mu ^{l}\left( E\right) =1\right \}
\end{equation*}%
is Borel in $\mathcal{N}_{n}(S_{-i}\times T_{-i})$. Note that $C_{n,m}^{1}$\
is the set of all $\bar{\mu}\in \mathcal{N}_{n}(S_{-i}\times T_{-i})$\ for
which condition (i) of Definition \ref{Definition assumption under an LPS}
holds for level $m$.

By the same argument, it follows that, for every $s_{-i}\in \mathrm{Proj}%
_{S_{-i}}\left( E\right) $, the set%
\begin{equation*}
C_{n,m}^{s_{-i}}:=\tbigcap_{l\leq m}\left \{ \bar{\mu}\in \mathcal{N}%
_{n}(S_{-i}\times T_{-i}):\mu ^{l}\left( \left \{ s_{-i}\right \} \times
T_{-i}\right) =0\right \}
\end{equation*}%
is Borel in $\mathcal{N}_{n}(S_{-i}\times T_{-i})$. Note that the set%
\begin{equation*}
C_{n,m}^{2}:=\tbigcap_{s_{-i}\in \mathrm{Proj}_{S_{-i}}\left( E\right)
}\left( \mathcal{N}_{n}(S_{-i}\times T_{-i})\backslash
C_{n,m}^{s_{-i}}\right)
\end{equation*}%
is the (measurable) set of all $\bar{\mu}\in \mathcal{N}_{n}(S_{-i}\times
T_{-i})$\ satisfying condition (ii) of Definition \ref{Definition assumption
under an LPS} for level $m$. Define $C_{n,m}:=C_{n,m}^{1}\cap C_{n,m}^{2}$;
clearly, $C_{n,m}$\ is a Borel subset of $\mathcal{N}_{n}(S_{-i}\times
T_{-i})$. Hence, the set of all $\bar{\mu}\in \mathcal{N}(S_{-i}\times
T_{-i})$\ under which $E$ is cautiously believed is given by $\cup _{n\in 
\mathbb{N}}\cup _{m\in \mathbb{N}}C_{n,m}$, so it is Borel in $\mathcal{N}%
(S_{-i}\times T_{-i})$.\hfill $\blacksquare $

\bigskip

By measurability of each belief map\ in a lexicographic type structure, we
obtain the following result.

\bigskip

\noindent \textbf{Corollary C.1 }\textit{Fix a type structure }$\mathcal{T}%
:=\langle S_{i},T_{i},\beta _{i}\rangle _{i\in I}$\textit{. For every }$i\in
I$\textit{, if }$E\subseteq S_{-i}\times T_{-i}$\textit{\ is a non-empty
event, then} $\mathbf{B}_{i}^{c}(E)$\textit{\ is a Borel subset of }$%
S_{i}\times T_{i}$.

\bigskip

We can state and prove the desired result.

\bigskip

\noindent \textbf{Lemma C.2} \textit{Fix a type structure }$\mathcal{T}%
:=\langle S_{i},T_{i},\beta _{i}\rangle _{i\in I}$\textit{. Then, for each }$%
i\in I$\textit{\ and} \textit{each }$m\geq 1$\textit{,}%
\begin{equation*}
R_{i}^{m+1}=R_{i}^{1}\cap \left( \tbigcap_{l\leq m}\mathbf{B}_{i}^{c}\left(
R_{-i}^{l}\right) \right) \text{,}
\end{equation*}%
\textit{and }$R_{i}^{m}$\textit{\ is Borel in }$S_{i}\times T_{i}$\textit{.}

\bigskip

\noindent \textbf{Proof}. The equality $R_{i}^{m+1}=R_{i}^{1}\cap \left(
\cap _{l\leq m}\mathbf{B}_{i}^{c}\left( R_{-i}^{l}\right) \right) $\ is
obvious. By Corollary D.2 in Catonini and De Vito (2020)\textbf{,} it
follows that, for each $i\in I$, the set $R_{i}^{1}:=R_{i}\cap C_{i}$ is
Borel in $S_{i}\times T_{i}$. By Corollary C.1, the set $\mathbf{B}%
_{i}^{c}\left( R_{-i}^{1}\right) $\ is Borel in $S_{i}\times T_{i}$. The
conclusion follows from an easy induction on $m$.\hfill $\blacksquare $

\bigskip

\noindent \textbf{Proof of Lemma \ref{Main Lemma finite type structure}}.
Let $M\geq 1$ be the smallest natural number such that $\prod_{i\in
I}S_{i}^{\infty }=\prod_{i\in I}S_{i}^{M}$.\footnote{%
Note that, if $S^{0}=S^{1}$, then $M$ is $1$ and not $0$. This will simplify
the exposition.}\ By Lemma E.1 in BFK, for every $n\in \left \{
1,...,M+1\right \} $ and $s_{i}\in S_{i}^{n}$, there exists $\mu
_{s_{i}}^{n}\in \mathcal{M}(S_{-i})$ such that $\mathrm{Supp}\mu
_{s_{i}}^{n}=S_{-i}^{n-1}$\ and%
\begin{equation*}
\pi _{i}(s_{i},\mu _{s_{i}}^{n})\geq \pi _{i}(s_{i}^{\prime },\mu
_{s_{i}}^{n})\text{, }\forall s_{i}^{\prime }\in S_{i}\text{.}
\end{equation*}%
We use this result to construct a finite type structure $\mathcal{T}%
:=\left
\langle S_{i},T_{i},\beta _{i}\right \rangle _{i\in I}$ as follows.

For each $i\in I$, let $T_{i}:=S_{i}^{1}$, and define each belief map $\beta
_{i}:T_{i}\rightarrow \mathcal{N}(S_{-i}\times T_{-i})$ as follows. Pick any 
$s_{i}\in T_{i}$. Fix also an arbitrary $\bar{s}_{-i}\in T_{-i}$, and define 
$\nu _{s_{i}}^{1}\in \mathcal{M(}S_{-i}\times T_{-i})$ as%
\begin{equation*}
\nu _{s_{i}}^{1}\left( \left \{ \left( s_{-i},\bar{s}_{-i}\right) \right \}
\right) :=\mu _{s_{i}}^{1}\left( \left \{ s_{-i}\right \} \right) \text{, }%
\forall s_{-i}\in S_{-i}\text{.}
\end{equation*}%
Next, let $m:=\max \left \{ k\leq M+1:s_{i}\in S_{i}^{k}\right \} $. (Note
that if $s_{i}\in S_{i}^{M}$, then $m=M+1$, because $S_{i}^{M}=S_{i}^{M+1}$%
.) So, if $m=1$, let $\beta _{i}(s_{i}):=\left( \nu _{s_{i}}^{1}\right) $.
Otherwise, for each $k=2,...,m$, define $\nu _{s_{i}}^{k}\in \mathcal{M(}%
S_{-i}\times T_{-i})$\ as%
\begin{equation*}
\nu _{s_{i}}^{k}\left( \left \{ \left( s_{-i},s_{-i}\right) \right \}
\right) :=\mu _{s_{i}}^{k}\left( \left \{ s_{-i}\right \} \right) \text{, }%
\forall s_{-i}\in S_{-i}^{k-1}\text{,}
\end{equation*}%
and let%
\begin{equation*}
\beta _{i}(s_{i}):=\left( \nu _{s_{i}}^{m},...,\nu _{s_{i}}^{1}\right) \text{%
.}
\end{equation*}%
Finiteness of each type set guarantees that each belief map is Borel
measurable (in fact, continuous). This completes the definition of the type
structure $\mathcal{T}$.

We now show that $\mathcal{T}$\ satisfies the required properties. To this
end, we find it convenient to define, for each $i\in I$ and $k=1,...,M$, the
following sets:%
\begin{equation*}
\Delta _{S_{i}^{k}\times T_{i}}:=\left \{ (s_{i},s_{i}^{\prime })\in
S_{i}^{k}\times T_{i}:s_{i}=s_{i}^{\prime }\right \} \text{.}
\end{equation*}%
That is, each set $\Delta _{S_{i}^{k}\times T_{i}}$\ is homeomorphic to the
diagonal of $S_{i}^{k}\times S_{i}^{k}$.\footnote{%
The diagonal of $S_{i}^{k}\times S_{i}^{k}$\ is the set $\left \{ \left(
s_{i},s_{i}^{\prime }\right) \in S_{i}^{k}\times
S_{i}^{k}:s_{i}=s_{i}^{\prime }\right \} $.} Next note that, for every $%
s_{i}\in S_{i}^{2}$, all the component measures of $\beta
_{i}(s_{i}):=\left( \nu _{s_{i}}^{m},...,\nu _{s_{i}}^{1}\right) $\ except
for $\nu _{s_{i}}^{1}$ are concentrated on those \textquotedblleft
diagonal\textquotedblright \ sets, namely%
\begin{equation*}
\mathrm{Supp}\nu _{s_{i}}^{k}=\Delta _{S_{-i}^{k-1}\times T_{-i}}\text{, }%
k=2,...,m\text{,}
\end{equation*}%
which implies $\mathrm{Supp}\nu _{s_{i}}^{k}\subseteq \mathrm{Supp}\nu
_{s_{i}}^{k-1}$ for $k\geq 3$.

The rest of the proof is by induction.

\textbf{Induction Hypothesis} ($n$): For each $i\in I$\textbf{, }$\mathrm{%
Proj}_{S_{i}}\left( R_{i}^{n}\right) =S_{i}^{n}$; moreover, $\Delta
_{S_{i}^{n}\times T_{i}}\subseteq R_{i}^{n}$ if $n\leq M$, and $\Delta
_{S_{i}^{M}\times T_{i}}\subseteq R_{i}^{n}$ if $n>M$.

\textbf{Basis Step} ($n=1$).\textbf{\ }Fix $i\in I$ and $s_{i}\in S_{i}^{1}$%
. Type $s_{i}$ is cautious, because%
\begin{equation*}
\mathrm{Suppmarg}_{S_{-i}}\nu _{s_{i}}^{1}=\mathrm{Supp}\mu
_{s_{i}}^{1}=S_{-i}\text{,}
\end{equation*}%
and the strategy-type pair $(s_{i},s_{i})$ is rational, in that%
\begin{equation*}
\overline{\mathrm{marg}}_{S_{-i}}\beta _{i}(s_{i})=\left( \mu
_{s_{i}}^{m},...,\mu _{s_{i}}^{1}\right) \text{.}
\end{equation*}%
This shows that $(s_{i},s_{i})\in R_{i}^{1}$. Therefore $\Delta
_{S_{i}^{1}\times T_{i}}\subseteq R_{i}^{1}$, which implies $%
S_{i}^{1}\subseteq \mathrm{Proj}_{S_{i}}\left( R_{i}^{1}\right) $.
Conversely, Proposition \ref{Proposition CR implies admissibility} yields $%
\mathrm{Proj}_{S_{i}}\left( R_{i}^{1}\right) \subseteq S_{i}^{1}$.

\textbf{Inductive Step} ($n+1$).\textbf{\ }For each $i\in I$, we have to
show that the following properties hold:

(1) $\mathrm{Proj}_{S_{i}}\left( R_{i}^{n+1}\right) =S_{i}^{n+1}$;

(2) $\Delta _{S_{i}^{n+1}\times T_{i}}\subseteq R_{i}^{n+1}$ if $n+1\leq M$,
and $\Delta _{S_{i}^{M}\times T_{i}}\subseteq R_{i}^{n+1}$ if $n+1>M$.

Fix $i\in I$ and $s_{i}\in S_{i}^{n+1}$. Let $k:=\min \left \{
n+1,M+1\right
\} $. Since $(s_{i},s_{i})\in \Delta _{S_{i}^{k-1}\times
T_{i}}$, by the induction hypothesis it follows that $(s_{i},s_{i})\in
R_{i}^{n}$. We show that $(s_{i},s_{i})\in \mathbf{B}_{i}^{c}\left(
R_{-i}^{n}\right) $; this will yield $(s_{i},s_{i})\in R_{i}^{n+1}$. Write $%
\beta _{i}(s_{i}):=\left( \nu _{s_{i}}^{m},...,\nu _{s_{i}}^{1}\right) $,
where $m\geq k$ because $s_{i}\in S_{i}^{k}$ and so, by construction, $\beta
_{i}(s_{i})$\ must have length at least $k$. To show that $R_{-i}^{n}$\ is
cautiously believed under $\beta _{i}(s_{i})$, recall that $\mathrm{Supp}\nu
_{s_{i}}^{l}=\Delta _{S_{-i}^{l-1}\times T_{-i}}\subseteq \Delta
_{S_{-i}^{k-1}\times T_{-i}}$ for each $l=k,...,m$. Since $\Delta
_{S_{-i}^{k-1}\times T_{-i}}\subseteq R_{-i}^{n}$ (induction hypothesis), it
follows that condition (i) of Definition \ref{Definition assumption under an
LPS} is satisfied at level $l=m-k+1$. Recall also that $\mathrm{Supp}\nu
_{s_{i}}^{k}=\Delta _{S_{-i}^{k-1}\times T_{-i}}$. By the induction
hypothesis, $\mathrm{Proj}_{S_{-i}}\left( R_{-i}^{n}\right) =S_{-i}^{k-1}=%
\mathrm{Proj}_{S_{-i}}\left( \Delta _{S_{-i}^{k-1}\times T_{-i}}\right) $.
Hence, $\beta _{i}(s_{i})$ satisfies condition (ii') of Proposition \ref%
{Lemma on properties of assumption one direction conjunction}.2. Thus $%
(s_{i},s_{i})\in \mathbf{B}_{i}^{c}\left( R_{-i}^{n}\right) $, as required.
So, we have shown that $S_{i}^{n+1}\subseteq \mathrm{Proj}_{S_{i}}\left(
R_{i}^{n+1}\right) $. For part (2), note the following fact: If $n+1\leq M$,
then for every $(s_{i},s_{i})\in \Delta _{S_{i}^{n+1}\times T_{i}}$ we have $%
s_{i}\in S_{i}^{n+1}$; analogously, if $n+1>M$, then for every $%
(s_{i},s_{i})\in \Delta _{S_{i}^{M}\times T_{i}}$ we have $s_{i}\in
S_{i}^{n+1}$. Therefore, by proving that $(s_{i},s_{i})\in R_{i}^{n+1}$ for
each $s_{i}\in S_{i}^{n+1} $, we have proven (2).

Conversely, pick any $(s_{i},s_{i}^{\prime })\in R_{i}^{n+1}\subseteq
R_{i}^{n}$.\ Then, by the induction hypothesis, $s_{i}\in S_{i}^{n}$. Let $%
\beta _{i}(s_{i}^{\prime }):=(\mu ^{1},...,\mu ^{n})$. Since $s_{i}^{\prime
} $ cautiously believes $R_{-i}^{n}$ at some level $l$, it follows from
Proposition \ref{Lemma on properties of assumption one direction conjunction}%
.2 and the induction hypothesis that 
\begin{equation*}
\tbigcup_{k\leq l}\mathrm{Suppmarg}_{S_{-i}}\mu ^{k}=S_{-i}^{n}\text{.}
\end{equation*}%
So, by Proposition 1 in Blume et al. (1991b), there exists $\nu \in \mathcal{%
M}(S_{-i})$, with $\mathrm{Supp}\nu =S_{-i}^{n}$, under which $s_{i}$ is
optimal. Therefore, by Remark \ref{Remark second Pearce lemma}, $s_{i}\in
S_{i}^{n+1}$. This shows that $\mathrm{Proj}_{S_{i}}\left(
R_{i}^{n+1}\right) \subseteq S_{i}^{n+1}$, establishing (1).\hfill $%
\blacksquare $

\bigskip

\noindent \textbf{Proof of Lemma \ref{Lemma conjecture referee}}. Fix a type 
$t_{i}$ that cautiously believes $E_{-i}$, and set $\beta _{i}\left(
t_{i}\right) :=\left( \beta _{i}^{1}\left( t_{i}\right) ,...,\beta
_{i}^{n}\left( t_{i}\right) \right) $. Let $t_{i}^{\ast }:=\varphi
_{i}\left( t_{i}\right) $. Note that bimeasurability of $\left( \varphi
_{i}\right) _{i\in I}$\ implies that $\left( \mathrm{Id}_{S_{-i}},\varphi
_{-i}\right) (E_{-i})$\ is an event in $S_{-i}\times T_{-i}^{\ast }$. We
show that $t_{i}^{\ast }$ cautiously believes $\left( \mathrm{Id}%
_{S_{-i}},\varphi _{-i}\right) (E_{-i})$, that is, $\beta _{i}^{\ast }\left(
t_{i}^{\ast }\right) =\widehat{\left( \mathrm{Id}_{S_{-i}},\varphi
_{-i}\right) }\left( \beta _{i}\left( t_{i}\right) \right) $ satisfies
conditions (i) and (ii) of Definition \ref{Definition assumption under an
LPS}.

First, note that%
\begin{equation*}
E_{-i}\subseteq \left( \mathrm{Id}_{S_{-i}},\varphi _{-i}\right) ^{-1}\left(
\left( \mathrm{Id}_{S_{-i}},\varphi _{-i}\right) \left( E_{-i}\right)
\right) \text{.}
\end{equation*}%
Hence, by definition of type morphism, it follows that, for all $l\leq n$,%
\begin{equation*}
\beta _{i}^{l}\left( t_{i}\right) \left( E_{-i}\right) \leq \beta
_{i}^{l}\left( t_{i}\right) \left( \left( \mathrm{Id}_{S_{-i}},\varphi
_{-i}\right) ^{-1}\left( \left( \mathrm{Id}_{S_{-i}},\varphi _{-i}\right)
\left( E_{-i}\right) \right) \right) =\beta _{i}^{\ast ,l}(t_{i}^{\ast
})\left( \left( \mathrm{Id}_{S_{-i}},\varphi _{-i}\right) \left(
E_{-i}\right) \right) \text{.}
\end{equation*}%
Since $E_{-i}$\ is cautiously believed under $\beta _{i}\left( t_{i}\right) $%
, it follows from condition (i) of Definition \ref{Definition assumption
under an LPS} that there exists $m\leq n$ such that $\beta _{i}^{l}\left(
t_{i}\right) \left( E_{-i}\right) =1$ for all $l\leq m$. Therefore, we have
that $\beta _{i}^{\ast ,l}\left( t_{i}^{\ast }\right) \left( \left( \mathrm{%
Id}_{S_{-i}},\varphi _{-i}\right) \left( E_{-i}\right) \right) =1$ for all $%
l\leq m$.\ Hence $\beta _{i}^{\ast }\left( t_{i}^{\ast }\right) $ satisfies
condition (i) of Definition \ref{Definition assumption under an LPS}.

Consider now an elementary cylinder $\hat{C}_{s_{-i}}:=\left \{
s_{-i}\right
\} \times T_{-i}^{\ast }$ satisfying $\left( \mathrm{Id}%
_{S_{-i}},\varphi _{-i}\right) \left( E_{-i}\right) \cap \hat{C}%
_{s_{-i}}\neq \emptyset $. First, note that%
\begin{eqnarray*}
\left( \left \{ s_{-i}\right \} \times T_{-i}\right) \cap E_{-i} &\subseteq
&\left( \left \{ s_{-i}\right \} \times T_{-i}\right) \cap \left( \left( 
\mathrm{Id}_{S_{-i}},\varphi _{-i}\right) ^{-1}\left( \left( \mathrm{Id}%
_{S_{-i}},\varphi _{-i}\right) \left( E_{-i}\right) \right) \right) \\
&=&\left( \left( \mathrm{Id}_{S_{-i}},\varphi _{-i}\right) ^{-1}\left( \hat{C%
}_{s_{-i}}\right) \right) \cap \left( \left( \mathrm{Id}_{S_{-i}},\varphi
_{-i}\right) ^{-1}\left( \left( \mathrm{Id}_{S_{-i}},\varphi _{-i}\right)
\left( E_{-i}\right) \right) \right) \\
&=&\left( \mathrm{Id}_{S_{-i}},\varphi _{-i}\right) ^{-1}\left( \hat{C}%
_{s_{-i}}\cap \left( \mathrm{Id}_{S_{-i}},\varphi _{-i}\right) \left(
E_{-i}\right) \right) \text{.}
\end{eqnarray*}%
Hence, by definition of type morphism, it follows that, for all $l\leq n$,%
\begin{eqnarray*}
\beta _{i}^{l}\left( t_{i}\right) \left( \left( \left \{ s_{-i}\right \}
\times T_{-i}\right) \cap E_{-i}\right) &\leq &\beta _{i}^{l}\left(
t_{i}\right) \left( \left( \mathrm{Id}_{S_{-i}},\varphi _{-i}\right)
^{-1}\left( \hat{C}_{s_{-i}}\cap \left( \mathrm{Id}_{S_{-i}},\varphi
_{-i}\right) \left( E_{-i}\right) \right) \right) \\
&=&\beta _{i}^{\ast ,l}\left( t_{i}^{\ast }\right) \left( \hat{C}%
_{s_{-i}}\cap \left( \mathrm{Id}_{S_{-i}},\varphi _{-i}\right) \left(
E_{-i}\right) \right) \text{.}
\end{eqnarray*}%
Since $E_{-i}$\ is cautiously believed under $\beta _{i}\left( t_{i}\right) $
at level $m\leq n$, and since $\hat{C}_{s_{-i}}\cap \left( \mathrm{Id}%
_{S_{-i}},\varphi _{-i}\right) \left( E_{-i}\right) \neq \emptyset $ implies 
$\left( \left \{ s_{-i}\right \} \times T_{-i}\right) \cap E_{-i}\neq
\emptyset $, by condition (ii) of Definition \ref{Definition assumption
under an LPS} there exists $k\leq m$\ such that $\beta _{i}^{k}\left(
t_{i}\right) \left( \left( \left \{ s_{-i}\right \} \times T_{-i}\right)
\cap E_{-i}\right) >0$. Therefore, we obtain%
\begin{equation*}
\beta _{i}^{\ast ,k}\left( t_{i}^{\ast }\right) \left( \hat{C}_{s_{-i}}\cap
\left( \mathrm{Id}_{S_{-i}},\varphi _{-i}\right) \left( E_{-i}\right)
\right) >0\text{.}
\end{equation*}%
Thus, $\beta _{i}^{\ast }\left( t_{i}^{\ast }\right) $ satisfies condition
(ii) of Definition \ref{Definition assumption under an LPS}.\hfill $%
\blacksquare $

\bigskip

\noindent \textbf{Proof of Theorem \ref{Theorem SAS}}. \textbf{Part (i)}:%
\textbf{\ }Fix a type structure $\mathcal{T}:=\langle S_{i},T_{i},\beta
_{i}\rangle _{i\in I}$. If\ $\tprod \nolimits_{i\in I}\mathrm{Proj}%
_{S_{i}}\left( R_{i}^{\infty }\right) =\emptyset $, then the result is
immediate. So in what follows we will assume that this set is non-empty. For
each $i\in I$ and $s_{i}\in \mathrm{Proj}_{S_{i}}\left( R_{i}^{\infty
}\right) $, there exists $t_{i}\in T_{i}$ such that $(s_{i},t_{i})\in
R_{i}^{\infty }$. Since $(s_{i},t_{i})\in R_{i}^{1}$, it follows that $s_{i}$
is a lexicographic best reply to $\overline{\mathrm{marg}}_{S_{-i}}\beta
_{i}(t_{i})\in \mathcal{N}^{+}(S_{-i})$. Therefore, by Proposition \ref%
{Proposition CR implies admissibility}, $s_{i}$ is admissible, hence
condition (a) of Definition \ref{Definition SAS} is satisfied.

Next note that, for each $k\geq 1$, type $t_{i}$ cautiously believes $%
R_{-i}^{k}$. So, it follows from Proposition \ref{Lemma on properties of
assumption one direction conjunction}.1 that $R_{-i}^{\infty }$ is
cautiously believed under $\beta _{i}(t_{i}):=(\mu ^{1},...,\mu ^{n})$ at
some level $m$. Moreover, Proposition \ref{Lemma on properties of assumption
one direction conjunction}.2 entails $\cup _{l\leq m}\mathrm{Suppmarg}%
_{S_{-i}}\mu ^{l}=\mathrm{Proj}_{S_{-i}}\left( R_{-i}^{\infty }\right) $.
Since $s_{i}$ is a lexicographic best reply to $\overline{\mathrm{marg}}%
_{S_{-i}}\beta _{i}(t_{i})$, Proposition 1 in Blume et al. (1991b) yields
the existence of some $\nu \in \mathcal{M}(S_{-i})$ under which $s_{i}$\ is
optimal and such that $\mathrm{Supp}\nu =\mathrm{Proj}_{S_{-i}}\left(
R_{-i}^{\infty }\right) $. Remark \ref{Remark second Pearce lemma}\ entails
that $s_{i}$ is admissible with respect to $S_{i}\times \mathrm{Proj}%
_{S_{-i}}\left( R_{-i}^{\infty }\right) $, establishing condition (b) of
Definition \ref{Definition SAS}.

Finally, by Corollary A1 in Brandenburger and Friedenberg (2010), every $%
s_{i}^{\prime }$ that supports $s_{i}$ is a lexicographic best reply to $%
\overline{\mathrm{marg}}_{S_{-i}}\beta _{i}(t_{i})$ as well. It follows that 
$(s_{i}^{\prime },t_{i})\in R_{i}^{\infty }$, and this in turn implies that $%
s_{i}^{\prime }\in \mathrm{Proj}_{S_{i}}\left( R_{i}^{\infty }\right) $,
establishing condition (c) of Definition \ref{Definition SAS}.

\textbf{Part (ii)}: Let $Q\in \mathcal{Q}$ be a non-empty SAS. Fix $i\in I$
and $s_{i}\in Q_{i}$. By conditions (a)\ and (b) of Definition \ref%
{Definition SAS}, and by Remark \ref{Remark second Pearce lemma}, there
exist $\nu _{s_{i}}^{2},\nu _{s_{i}}^{1}\in \mathcal{M}(S_{-i})$ such that $%
\mathrm{Supp}\nu _{s_{i}}^{2}=S_{-i}$ and $\mathrm{Supp}\nu
_{s_{i}}^{1}=Q_{-i}$, and such that $s_{i}$ is optimal under $\nu
_{s_{i}}^{2}$ and $\nu _{s_{i}}^{1}$. Hence, $s_{i}$ is a lexicographic best
reply to $(\nu _{s_{i}}^{1},\nu _{s_{i}}^{2})\in \mathcal{N}^{+}(S_{-i})$.
Moreover, as in BFK (p. 328), we can choose $\nu _{s_{i}}^{2}$ and $\nu
_{s_{i}}^{1}$ in such a way that every strategy $s_{i}^{\prime }$ is optimal
under $\nu _{s_{i}}^{2}$ and $\nu _{s_{i}}^{1}$ if and only if $%
s_{i}^{\prime }$ is supported by $s_{i}$. Now we construct a finite type
structure $\mathcal{T}:=\langle S_{i},T_{i},\beta _{i}\rangle _{i\in I}$ as
follows.

For each $i\in I$, let $T_{i}:=Q_{i}$. For every $s_{i}\in T_{i}$, define $%
\mu _{s_{i}}^{1},\mu _{s_{i}}^{2}\in \mathcal{M}(S_{-i}\times T_{-i})$ as%
\begin{eqnarray*}
\mu _{s_{i}}^{1}\left( \left \{ \left( s_{-i},s_{-i}\right) \right \}
\right) &:&=\nu _{s_{i}}^{1}(\left \{ s_{-i}\right \} ),\forall s_{-i}\in
Q_{-i}\text{,} \\
\mu _{s_{i}}^{2}\left( \left \{ \left( s_{-i},\bar{s}_{-i}\right) \right \}
\right) &:&=\nu _{s_{i}}^{2}(\left \{ s_{-i}\right \} ),\forall s_{-i}\in
S_{-i}\text{,}
\end{eqnarray*}%
where $\bar{s}_{-i}\in T_{-i}$\ is arbitrarily chosen. Let $\beta
_{i}(s_{i}):=(\mu _{s_{i}}^{1},\mu _{s_{i}}^{2})$. Finiteness of each type
set guarantees that each belief map is measurable (in fact, continuous).
This completes the definition of the type structure $\mathcal{T}$.

We now show that $\mathcal{T}$\ satisfies the required properties. Note that
each type $s_{i}\in T_{i}$ is cautious because $\mathrm{Supp}\nu
_{s_{i}}^{2}=S_{-i}$; hence, $\mathcal{T}$\ is a cautious type structure.
For every $i\in I$ and $s_{i}\in Q_{i}$, strategy-type pair $(s_{i},s_{i})$
is cautiously rational by construction; for every $s_{i}^{\prime }\notin
Q_{i}$, condition (c) of Definition \ref{Definition SAS} implies that $%
s_{i}^{\prime }$ does not support $s_{i}$, so by construction the pair $%
(s_{i}^{\prime },s_{i})$ is not rational. Hence, $\mathrm{Proj}%
_{S_{i}}\left( R_{i}^{1}\right) =Q_{i}$. Now, suppose by way of induction
that $(s_{i},s_{i})\in R_{i}^{m}$ for each $i\in I$ and each $s_{i}\in Q_{i}$%
. We show that type $s_{i}$ cautiously believes $R_{-i}^{m}$, establishing
that $(s_{i},s_{i})\in R_{i}^{m+1}$; this will yield $(s_{i},s_{i})\in
R_{i}^{\infty }$. Note that $\mathrm{Supp}\mu _{s_{i}}^{1}=\left \{
(s_{-i},s_{-i}):s_{-i}\in Q_{-i}\right \} \subseteq R_{-i}^{m}$, where the
inclusion follows from the induction hypothesis. Moreover, since $\mathrm{%
Suppmarg}_{S_{-i}}\mu _{s_{i}}^{1}=\mathrm{Supp}\nu _{s_{i}}^{1}=Q_{-i}$,
Proposition \ref{Lemma on properties of assumption one direction conjunction}%
.2\ entails that $R_{-i}^{m}$ is cautiously believed under $\beta
_{i}(s_{i}) $ at level $1$. Therefore, we conclude that $\mathrm{Proj}%
_{S_{i}}\left( R_{i}^{\infty }\right) =Q_{i}$.\hfill $\blacksquare $

\bigskip

For the proof of Theorem \ref{Theorem insufficiency belief-completeness}, we
need an auxiliary technical fact.

\bigskip

\noindent \textbf{Lemma C.3} \textit{Fix two sequences of pairwise disjoint
topological spaces }$\left( X_{n}\right) _{n\in \mathbb{N}}$\textit{\ and }$%
\left( Y_{n}\right) _{n\in \mathbb{N}}$\textit{.} \textit{Let }$X:=\cup
_{n\in \mathbb{N}}X_{n}$\textit{\ and }$Y:=\cup _{n\in \mathbb{N}}Y_{n}$. 
\textit{Suppose that, for each }$n\in \mathbb{N}$\textit{, there is a map }$%
f_{n}:X_{n}\rightarrow Y_{n}$\textit{. If each map }$f_{n}$\textit{\ is
continuous (resp. surjective), then the union map }$\cup _{n\in \mathbb{N}%
}f_{n}:X\rightarrow Y$\textit{\ is continuous (resp. surjective).}

\bigskip

\noindent \textbf{Proof}. Let $O$ be open in $Y$. By definition of direct
sum topology, the set $O$\ can be written as $O=\cup _{n\in \mathbb{N}}O_{n}$%
, where each $O_{n}:=O\cap Y_{n}$\ is open in $Y_{n}$ (see Engelking 1989,
p. 74). Thus,%
\begin{equation*}
\left( \tbigcup_{n\in \mathbb{N}}f_{n}\right) ^{-1}\left( O\right)
=\tbigcup_{n\in \mathbb{N}}f_{n}^{-1}\left( O_{n}\right) \text{.}
\end{equation*}%
So, if each $f_{n}$\ is continuous, then each $f_{n}^{-1}\left( O_{n}\right) 
$\ is open, and this in turn implies that $\left( \cup _{n\in \mathbb{N}%
}f_{n}\right) ^{-1}\left( O\right) $\ is open. The conclusion that $\cup
_{n\in \mathbb{N}}f_{n}$\ is surjective if each $f_{n}$\ is surjective is
immediate by inspection of the definitions.\hfill $\blacksquare $

\bigskip

\noindent \textbf{Proof of Theorem \ref{Theorem insufficiency
belief-completeness}}. The desired type structure $\mathcal{T}:=\langle
S_{i},T_{i},\beta _{i}\rangle _{i\in I}$ is constructed as follows. For each 
$i\in I$, let $T_{i}$ be the Baire space $\mathbb{N}_{0}^{\mathbb{N}_{0}}$,%
\footnote{%
Here $\mathbb{N}_{0}$\ denotes the set $\left \{ 0,1,2,...\right \} $, i.e., 
$\mathbb{N}_{0}:=\mathbb{N}\cup \left \{ 0\right \} $. The Baire space is
sometimes defined as the set $\mathbb{N}^{\mathbb{N}}$\ of all infinite
sequences of natural numbers. This difference is immaterial for all the
relevant topological properties we are going to use in this proof.} so that
each $t_{i}\in T_{i}$\ is an infinite sequence of non-negative integers. The
set $\mathbb{N}_{0}$\ is endowed with the discrete topology, and $\mathbb{N}%
_{0}^{\mathbb{N}_{0}}$ is endowed with the product topology. The basic open
sets of $\mathbb{N}_{0}^{\mathbb{N}_{0}}$ are sets of the form%
\begin{equation*}
O_{k}:=\left \{ \left( n_{1},n_{2},...\right) \in \mathbb{N}_{0}^{\mathbb{N}%
_{0}}:(n_{1},...,n_{k})=\left( o_{1},...,o_{k}\right) \right \}
\end{equation*}%
for each $k\in \mathbb{N}_{0}$\ and $(o_{1},...,o_{k})\in \left( \mathbb{N}%
_{0}\right) ^{k}$. With this topology, a basic open set is also closed, so
sets of the form $O_{k}$\ constitute a clopen basis. The space $\mathbb{N}%
_{0}^{\mathbb{N}_{0}}$ is Polish and uncountable, but not compact.

For each $i\in I$, we partition $T_{i}$ into a countable family of non-empty
Borel subsets. For each $k\geq 0$, let%
\begin{equation*}
\mathrm{T}_{i}^{k}:=\left \{ \left( n_{1},n_{2},...\right) \in \mathbb{N}%
_{0}^{\mathbb{N}_{0}}:n_{1}=k\right \} \text{.}
\end{equation*}%
Each $\mathrm{T}_{i}^{k}$ is a subbasic clopen subset of $T_{i}$; moreover,
each $\mathrm{T}_{i}^{k}$ is homeomorphic to the Baire space. It is clear
that $T_{i}=\cup _{k\geq 0}\mathrm{T}_{i}^{k}$, and all the $\mathrm{T}%
_{i}^{k}$'s are pairwise disjoint.

The next step is to construct the belief maps in such a way that, for all $%
k\geq 0$, $t_{i}\in \mathrm{T}_{i}^{k}$ and $s_{i}\in S_{i}$, the pair $%
(s_{i},t_{i})$ \emph{does not} belong to $R_{i}^{k+1}$. For each $i\in I$,
we construct a countable partition of $\mathcal{N}(S_{-i}\times T_{-i})$\
that \textquotedblleft mirrors\textquotedblright \ the above partition of $%
T_{i}$. This is done as follows: For each $i\in I$, let%
\begin{equation*}
\Lambda _{i}^{0}:=\mathcal{N}(S_{-i}\times T_{-i})\backslash \mathcal{C}%
_{i}^{0}\text{,}
\end{equation*}%
where $\mathcal{C}_{i}^{0}$ is the set of all LPS's $\bar{\mu}_{i}\in 
\mathcal{N}(S_{-i}\times T_{-i})$ such that $\overline{\mathrm{marg}}%
_{S_{-i}}\bar{\mu}_{i}\in \mathcal{N}^{+}\left( S_{-i}\right) $. Since $%
\left \vert S_{i}\right \vert \geq 2$ for each $i\in I$, it follows that $%
\Lambda _{i}^{0}\neq \emptyset $.

Next, let%
\begin{equation*}
\Lambda _{i}^{1}:=\left \{ \bar{\mu}_{i}\in \mathcal{C}_{i}^{0}:\mu
_{i}^{1}(S_{-i}\times \mathrm{T}_{-i}^{0})>0\right \} \text{,}
\end{equation*}%
and, for each $k\geq 2$,%
\begin{equation*}
\Lambda _{i}^{k}:=\tbigcap_{m\in \left \{ 1,...,k-1\right \} }\left \{ \bar{%
\mu}_{i}\in \mathcal{C}_{i}^{0}:\mu _{i}^{1}(S_{-i}\times \mathrm{T}%
_{-i}^{m-1})=0\right \} \cap \left \{ \bar{\mu}_{i}\in \mathcal{C}%
_{i}^{0}:\mu _{i}^{1}(S_{-i}\times \mathrm{T}_{-i}^{k-1})>0\right \} \text{.}
\end{equation*}%
In words: $\Lambda _{i}^{1}$ is the set of all LPS's on $S_{-i}\times T_{-i}$%
\ such that the marginal on $S_{-i}$\ has full support and the first
component measure assigns strictly positive probability to $S_{-i}\times 
\mathrm{T}_{-i}^{0}$; $\Lambda _{i}^{2}$ is the set of all LPS's on $%
S_{-i}\times T_{-i}$\ such that the marginal on $S_{-i}$\ has full support
and the first component measure assigns probability $0$ to $S_{-i}\times 
\mathrm{T}_{-i}^{0}$, and strictly positive probability to $S_{-i}\times 
\mathrm{T}_{-i}^{1}$; and so on.

It is immediate to check that $\mathcal{N}(S_{-i}\times T_{-i})=\cup _{k\geq
0}\Lambda _{i}^{k}$ and all the $\Lambda _{i}^{k}$'s\ are non-empty,
pairwise disjoint sets; so the countable family of all $\Lambda _{i}^{k}$'s\
is a partition of $\mathcal{N}(S_{-i}\times T_{-i})$.

\bigskip

\noindent \textbf{Claim C.1 }\textit{For each }$k\geq 0$, $\Lambda _{i}^{k}$ 
\textit{is a Borel subset of }$\mathcal{N}(S_{-i}\times T_{-i})$.

\bigskip

\noindent \textbf{Proof}. Since $\mathcal{C}_{i}^{0}$\ is Borel (see Lemma
D.2 in Catonini and De Vito, 2020), so is $\Lambda _{i}^{0}$. For each $%
k\geq 1$, let%
\begin{equation*}
\mathrm{P}_{i}^{k}:=\left \{ \bar{\mu}_{i}\in \mathcal{C}_{i}^{0}:\mu
_{i}^{1}(S_{-i}\times \mathrm{T}_{-i}^{k-1})>0\right \} \text{.}
\end{equation*}%
Note that $\Lambda _{i}^{1}=\mathrm{P}_{i}^{1}$, and for each $k\geq 2$, $%
\Lambda _{i}^{k}$ is the intersection of $\mathrm{P}_{i}^{k}$ with the
complements of $\mathrm{P}_{i}^{1},...,\mathrm{P}_{i}^{k-1}$. Thus, in order
to show that each $\Lambda _{i}^{k}$ is Borel in $\mathcal{N}(S_{-i}\times
T_{-i})$, it is sufficient to show that each $\mathrm{P}_{i}^{k}$ is Borel
in $\mathcal{N}(S_{-i}\times T_{-i})$. Let%
\begin{equation*}
\mathrm{M}_{i}^{k}:=\left \{ \mu \in \mathcal{M}\left( S_{-i}\times \mathrm{T%
}_{-i}\right) :\mu (S_{-i}\times \mathrm{T}_{-i}^{k-1})>0\right \} \text{.}
\end{equation*}%
By Theorem 17.24 in Kechris (1995), if $X$\ is a Polish space, then the
Borel $\sigma $-field on $\mathcal{M}\left( X\right) $\ is generated by sets
of the form $\left \{ \mu \in \mathcal{M}\left( X\right) :\mu \left(
E\right) \geq p\right \} $, where $E\in \Sigma _{X}$ and $p\in \mathbb{Q\cap 
}\left[ 0,1\right] $. Hence, for every $E\in \Sigma _{X}$, the set $\left \{
\mu \in \mathcal{M}\left( X\right) :\mu \left( E\right) >0\right \} $ is
Borel, since it can be written as $\cap _{n\in \mathbb{N}}\left \{ \mu \in 
\mathcal{M}\left( X\right) :\mu \left( E\right) \geq \frac{1}{n}\right \} $.
This implies that $\mathrm{M}_{i}^{k}$ is Borel in $\mathcal{M}\left(
S_{-i}\times \mathrm{T}_{-i}\right) $. Moreover, for each $n\in \mathbb{N}$,
the canonical projection map%
\begin{equation*}
\begin{array}{cccc}
\mathrm{Proj}_{1,n}: & \mathcal{N}_{n}(S_{-i}\times T_{-i}) & \rightarrow & 
\mathcal{M}\left( S_{-i}\times \mathrm{T}_{-i}\right) \text{,} \\ 
& \left( \mu _{i}^{1},...,\mu _{i}^{n}\right) & \mapsto & \mu _{i}^{1}\text{,%
}%
\end{array}%
\end{equation*}%
is continuous, hence the set $\mathrm{Proj}_{1,n}^{-1}\left( \mathrm{M}%
_{i}^{k}\right) $\ is Borel in $\mathcal{N}_{n}(S_{-i}\times T_{-i})$. With
this, the conclusion follows from the observation that $\mathrm{P}_{i}^{k}$
can be written as%
\begin{equation*}
\mathrm{P}_{i}^{k}=\left( \tbigcup_{n\in \mathbb{N}}\mathrm{Proj}%
_{1,n}^{-1}\left( \mathrm{M}_{i}^{k}\right) \right) \cap \mathcal{C}_{i}^{0}%
\text{.}
\end{equation*}%
\hfill $\square $

\bigskip

Recall that every Borel subset of a Polish space is a Lusin space, when
endowed with the relative topology. Moreover, every Lusin space is also
analytic (see Cohn 2003, Proposition 8.6.13). Thus, by Claim C.1, each $%
\Lambda _{i}^{k}$ is analytic. Since each $\mathrm{T}_{i}^{k}$ is
homeomorphic to the Baire space, it follows from Corollary 8.2.8 in Cohn
(2003; see also Kechris 1995, p. 85)\ that, for every $k\geq 0$, there
exists a surjective continuous map $\beta _{i}^{\left[ k\right] }:\mathrm{T}%
_{i}^{k}\rightarrow \Lambda _{i}^{k}$. For each $i\in I$, let $\beta _{i}$\
be the union of the $\beta _{i}^{\left[ k\right] }$'s, i.e., $\beta
_{i}:=\cup _{k\geq 0}\beta _{i}^{\left[ k\right] }:T_{i}\rightarrow \mathcal{%
N}(S_{-i}\times T_{-i})$. The map is well defined because the $\mathrm{T}%
_{i}^{k}$'s are pairwise disjoint. By Lemma C.3, $\beta _{i}$ is a
continuous (and so Borel) surjective map. This completes the definition of
the type structure $\mathcal{T}$.

We now show that $\mathcal{T}$ satisfies the required properties.

\bigskip

\noindent \textbf{Claim C.2 }\textit{For each }$i\in I$\textit{\ and for
each }$k\geq 0$\textit{,}%
\begin{equation*}
(S_{i}\times \mathrm{T}_{i}^{k})\cap R_{i}^{k+1}=\emptyset \text{.}
\end{equation*}

\bigskip

\noindent \textbf{Proof}. By induction on $k\geq 0$.

(Basis step: $k=0$) Fix $i\in I$ and $(s_{i},t_{i})\in S_{i}\times T_{i}$
with $t_{i}\in \mathrm{T}_{i}^{0}$. We clearly have $(s_{i},t_{i})\not \in
R_{i}^{1}$ because $\beta _{i}(t_{i})\in \Lambda _{i}^{0}$, hence $t_{i}$ is
not cautious. Therefore $(S_{i}\times \mathrm{T}_{i}^{0})\cap
R_{i}^{1}=\emptyset $.

(Inductive step: $k\geq 1$) Suppose we have already shown that $(S_{i}\times 
\mathrm{T}_{i}^{k-1})\cap R_{i}^{k}=\emptyset $ for each $i\in I$. Fix $i\in
I$ and $(s_{i},t_{i})\in S_{i}\times T_{i}$ with $t_{i}\in \mathrm{T}%
_{i}^{k} $. Thus $\beta _{i}(t_{i}):=(\mu _{i}^{1},...,\mu _{i}^{n})\in
\Lambda _{i}^{k}$, hence $\mu _{i}^{1}(S_{-i}\times \mathrm{T}_{-i}^{k-1})>0$%
.\ Since, by the induction hypothesis, $(S_{-i}\times \mathrm{T}%
_{-i}^{k-1})\cap R_{-i}^{k}=\emptyset $, it must be the case that $\mu
_{i}^{1}(R_{-i}^{k})<1$. Therefore $R_{-i}^{k}$\ is not cautiously believed
under $\beta _{i}(t_{i})$; this implies $(s_{i},t_{i})\not \in \mathbf{B}%
_{i}^{c}(R_{-i}^{k})$. Hence $(s_{i},t_{i})\not \in R_{i}^{k+1}$.\hfill $%
\square $

\bigskip

To conclude the proof, pick any $(s_{i},t_{i})\in S_{i}\times T_{i}$. Then
there exists $k\geq 0$\ such that $t_{i}\in \mathrm{T}_{i}^{k}$. By Claim
C.2, it follows that $(s_{i},t_{i})\not \in R_{i}^{k+1}$. Since $%
R_{i}^{\infty }:=\cap _{m\geq 1}R_{i}^{m}$, this shows that $R_{i}^{\infty
}=\emptyset $, as required.\hfill $\blacksquare $


\begin{thebibliography}{99}
\bibitem{AlBorder} \textsc{Aliprantis, C.D., and K.C. Border} (1999): 
\textit{Infinite Dimensional Analysis}. Berlin: Springer Verlag.

\bibitem{AD} \textsc{Asheim, G., and M. Dufwenberg} (2003):
\textquotedblleft Admissibility and Common Belief,\textquotedblright \ 
\textit{Games and Economic Behavior,} 42, 208-234.

\bibitem{Barelli Galanis 2013} \textsc{Barelli, P., and S. Galanis} (2013):
\textquotedblleft Admissibility and Event-Rationality,\textquotedblright \ 
\textit{Games and Economic Behavior}, 77, 21--40.

\bibitem{Batti97jet} \textsc{Battigalli, P.} (1997): \textquotedblleft On
Rationalizability in Extensive Games,\textquotedblright \  \textit{Journal of
Economic Theory}, 74, 40-61\textit{.}

\bibitem{battifriede2012TE} \textsc{Battigalli, P., and A. Friedenberg}
(2012a): \textquotedblleft Forward Induction Reasoning
Revisited,\textquotedblright \  \textit{Theoretical Economics}, 7, 57-98.

\bibitem{BattiFriedenberg2009} \textsc{Battigalli, P., and A. Friedenberg}
(2012b): \textquotedblleft Context-Dependent Forward Induction
Reasoning,\textquotedblright \ IGIER Working Paper 351, Bocconi University.

\bibitem{BattiFriedSini2021} \textsc{Battigalli, P., A. Friedenberg and M.
Siniscalchi} (2021): \textit{Strategic Uncertainty: An Epistemic Approach to
Game Theory}. (Working Title).

\bibitem{BS} \textsc{Battigalli, P., and M. Siniscalchi} (2002):
\textquotedblleft Strong Belief and Forward Induction
Reasoning,\textquotedblright \  \textit{Journal of Economic Theory}, 106,
356-391.

\bibitem{BBD1} \textsc{Blume, L., A. Brandenburger, and E. Dekel} (1991a):
\textquotedblleft Lexicographic Probabilities and Choice under
Uncertainty,\textquotedblright \  \textit{Econometrica,}\textbf{\ }59, 61-79.

\bibitem{BBD} \textsc{Blume, L., A. Brandenburger, and E. Dekel} (1991b):
\textquotedblleft Lexicographic Probabilities and Equilibrium
Refinements,\textquotedblright \  \textit{Econometrica,}\textbf{\ }59, 81-98.

\bibitem{Brandenburger 1992} \textsc{Brandenburger, A.} (1992):
\textquotedblleft Lexicographic Probabilities and Iterated
Admissibility,\textquotedblright \ in \textit{Economic Analysis of Markets
and Games}, ed. by P. Dasgupta, D. Gale, O. Hart, and E. Maskin. Cambridge
MA: MIT Press, 282-290.

\bibitem{Brandenburger 03} \textsc{Brandenburger, A.} (2003):
\textquotedblleft On the Existence of a `Complete' Possibility
Structure,\textquotedblright \ in \textit{Cognitive Processes and Economic
Behavior}, ed. by M. Basili, N. Dimitri and I. Gilboa. New York: Routledge,
30-34.

\bibitem{Brandenburger Friedenberg 2010} \textsc{Brandenburger, A., and A.
Friedenberg} (2010): \textquotedblleft Self-Admissible
Sets,\textquotedblright \  \textit{Journal of Economic Theory}, 145, 785-811.

\bibitem{BFK} \textsc{Brandenburger, A., A. Friedenberg, and H.J. Keisler}
(2008): \textquotedblleft Admissibility in Games,\textquotedblright \ 
\textit{Econometrica}, 76, 307-352.

\bibitem{BFK 2012} \textsc{Brandenburger, A., A. Friedenberg, and H.J.
Keisler} (2012): \textquotedblleft Fixed Points in Epistemic Game
Theory,\textquotedblright \ in \textit{Proceedings of the 2008 Clifford
Lectures, AMS Proceedings of Symposia in Applied Mathematics}.

\bibitem{BFK2023} \textsc{Brandenburger, A., A. Friedenberg, and H.J. Keisler%
} (2023): \textquotedblleft The Relationship between Strong Belief and
Assumption,\textquotedblright \  \textit{Synthese}, 201, 175.
https://doi.org/10.1007/s11229-023-04167-6.

\bibitem{C} \textsc{Catonini, E.} (2013): \textquotedblleft Common
Assumption of Cautious Rationality and Iterated
Admissibility,\textquotedblright \ PhD thesis, Bocconi University.

\bibitem{CDV4} \textsc{Catonini, E., and N. De Vito} (2017):
\textquotedblleft A Comment on `Admissibility and
Assumption',\textquotedblright \ working paper.

\bibitem{CDV} \textsc{Catonini, E., and N. De Vito} (2018):
\textquotedblleft Hierarchies of Lexicographic Beliefs,\textquotedblright \
working paper.

\bibitem{CDV5} \textsc{Catonini, E., and N. De Vito} (2020):
\textquotedblleft Weak Belief and Permissibility,\textquotedblright \ 
\textit{Games and Economic Behavior}, 120, 154-179.

\bibitem{Cohn2013} \textsc{Cohn, D.L.} (2013): \textit{Measure Theory}.
Boston: Birkhauser.

\bibitem{DS} \textsc{Dekel, E., and M. Siniscalchi} (2015):
\textquotedblleft Epistemic Game Theory,\textquotedblright \ in\textit{\
Handbook of Game Theory with Economic Applications, Volume 4}, ed. by P.
Young and S. Zamir. Amsterdam: North-Holland, 619-702.

\bibitem{DFS} \textsc{Dekel, E., A. Friedenberg, and M. Siniscalchi} (2016):
\textquotedblleft Lexicographic Beliefs and Assumption,\textquotedblright \ 
\textit{Journal of Economic Theory}, 163, 955-985.

\bibitem{DeVito2023} \textsc{De Vito, N}. (2023): \textquotedblleft Directed
Lexicographic Rationalizability,\textquotedblright \  \textit{Economics
Letters}, 227, 111134.

\bibitem{Engelking} \textsc{Engelking, R.} (1989): \textit{General Topology}%
. Berlin: Heldermann.

\bibitem{Friedenberg} \textsc{Friedenberg, A.} (2010): \textquotedblleft
When Do Type Structures Contain All Hierarchies of
Beliefs?,\textquotedblright \  \textit{Games and Economic Behavior}, 68,
108-129.

\bibitem{Friedenberg Keisler 2021} \textsc{Friedenberg, A., and H.J. Keisler}
(2021): \textquotedblleft Iterated Dominance Revisited,\textquotedblright \ 
\textit{Economic Theory}, 72, 377-421.

\bibitem{Halpern2010} \textsc{Halpern, J.Y.} (2010): \textquotedblleft
Lexicographic Probability, Conditional Probability, and Nonstandard
Probability,\textquotedblright \  \textit{Games and Economic Behavior}, 68,
155-179.

\bibitem{Halpern Pass 2019} \textsc{Halpern, J.Y., and R. Pass} (2019):
\textquotedblleft A Conceptually Well-Founded Characterization of Iterated
Admissibility Using an \textquotedblleft All I Know\textquotedblright \
Operator,\textquotedblright \ in\textit{\ Theoretical Aspects of Rationality
and Knowledge, Proceedings Seventeenth Conference}, ed. by L.S. Moss,
221-232.

\bibitem{HS} \textsc{Heifetz, A., and D. Samet} (1998): \textquotedblleft
Topology-Free Typology of Beliefs,\textquotedblright \  \textit{Journal of
Economic Theory}, 82, 324-341.

\bibitem{HMS2019} \textsc{Heifetz, A., M. Meier and B. Schipper} (2019):
\textquotedblleft Comprehensive Rationalizability,\textquotedblright \ 
\textit{Games and Economic Behavior}, 116, 185-202.

\bibitem{K} \textsc{Kechris, A.} (1995): \textit{Classical Descriptive Set
Theory. }Berlin: Springer Verlag.

\bibitem{KL} \textsc{Keisler, H.J., and B.S. Lee} (2023): \textquotedblleft
Common Assumption of Rationality,\textquotedblright \ working paper,
University of Toronto.

\bibitem{Lee 2016} \textsc{Lee, B.S.} (2016a): \textquotedblleft
Admissibility and Assumption,\textquotedblright \  \textit{Journal of
Economic Theory}, 163, 42-72.

\bibitem{Lee 2016b} \textsc{Lee, B.S.} (2016b): \textquotedblleft
Generalizing Type Spaces,\textquotedblright \ working paper, University of
Toronto.

\bibitem{Lo 1999} \textsc{Lo, K.C.} (1999): \textquotedblleft Nash
Equilibrium without Mutual Knowledge of Rationality,\textquotedblright \ 
\textit{Economic Theory}, 14, 621-633.

\bibitem{Moulin1979} \textsc{Moulin, H}. (1984): \textquotedblleft Dominance
Solvable Voting Schemes,\textquotedblright \  \textit{Econometrica}, 47,
1337-51.

\bibitem{P} \textsc{Pearce, D.} (1984): \textquotedblleft Rationalizable
Strategic Behavior and the Problem of Perfection,\textquotedblright \ 
\textit{Econometrica, }52, 1029-1050.

\bibitem{Perea 2012book} \textsc{Perea, A. }(2012): \textit{Epistemic Game
Theory: Reasoning and Choice}, CUP Press.

\bibitem{SA} \textsc{Samuelson, L.} (1992): \textquotedblleft Dominated
Strategies and Common Knowledge,\textquotedblright \textit{\ Games and
Economic Behavior}, 4, 284-313.

\bibitem{Shimoji2004} \textsc{Shimoji, M.} (2004): \textquotedblleft On the
Equivalence of Weak Dominance and Sequential Best
Response,\textquotedblright \  \textit{Games and Economic Behavior}, 48,
385-402.

\bibitem{Srivastava1998} \textsc{Srivastava, S.M.} (1998): \textit{A Course
on Borel Sets}. New York: Springer-Verlag.

\bibitem{Stahl 1995} \textsc{Stahl, D.O.} (1995): \textquotedblleft
Lexicographic Rationalizability and Iterated
Admissibility,\textquotedblright \  \textit{Economics Letters}, 47, 155-159.

\bibitem{Y} \textsc{Yang, C.} (2015): \textquotedblleft Weak Assumption and
Iterative Admissibility,\textquotedblright \  \textit{Journal of Economic
Theory}, 158, 87-101.

\bibitem{Ziegler Zuazo-Garin 2020} \textsc{Ziegler, G., and P. Zuazo-Garin}
(2020): \textquotedblleft Strategic Cautiousness as an Expression of
Robustness to Ambiguity,\textquotedblright \  \textit{Games and Economic
Behavior}, 119, 197-215.
\end{thebibliography}
\end{document}


\title{Cautious Belief and Iterated Admissibility\\
(Supplementary Appendix)}
\author{Emiliano Catonini\textbf{\medskip \thanks{%
New York University in Shanghai, emiliano.catonini@gmail.com}} \and Nicodemo
De Vito\textbf{\thanks{%
Bocconi University, nicodemo.devito@unibocconi.it}}}
\date{May 2023}
\maketitle

\begin{abstract}
This supplementary appendix contains supplemental material and elaborations
upon the main results in the paper. Notation is the same as in the main
text, unless otherwise stated.
\end{abstract}

The following appendices provide a formal analysis of some results mentioned
in the main text that are not stated there as formal theorems. Appendix D.1
provides an alternative foundation of cautious belief. Appendix D.2 explores
the relationship between the epistemic assumptions used for the
characterization results of the paper.

\section*{Appendix D}

\subsection*{D.1 Alternative foundation for cautious belief}

We begin our analysis by introducing a notion of weak dominance for acts,
which is a slight generalization of the notion used in Condition (*) in
Definition A.2.

\bigskip

\noindent \textbf{Definition D.1 }\textit{Fix a non-empty event }$E\subseteq
\Omega $\textit{\ and }$f,g\in \mathrm{ACT}\left( \Omega \right) $.\textit{\
Say }$f$\textit{\  \textbf{weakly dominates} }$g$\textit{\ on }$E$\textit{,
and write }$f$\textrm{WD}$_{E}^{\bar{\mu}}g$, \textit{if}

\begin{description}
\item[1] $f\left( \omega \right) \geq g\left( \omega \right) $ \textit{for
all }$\omega \in E$\textit{, and}

\item[2] \textit{there exists a relevant part }$E_{s_{-i}}\subseteq E$%
\textit{\ which is \emph{non-null} under }$\succsim ^{\bar{\mu}}$ \textit{%
and satisfies }$f\left( \omega \right) >g\left( \omega \right) $ \textit{for
all }$\omega \in E_{s_{-i}}$.
\end{description}

\bigskip

The following definition is taken from Dekel et al. (2016, Definition 4.1):

\bigskip

\noindent \textbf{Definition D.2 }\textit{Fix a non-empty event }$E\subseteq
\Omega $\textit{\ and }$f,g\in \mathrm{ACT}\left( \Omega \right) $.\textit{\
Say }$f$\textit{\  \textbf{P-weakly dominates} }$g$\textit{\ on }$E$\textit{,
and write }$f$\textrm{PWD}$_{E}^{\bar{\mu}}g$\textit{,\ if}

\begin{description}
\item[1] $f\left( \omega \right) \geq g\left( \omega \right) $ \textit{for
all }$\omega \in E$\textit{, and}

\item[2] \textit{there exists some Borel set }$F\subseteq E$\textit{\ which
is non-null under }$\succsim ^{\bar{\mu}}$ \textit{and satisfies }$f\left(
\omega \right) >g\left( \omega \right) $ \textit{for all }$\omega \in F$.
\end{description}

\bigskip

Condition 2 in Definition D.2 requires strict preference on a subset of $E$
that is \textquotedblleft significant.\textquotedblright \ As Dekel et al.
(2016, Section 4.1) put it, P-weak dominance employs a \textit{P}%
reference-based notion of significance, which coincides with the usual
(topological) notion of weak dominance if $\Omega $\ is a finite set.

Clearly, if $f$\textit{\ }weakly dominates $g$ on $E$\textit{,} then $f$%
\textrm{PWD}$_{E}^{\bar{\mu}}g$.\ We claim that the converse is true if $%
f,g\in $\textrm{ACT}$^{S_{-i}}\left( \Omega \right) $. To see this, let $f$%
\textrm{PWD}$_{E}^{\bar{\mu}}g$\ with $f,g\in $\textrm{ACT}$^{S_{-i}}\left(
\Omega \right) $. Let $F\subseteq E$\ be a non-null event such that
Condition (2) of Definition D.2 holds. There exists a relevant part of $F$,
viz. $F_{s_{-i}}:=F\cap \left( \left \{ s_{-i}\right \} \times T_{-i}\right) 
$, which is non-null under $\succsim ^{\bar{\mu}}$; for, if every relevant
part of $F$ were null under $\succsim ^{\bar{\mu}}$, then also $F$ would be
null. Certainly, $f\left( \omega \right) >g\left( \omega \right) $ for all $%
\omega \in F_{s_{-i}}$. But note that $F_{s_{-i}}\subseteq E_{s_{-i}}$, and
we have that $f\left( \omega \right) >g\left( \omega \right) $ for all $%
\omega \in E_{s_{-i}}$. This is so because every $f\in $\textrm{ACT}$%
^{S_{-i}}\left( \Omega \right) $ is constant on the cylinder $\left \{
s_{-i}\right \} \times T_{-i}$.

\bigskip

\noindent \textbf{Definition D.3 }\textit{Fix }$\bar{\mu}:=(\mu ^{1},...,\mu
^{n})\in \mathcal{N}(\Omega )$\textit{\ and a non-empty event} $E\subseteq
\Omega $\textit{.}

\begin{description}
\item[1] \textit{\textbf{Relevance}: Say that }$E$\textit{\ is relevant
under }$\succsim ^{\bar{\mu}}$\textit{\ if every relevant part }$E_{s_{-i}}$%
\textit{\ is non-null under }$\succsim ^{\bar{\mu}}$\textit{.}

\item[2] \textit{\textbf{Weak Dominance (WD) Determination}: Say that }$E$%
\textit{\ WD-determines }$\succsim ^{\bar{\mu}}$\  \textit{if, for all }$%
f,g\in $\textrm{ACT}$^{S_{-i}}\left( \Omega \right) $\textit{, }$f\mathrm{WD}%
_{E}^{\bar{\mu}}g$\textit{\ implies }$f\succ ^{\bar{\mu}}g$\textit{.}

\item[3] \textit{\textbf{PWD Determination}}: \textit{Say that }$E$\textit{\
PWD-determines }$\succsim ^{\bar{\mu}}$\  \textit{if, for all }$f,g\in $%
\textrm{ACT}$\left( \Omega \right) $\textit{, }$f\mathrm{PWD}_{E}^{\bar{\mu}%
}g$\textit{\ implies }$f\succ ^{\bar{\mu}}g$\textit{.}
\end{description}

\bigskip

\noindent \textbf{Remark D.1 }\textit{Relevance implies that }$E$\textit{\
is non-null under }$\succsim ^{\bar{\mu}}$\textit{.}

\bigskip

\noindent \textbf{Remark D.2 }\textit{If }$E$\textit{\ is Savage-null under }%
$\succsim ^{\bar{\mu}}$\textit{, then WD and PWD Determination hold
vacuously. As already observed, PWD and WD Determination are equivalent for
pairs of acts belonging to }\textrm{ACT}$^{S_{-i}}\left( \Omega \right) $%
\textit{. If }$E$\textit{\ is relevant under }$\succsim ^{\bar{\mu}}$\textit{%
, then, by Remark D.1, PWD and WD Determination do not hold vacuously.}

\bigskip

Now we state and prove the characterization result for cautious belief:

\bigskip

\noindent \textbf{Theorem D.1} \textit{Fix} $\bar{\mu}:=(\mu ^{1},...,\mu
^{n})\in \mathcal{N}(\Omega )$ \textit{and a non-empty event }$E\subseteq
\Omega $\textit{. Then }$E$\textit{\ is cautiously believed under }$\bar{\mu}
$\textit{\ if and only if }$\succsim ^{\bar{\mu}}$\textit{\ satisfies
Relevance and WD Determination.}

\bigskip

\noindent \textbf{Proof}. Suppose first that $E$\ is cautiously believed
under $\bar{\mu}$ at some level $m$. Then Condition (ii) of Definition A.5
implies that every relevant part of $E$ is non-null under $\succsim ^{\bar{%
\mu}}$, hence Relevance is satisfied. Pick acts $f,g\in $\textrm{ACT}$%
^{S_{-i}}\left( \Omega \right) $ such that $f\mathrm{WD}_{E}^{\bar{\mu}}g$.
By Condition (i) of Definition A.5, $\mu ^{l}\left( E\right) =1$\ for all $%
l\leq m$. Hence%
\begin{equation*}
\dint fd\mu ^{l}=\dint_{E}fd\mu ^{l}\geq \dint_{E}gd\mu ^{l}=\dint gd\mu ^{l}
\end{equation*}%
for every $l\leq m$. By Condition (ii) of Definition A.5, there exists some $%
l\leq m$\ such that $\mu ^{l}\left( E_{s_{-i}}\right) >0$. This implies that
there is $l\leq m$\ such that%
\begin{equation*}
\dint fd\mu ^{l}=\dint_{E}fd\mu ^{l}>\dint_{E}gd\mu ^{l}=\dint gd\mu ^{l}%
\text{,}
\end{equation*}%
hence $f\succ ^{\bar{\mu}}g$. As $f$ and $g$\ are arbitrary, this shows that
WD Determination holds.

Conversely, suppose that $\succsim ^{\bar{\mu}}$\ satisfies Relevance. This
entails that $E$\ is non-null under $\succsim ^{\bar{\mu}}$.

\bigskip

\noindent \textbf{Claim D.1} \textit{Let }$E$\textit{\ be non-null under }$%
\succsim ^{\bar{\mu}}$. \textit{If }$E$\textit{\ WD-determines }$\succsim ^{%
\bar{\mu}}$\textit{, then }$\mu ^{1}\left( E\right) =1$\textit{.}

\bigskip

\noindent \textit{Proof}. We prove the claim by contraposition. Suppose that 
$\mu ^{1}\left( E\right) <1$. This implies that $\mu ^{1}\left( \Omega
\backslash E\right) >0$, and so there exists $\varepsilon \in \left( 0,\mu
^{1}\left( \Omega \backslash E\right) \right) $. Consider acts $f,g\in $%
\textrm{ACT}$^{S_{-i}}\left( \Omega \right) $\ such that $f:=\overrightarrow{%
\varepsilon }$\ and $g:=\left( \overrightarrow{0}_{E},\overrightarrow{1}%
_{\Omega \backslash E}\right) $. As $E$\ is non-null under $\succsim ^{\bar{%
\mu}}$, there is a non-null relevant part of it, and we have $f\mathrm{WD}%
_{E}^{\bar{\mu}}g$. Since $\dint gd\mu ^{1}=\mu ^{1}\left( \Omega \backslash
E\right) >\varepsilon =\dint fd\mu ^{1}$, it follows that $g\succ ^{\bar{\mu}%
}f$. Thus, WD Determination\ fails.\hfill $\square $

\bigskip

By Claim D.1, there exists $k\geq 1$\ such that $\mu ^{l}\left( E\right) =1$%
\ for all $l\leq k$.\ Let $m$\ denote the maximum level of LPS $(\mu
^{1},...,\mu ^{n})$\ such that $\mu ^{l}\left( E\right) =1$\ for all $l\leq
m $. Since Relevance holds, each relevant part $E_{s_{-i}}$\ satisfies $\mu
^{l}\left( E_{s_{-i}}\right) >0$\ for some $l\leq n$. We now show that if $%
\mu ^{l}\left( E_{s_{-i}}\right) >0$\ for some $l>m$, then there exists $%
k\leq m$\ such that $\mu ^{k}(E_{s_{-i}})>0$.

The statement is vacuously true if $m=n$; in such a case, $\Omega \backslash
E$\ is Savage-null under $\succsim ^{\bar{\mu}}$. So, suppose that $m<n$,
and let $m+1:=\mathcal{I}_{\bar{\mu}}\left( \Omega \backslash E\right) $, so
that $\mu ^{m+1}(E)<1$. We claim that if $E$\ WD-determines $\succsim ^{\bar{%
\mu}}$, then there exists $k\leq m$\ such that $\mu ^{k}(E_{s_{-i}})>0$.
This is the consequence of the following

\bigskip

\noindent \textbf{Claim D.2} \textit{Let }$E_{s_{-i}}$\textit{\ be a
non-null, relevant part of }$E$\textit{\ such that }$\mu ^{l}\left(
E_{s_{-i}}\right) >0$\textit{\ for some }$l\geq m+1$\textit{. If }$\mu
^{k}(E_{s_{-i}})=0$\textit{\ for all }$k<m+1$\textit{, then there are acts }$%
f,g\in $\textrm{ACT}$^{S_{-i}}\left( \Omega \right) $\textit{\ such that }$f%
\mathrm{WD}_{E}^{\bar{\mu}}g$\textit{\ and }$g\succsim ^{\bar{\mu}}f$\textit{%
.}

\bigskip

\noindent \textit{Proof}. Define acts $f,g\in $\textrm{ACT}$^{S_{-i}}\left(
\Omega \right) $ as follows:%
\begin{eqnarray*}
f(\omega ) &:&=\left \{ 
\begin{tabular}{ll}
$\mu ^{m+1}\left( \Omega \backslash E\right) \text{,}$ & if $\omega \in
E_{s_{-i}}\text{,}$ \\ 
$0\text{,}$ & if $\omega \in \Omega \backslash E_{s_{-i}}$,%
\end{tabular}%
\right. \\
g &:&=(\overrightarrow{0}_{E},\overrightarrow{1}_{\Omega \backslash E})\text{%
.}
\end{eqnarray*}%
\bigskip It is easy to observe that $f\mathrm{WD}_{E}^{\bar{\mu}}g$. To see
that $g\succsim ^{\bar{\mu}}f$, note that%
\begin{equation*}
\int fd\mu ^{m+1}=\mu ^{m+1}\left( \Omega \backslash E\right) \mu
^{m+1}\left( E_{s_{-i}}\right) <\mu ^{m+1}\left( \Omega \backslash E\right)
=\int gd\mu ^{m+1}\text{,}
\end{equation*}%
where the strict inequality follows from the fact that $\mu ^{m+1}\left(
E_{s_{-i}}\right) \leq \mu ^{m+1}(E)<1$; and, for each $k\leq m$,%
\begin{equation*}
\int fd\mu ^{k}=\mu ^{m+1}\left( \Omega \backslash E\right) \mu ^{k}\left(
E_{s_{-i}}\right) =0=\int gd\mu ^{k}\text{,}
\end{equation*}%
where the second equality follows by the assumption.\hfill $\square $

\bigskip

Summing up: Relevance implies that $E$\ is non-null under $\succsim ^{\bar{%
\mu}}$. By Claim D.1, WD determination yields the existence of some $m\leq n$%
\ such that $\mu ^{l}(E)=1$ for all $l\leq m$. Furthermore, Relevance
implies that each relevant part $E_{s_{-i}}$\ satisfies $\mu ^{l}\left(
E_{s_{-i}}\right) >0$\ for some $l\leq n$. By Claim D.2, WD Determination
implies that there exists $l\leq m$\ such that $\mu ^{l}\left(
E_{s_{-i}}\right) >0$. The conclusion follows.\hfill $\blacksquare $

\bigskip

Theorem D.1 shows how cautious belief is weaker than PWD-assumption, a
concept of belief put forward by Dekel et al. (2016). The notion of
PWD-assumption of an event $E$ is based on two axioms: (i) PWD
Determination, and (ii) Non-Triviality, which states that every \textit{part}
of $E$ (i.e., every relatively open subset of $E$) is non-null. Relevance
implies Non-Triviality, since---technically---a relevant part of $E$ is a
relatively open subset of $E$; and, as already observed (Remark D.2), PWD
and WD Determination coincide on \textrm{ACT}$^{S_{-i}}\left( \Omega \right) 
$.

Dekel et al. (2016) introduce PWD-assumption as an extention of Asheim and
Dufwenberg (2003)'s notion of \textbf{full belief} to uncountable, abstract
state spaces $\Omega $. We briefly connect our analysis to the notion of
full belief as in Asheim and Dufwenberg (2003) and Asheim and Sovik (2005),
but restricted to LPS's on strategies.

Fix $\bar{\mu}=(\mu ^{1},...,\mu ^{n})\in \mathcal{N}(S_{-i})$ and a
non-empty event $E\subseteq S_{-i}$. Say that $E$\ is \textbf{fully believed}
under $\bar{\mu}$ if there exists $m\leq n$\ such that $\cup _{l=1}^{m}$%
\textrm{Supp}$\mu ^{l}=E$.

The following result is immediate by Proposition 2.2 in the main text:

\bigskip

\noindent \textbf{Proposition D.1} \textit{Fix} $\bar{\mu}:=(\mu
^{1},...,\mu ^{n})\in \mathcal{N}(S_{-i}\times T_{-i})$ \textit{and a
non-empty event }$E\subseteq \Omega $\textit{. If }$E$\textit{\ is
cautiously believed under }$\bar{\mu}$\textit{, then }\textrm{Proj}$%
_{S_{-i}}\left( E\right) $\textit{\ is fully believed under }$\overline{%
\mathrm{marg}}_{S_{-i}}\bar{\mu}$\textit{.}

\bigskip

We conclude with the following observation. Fix an abstract, finite space $%
\Omega $. The version of full belief in Asheim and Sovik (2005) corresponds
to the following preference-based definition:

\begin{itemize}
\item An event $E\subseteq \Omega $ is fully believed if the preferences are
admissible on $E$ (i.e., if one act weakly dominates another conditional on $%
E$, then the former is strictly preferred to the latter).
\end{itemize}

This notion of fully belief has the following characterization in terms of
the infinitely more likely relation:

\begin{itemize}
\item An event $E\subseteq \Omega $ is fully believed if and only if each
state in $E$ is infinitely more likely than each state in $\Omega \backslash
E$.
\end{itemize}

It should be noted that the notion of infinitely more likely is that of
Blume et al. (1989a, Definition 5.1). But, this notion\textit{\ }is used
only for pairs of states (i.e., elements in $\Omega $),\textit{\ not for
pairs of events with more than one state}. For pairs of states (i.e.,
singletons), the definition of infinitely more likely of Blume et al.
(1989a) is equivalent to Lo (1999)'s definition---see Blume et al. (1989a,
Footnote 8). This is not true for events, since, as remarked by Asheim and
Sovik (2005, p. 73), Blume et al. (1989a)'s notion of infinitely more likely
may fail disjunction.

The result of Proposition D.1 states that every \textquotedblleft
state\textquotedblright \ $s_{-i}\in $\textrm{Proj}$_{S_{-i}}\left( E\right) 
$ is infinitely more likely than every \textquotedblleft
state\textquotedblright \ $s_{-i}^{\prime }\notin $\textrm{Proj}$%
_{S_{-i}}\left( E\right) $; in this case, there is no difference between the
above definitions of infinitely more likely. Yet, this result implies that 
\textrm{Proj}$_{S_{-i}}\left( E\right) $\ is infinitely more likely than its
complement in Lo (1999)'s sense, but not in the sense of Blume et al.
(1989a).

\subsection*{D.2 On the relationship between the epistemic assumptions}

We first exhibit an example which shows that events $\tprod_{i\in
I}R_{i}^{\infty }$\ and $\tprod_{i\in I}\hat{R}_{i}^{\infty }$ can be
different, and the SAS's that characterize their behavioral implications are
disjoint.

\bigskip

\noindent \textbf{Example D.1 }Consider the game in Example 1 in the main
text, which is reproduced here for convenience.

\begin{equation*}
\begin{tabular}{|c|c|c|}
\hline
$a\backslash b$ & $\ell $ & $r$ \\ \hline
$u$ & $2,2$ & $2,2$ \\ \hline
$m$ & $3,1$ & $0,0$ \\ \hline
$d$ & $0,0$ & $1,3$ \\ \hline
\end{tabular}%
\end{equation*}

There are three non-empty SAS's: $\left \{ u\right \} \times \left \{
r\right \} $, $\left \{ u\right \} \times \left \{ \ell ,r\right \} $\ and $%
\left \{ m\right \} \times \left \{ \ell \right \} $. The SAS $\left \{
m\right \} \times \left \{ \ell \right \} $\ is the IA set.

Append to this game a \textit{non-cautious} type structure%
\begin{equation*}
\mathcal{T}:=\langle S_{a},S_{b},T_{a},T_{b},\beta _{a},\beta _{b}\rangle
_{i\in I}\text{,}
\end{equation*}%
with type spaces $T_{a}:=\left \{ t_{a}^{1},t_{a}^{2},t_{a}^{\ast }\right \} 
$ and $T_{b}:=\left \{ t_{b}^{1},t_{b}^{2},t_{b}^{\ast }\right \} $. Types $%
t_{a}^{\ast }$\ and\ $t_{b}^{\ast }$\ are not cautious, as they are
associated with a deterministic probability measure on $S_{b}\times T_{b}$\
and $S_{a}\times T_{a}$, respectively; formally,\footnote{%
For any $x\in X$, we let $\delta _{x}$ denote the Dirac measure supported by 
$x$.}%
\begin{eqnarray*}
\beta _{a}(t_{a}^{\ast }) &:&=\left( \delta _{\left( \ell ,t_{b}^{1}\right)
}\right) \text{,} \\
\beta _{b}(t_{b}^{\ast }) &:&=\left( \delta _{\left( m,t_{a}^{1}\right)
}\right) \text{.}
\end{eqnarray*}%
All the remaining types are cautious, and the belief maps are as follows.
Consider first types $t_{a}^{1}$\ and $t_{b}^{1}$:%
\begin{eqnarray*}
\beta _{a}(t_{a}^{1}) &:&=\left( \mu _{a}^{1},\mu _{a}^{2}\right) :=\left( 
\frac{4}{5}\delta _{\left( \ell ,t_{b}^{1}\right) }+\frac{1}{5}\delta
_{\left( r,t_{b}^{2}\right) },\delta _{\left( r,t_{b}^{\ast }\right)
}\right) \text{,} \\
\beta _{b}(t_{b}^{1}) &:&=\left( \mu _{b}^{1},\mu _{b}^{2}\right) :=\left( 
\frac{1}{2}\delta _{\left( m,t_{a}^{1}\right) }+\frac{1}{2}\delta _{\left(
u,t_{a}^{2}\right) },\delta _{\left( d,t_{a}^{\ast }\right) }\right) \text{.}
\end{eqnarray*}%
Types $t_{a}^{2}$\ and $t_{b}^{2}$ are associated with the following LPS's:%
\begin{eqnarray*}
\beta _{a}(t_{a}^{2}) &:&=\left( \nu _{a}^{1},\nu _{a}^{2}\right) :=\left(
\delta _{\left( r,t_{b}^{2}\right) },\delta _{\left( \ell ,t_{b}^{2}\right)
}\right) \text{,} \\
\beta _{b}(t_{b}^{2}) &:&=\left( \nu _{b}^{1},\nu _{b}^{2}\right) :=\left(
\delta _{\left( u,t_{a}^{2}\right) },\frac{1}{2}\delta _{\left(
m,t_{a}^{2}\right) }+\frac{1}{2}\delta _{\left( d,t_{a}^{2}\right) }\right) 
\text{.}
\end{eqnarray*}

The event corresponding to transparency of cautiousness in $\mathcal{T}$ is%
\begin{equation*}
C^{\infty }=C_{a}^{\infty }\times C_{b}^{\infty }=\left( S_{a}\times \left
\{ t_{a}^{2}\right \} \right) \times \left( S_{b}\times \left \{
t_{b}^{2}\right \} \right) \text{.}
\end{equation*}%
Indeed, cautious type $t_{a}^{1}$\ (resp. $t_{b}^{1}$) does not certainly
believe event $C_{b}=S_{b}\times \left \{ t_{b}^{1},t_{b}^{2}\right \} $
(resp. $C_{b}=S_{b}\times \left \{ t_{b}^{1},t_{b}^{2}\right \} $), since
measure $\mu _{a}^{2}$\ (resp. $\mu _{b}^{2}$) assigns probability $1$ to $%
\left( r,t_{b}^{\ast }\right) \notin C_{b}$\ (resp. $\left( d,t_{a}^{\ast
}\right) \notin C_{a}$). With this, we obtain%
\begin{eqnarray*}
\hat{R}_{a}^{1} &:&=R_{a}\cap C_{a}^{\infty }=\left \{ \left(
u,t_{a}^{2}\right) \right \} \text{,} \\
\hat{R}_{b}^{1} &:&=R_{b}\cap C_{b}^{\infty }=\left \{ \left(
r,t_{b}^{2}\right) \right \} \text{.}
\end{eqnarray*}%
In particular, $\left( \ell ,t_{b}^{2}\right) \notin \hat{R}_{b}^{1}$: both $%
\ell $ and $r$ are optimal under $\mathrm{marg}_{S_{a}}\nu _{b}^{1}=\delta
_{u}$, and the expected payoffs of playing $\ell $ and $r$ under $\mathrm{%
marg}_{S_{a}}\nu _{b}^{2}=\frac{1}{2}\delta _{m}+\frac{1}{2}\delta _{d}$ are 
$1/2$ and $3/2$, respectively. Note that $\hat{R}_{b}^{1}$\ and $\hat{R}%
_{a}^{1}$ are cautiously believed (at level $1$) under $\beta
_{a}(t_{a}^{2}) $ and $\beta _{b}(t_{b}^{2})$; hence $\hat{R}_{a}^{2}=\hat{R}%
_{a}^{1}$\ and $\hat{R}_{b}^{2}=\hat{R}_{b}^{1}$. By induction, it follows
that $\hat{R}_{a}^{\infty }=\hat{R}_{a}^{1}$\ and $\hat{R}_{b}^{\infty }=%
\hat{R}_{b}^{1}$; therefore,%
\begin{equation*}
\mathrm{Proj}_{S_{a}}\left( \hat{R}_{a}^{\infty }\right) \times \mathrm{Proj}%
_{S_{b}}\left( \hat{R}_{b}^{\infty }\right) =\left \{ u\right \} \times
\left \{ r\right \} \text{.}
\end{equation*}

Now, consider the event \textquotedblleft cautious
rationality.\textquotedblright \ We have that%
\begin{eqnarray*}
R_{a}^{1} &:&=R_{a}\cap C_{a}=\left \{ \left( m,t_{a}^{1}\right) ,\left(
u,t_{a}^{2}\right) \right \} \text{,} \\
R_{b}^{1} &:&=R_{b}\cap C_{b}=\left \{ \left( \ell ,t_{b}^{1}\right) ,\left(
r,t_{b}^{2}\right) \right \} \text{.}
\end{eqnarray*}%
In particular, $m$\ is the unique strategy which is optimal under $\beta
_{a}(t_{a}^{1})$, as the expected payoff of playing $m$ under $\mathrm{marg}%
_{S_{b}}\mu _{a}^{1}=\frac{4}{5}\delta _{\ell }+\frac{1}{5}\delta _{r}$ is $%
12/5$, while the expected payoffs of playing $u$ and $d$ are $2$\ and $1/5$,
respectively. Similarly, $\ell $\ is the unique strategy which is optimal
under $\beta _{b}(t_{b}^{1})$, as the expected payoff of playing $\ell $
under $\mathrm{marg}_{S_{a}}\mu _{b}^{1}=\frac{1}{2}\delta _{u}+\frac{1}{2}%
\delta _{m}$ is $3/2$, while the expected payoff of playing $r$ is $1$.

Next note that types $t_{a}^{1}$\ cautiously believes $R_{b}^{1}$, but an
analogous conclusion does not hold for type $t_{a}^{2}$. To see this,
observe that the only possibility is that $R_{b}^{1}$\ is cautiously
believed under $\beta _{a}(t_{a}^{2})$\ at level $1$, because $\nu
_{a}^{2}\left( R_{b}^{1}\right) =0$. Since%
\begin{equation*}
\mathrm{Suppmarg}_{S_{b}}\nu _{a}^{1}=\left \{ r\right \} \neq \left \{ \ell
,r\right \} =\mathrm{Proj}_{S_{b}}\left( R_{b}^{1}\right) \text{,}
\end{equation*}%
it follows from Proposition 2.2 that $t_{a}^{2}$\ does not cautiously
believe $R_{b}^{1}$. Moreover, it is easily checked that type $t_{b}^{1}$\
cautiously believes $R_{a}^{1}$. By contrast, $t_{b}^{2}$ does not
cautiously believes $R_{a}^{1}$: the same argument as above shows that $%
R_{a}^{1}$\ could be cautiously believed under $\beta _{b}(t_{b}^{2})$\ only
at level $1$, because $\nu _{b}^{2}\left( R_{a}^{1}\right) =0$. Since%
\begin{equation*}
\mathrm{Suppmarg}_{S_{a}}\nu _{b}^{1}=\left \{ u\right \} \neq \left \{
u,m\right \} =\mathrm{Proj}_{S_{a}}\left( R_{a}^{1}\right) \text{,}
\end{equation*}%
it follows from Proposition 2.2 that $t_{b}^{2}$\ does not cautiously
believe $R_{a}^{1}$.

Thus,%
\begin{eqnarray*}
R_{a}^{2} &:&=R_{a}^{1}\cap \mathbf{B}_{a}^{c}\left( R_{b}^{1}\right) =\left
\{ \left( m,t_{a}^{1}\right) \right \} \text{,} \\
R_{b}^{2} &:&=R_{b}^{1}\cap \mathbf{B}_{b}^{c}\left( R_{a}^{1}\right) =\left
\{ \left( \ell ,t_{b}^{1}\right) \right \} \text{.}
\end{eqnarray*}%
It follows that $R_{a}^{\infty }=R_{a}^{2}$\ and $R_{b}^{\infty }=R_{b}^{2}$%
; hence,%
\begin{equation*}
\mathrm{Proj}_{S_{a}}\left( R_{a}^{\infty }\right) \times \mathrm{Proj}%
_{S_{b}}\left( R_{b}^{\infty }\right) =\left \{ m\right \} \times \left \{
\ell \right \} \text{.}
\end{equation*}%
\hfill $\blacktriangledown $

\bigskip

Fix a finite game $G:=\left \langle I,(S_{i},\pi _{i})_{i\in
I}\right
\rangle $ and an associated non-cautious type structure $\mathcal{T%
}$. Example D.1 shows that events $\tprod_{i\in I}R_{i}^{\infty }$\ and $%
\tprod_{i\in I}\hat{R}_{i}^{\infty }$ can be different, if $\mathcal{T}$\ is
a \textit{finite} type structure. With this in mind, we show the following
result: if $\mathcal{T}$\ is belief-complete, then (i) $\tprod_{i\in I}\hat{R%
}_{i}^{\infty }$\ is a subset of $\tprod_{i\in I}R_{i}^{\infty }$; and (ii)
the behavioral implications of $\tprod_{i\in I}\hat{R}_{i}^{\infty }$\ and $%
\tprod_{i\in I}R_{i}^{\infty }$ are the same.

To this end, we first state and prove a technical lemma that will be useful
for the proof. So, we need some preliminary definitions and results. In what
follows, fix a finite, discrete space $X$, and a Polish space $Y$.

We say that LPS $\bar{\mu}\in \mathcal{N}\left( X\times Y\right) $\  \textbf{%
cautiously believes} a nonempty event $E\subseteq X\times Y$\ if $E$\ is
cautiously believed under $\bar{\mu}$. We say that $\bar{\mu}$\ cautiously
believes a sequence of nonempty events $\left( E_{m}\right) _{m=1}^{n}$\ in $%
X\times Y$\ if, for each $m=1,...,n$, $\bar{\mu}$\ cautiously believes $%
E_{m} $.

We say that $\overline{\nu }:=\left( \nu ^{1},...,\nu ^{n}\right) \in 
\mathcal{N}\left( X\right) $ \textbf{cautiously believes}\ a nonempty set
(event) $F\subseteq X$\ if there is $m\leq n$\ such that $\cup _{m\leq n}%
\mathrm{Supp}\nu ^{n}=F$. We say that $\overline{\nu }$\ cautiously believes
a sequence of nonempty events $\left( F_{m}\right) _{m=1}^{n}$\ in $X$\ if,
for each $m=1,...,n$, $\overline{\nu }$\ fully believes $F_{m}$.

\bigskip

\noindent \textbf{Remark D.3} \textit{Fix a Polish space }$Y$\textit{\ and a
finite, discrete space }$X$\textit{. For any nonempty event }$E\subseteq
X\times Y$\textit{, there exists a continuous function }$f:\mathrm{Proj}%
_{X}\left( E\right) \rightarrow Y$\textit{\ such that }$\left( x,f\left(
x\right) \right) \in E$\textit{\ for every }$x\in \mathrm{Proj}_{X}\left(
E\right) $.

\bigskip

The result in Remark D.3\ holds because $\mathrm{Proj}_{X}$\ is a surjective
map; so, by the Axiom of Choice, we can define a function $f:\mathrm{Proj}%
_{X}\left( E\right) \rightarrow Y$ which is obviously continuous with
respect to the (relative) discrete topology on $\mathrm{Proj}_{X}\left(
E\right) $.

\bigskip

\noindent \textbf{Lemma D.1} \textit{Fix a finite, decreasing sequence of
events }$\left( E_{m}\right) _{m=1}^{n}$\textit{\ in }$X\times Y$\textit{.
Then, for each LPS }$\overline{\nu }\in \mathcal{N}\left( X\right) $\textit{%
\ that cautiously believes }$\left( \mathrm{Proj}_{X}\left( E_{m}\right)
\right) _{m=1}^{n}$\textit{, there is an LPS }$\bar{\mu}\in \mathcal{N}%
\left( X\times Y\right) $\textit{\ that cautiously believes }$\left(
E_{m}\right) _{m=1}^{n}$\textit{\ such that }$\overline{\mathrm{marg}}_{X}%
\bar{\mu}=\overline{\nu }$\textit{.}

\bigskip

\noindent \textbf{Proof. }Define an ordered partition $\left(
P_{0},P_{1},...,P_{n}\right) $\ of $X$ as follows:%
\begin{eqnarray*}
P_{0} &:&=X\backslash \mathrm{Proj}_{X}\left( E_{1}\right) \text{,} \\
P_{m} &:&=\mathrm{Proj}_{X}\left( E_{m}\right) \backslash \mathrm{Proj}%
_{X}\left( E_{m+1}\right) \text{, }m=1,...,n-1\text{,} \\
P_{n} &:&=\mathrm{Proj}_{X}\left( E_{n}\right) \text{.}
\end{eqnarray*}%
For each $m=1,...,n-1$, let%
\begin{equation*}
F_{m}:=E_{m}\backslash \left( \mathrm{Proj}_{X}\left( E_{m+1}\right) \times
Y\right) \text{.}
\end{equation*}

\bigskip

\noindent \textbf{Claim D.3} \textit{For each }$m=1,...,n-1$\textit{,}%
\begin{equation*}
F_{m}\subseteq E_{m}\backslash E_{m+1}\text{\textit{,}}
\end{equation*}%
\textit{and}%
\begin{equation*}
\mathrm{Proj}_{X}\left( F_{m}\right) =P_{m}\text{\textit{.}}
\end{equation*}

\noindent \textit{Proof.} Fix any $m=1,...,n-1$. Since%
\begin{equation*}
E_{m+1}\subseteq \mathrm{Proj}_{X}\left( E_{m+1}\right) \times Y\text{,}
\end{equation*}%
it follows that%
\begin{equation*}
F_{m}=E_{m}\backslash \left( \mathrm{Proj}_{X}\left( E_{m+1}\right) \times
Y\right) \subseteq E_{m}\backslash E_{m+1}\text{.}
\end{equation*}%
Next, the following equality and set inclusions are immediate:%
\begin{eqnarray*}
\mathrm{Proj}_{X}\left( \mathrm{Proj}_{X}\left( E_{m+1}\right) \times
Y\right) &=&\mathrm{Proj}_{X}\left( E_{m+1}\right) \text{,} \\
\mathrm{Proj}_{X}\left( E_{m}\right) \backslash \mathrm{Proj}_{X}\left(
E_{m+1}\right) &\subseteq &\mathrm{Proj}_{X}\left( E_{m}\backslash
E_{m+1}\right) \text{;}
\end{eqnarray*}%
combining them, we obtain%
\begin{equation*}
\mathrm{Proj}_{X}\left( E_{m}\right) \backslash \mathrm{Proj}_{X}\left( 
\mathrm{Proj}_{X}\left( E_{m+1}\right) \times Y\right) \subseteq \mathrm{Proj%
}_{X}\left( E_{m}\backslash \left( \mathrm{Proj}_{X}\left( E_{m+1}\right)
\times Y\right) \right) \text{.}
\end{equation*}%
With this, we get%
\begin{eqnarray*}
\mathrm{Proj}_{X}\left( F_{m}\right) &=&\mathrm{Proj}_{X}\left(
E_{m}\backslash \left( \mathrm{Proj}_{X}\left( E_{m+1}\right) \times
Y\right) \right) \\
&\subseteq &\mathrm{Proj}_{X}\left( E_{m}\right) \backslash \mathrm{Proj}%
_{X}\left( \mathrm{Proj}_{X}\left( E_{m+1}\right) \times Y\right) \\
&=&\mathrm{Proj}_{X}\left( E_{m}\right) \backslash \mathrm{Proj}_{X}\left(
E_{m+1}\right) \\
&=&P_{m}\text{,}
\end{eqnarray*}%
and%
\begin{eqnarray*}
P_{m} &=&\mathrm{Proj}_{X}\left( E_{m}\right) \backslash \mathrm{Proj}%
_{X}\left( E_{m+1}\right) \\
&=&\mathrm{Proj}_{X}\left( E_{m}\right) \backslash \mathrm{Proj}_{X}\left( 
\mathrm{Proj}_{X}\left( E_{m+1}\right) \times Y\right) \\
&\subseteq &\mathrm{Proj}_{X}\left( E_{m}\backslash \left( \mathrm{Proj}%
_{X}\left( E_{m+1}\right) \times Y\right) \right) \\
&=&\mathrm{Proj}_{X}\left( F_{m}\right) \text{.}
\end{eqnarray*}%
\hfill $\square $

\bigskip

In light of Remark D.3 and Claim D.3, we can define a map $g:X\rightarrow Y$%
\ as follows.

\begin{itemize}
\item First, let $g_{0}:P_{0}\rightarrow Y$\ be a constant function such
that $g_{0}\left( P_{0}\right) \cap E_{1}=\emptyset $.

\item Next, for each $m=1,...,n-1$, let $g_{m}:P_{m}\rightarrow Y$\ be a
function such that $\left( x,g_{m}\left( x\right) \right) \in F_{m}\subseteq
E_{m}\backslash E_{m+1}$\ for every $x\in \mathrm{Proj}_{X}\left(
F_{m}\right) =P_{m}$.

\item Finally, let $g_{n}:P_{n}\rightarrow Y$ a function such that $\left(
x,g_{n}\left( x\right) \right) \in E_{n}$\ for every $x\in \mathrm{Proj}%
_{X}\left( E_{n}\right) =P_{n}$.
\end{itemize}

With this, let $g:X\rightarrow Y$\ be defined as the union of the maps $%
g_{0},...,g_{n}$, i.e., $g\left( x\right) =g_{m}\left( x\right) $ for $x\in
P_{m}$ ($m=0,1,...,n$).\footnote{%
The map is well defined because, by construction, the domains and codomains
of the maps $g_{0},...,g_{n}$\ are pairwise disjoint sets.}

\bigskip

\noindent \textbf{Claim D.4} \textit{The map }$g:X\rightarrow Y$\textit{\
satisfies the following property: for all }$m=1,...,n$\textit{,}%
\begin{equation*}
x\in \mathrm{Proj}_{X}\left( E_{m}\right) \Rightarrow \left( x,g\left(
x\right) \right) \in E_{m}\text{\textit{.}}
\end{equation*}

\bigskip

\noindent \textit{Proof.} Note that, since $\left( E_{m}\right) _{m=1}^{n}$\
is decreasing, $\left( x,g\left( x\right) \right) \in E_{m}$\ for every $%
x\in P_{k}$, $k\geq m$. By construction, we have%
\begin{equation*}
\mathrm{Proj}_{X}\left( E_{m}\right) =\tbigcup_{k\geq m}P_{k}\text{.}
\end{equation*}%
Hence, $\left( x,g\left( x\right) \right) \in E_{m}$ provided that $x\in 
\mathrm{Proj}_{X}\left( E_{m}\right) $.\hfill $\square $

\bigskip

Define the map $\psi :X\rightarrow X\times Y$\ as $\psi :=\left( \mathrm{Id}%
_{X},g\right) $. Clearly, $\psi $\ is injective, and it satisfies $\mathrm{%
Proj}_{X}\circ \psi =\mathrm{Id}_{X}$---that is, $\psi $\ is a right inverse
of $\mathrm{Proj}_{X}$. Moreover, it follows from Claim D.4 that $\psi
\left( \mathrm{Proj}_{X}\left( E_{m}\right) \right) \subseteq E_{m}$\ for
every $m=1,...,n$. This in turn implies that%
\begin{equation}
\mathrm{Proj}_{X}\left( E_{m}\right) \subseteq \psi ^{-1}\left( E_{m}\right)
\tag{D1}  \label{set inclusion lemma extension LPS}
\end{equation}%
for every $m=1,...,n$.

Fix any $\overline{\nu }:=\left( \nu ^{1},...,\nu ^{l}\right) \in \mathcal{N}%
\left( X\right) $ such that $\overline{\nu }$ cautiously believes $\left( 
\mathrm{Proj}_{X}\left( E_{m}\right) \right) _{m=1}^{n}$. Define $\bar{\mu}%
\in \mathcal{N}\left( X\times Y\right) $\ by $\bar{\mu}:=\widehat{\psi }%
\left( \overline{\nu }\right) $. It turns out that $\bar{\mu}$\ is the
desired LPS such that $\overline{\mathrm{marg}}_{X}\bar{\mu}=\overline{\nu }$%
, since%
\begin{eqnarray*}
\overline{\mathrm{marg}}_{X}\bar{\mu} &=&\left( \mathrm{marg}_{X}\widetilde{%
\psi }\left( \nu ^{k}\right) \right) _{k=1}^{l} \\
&=&\left( \nu ^{k}\left( \psi ^{-1}\left( \mathrm{Proj}_{X}^{-1}\left( \cdot
\right) \right) \right) \right) _{k=1}^{l} \\
&=&\left( \nu ^{k}\left( \left( \mathrm{Proj}_{X}\circ \psi \right)
^{-1}\left( \cdot \right) \right) \right) _{k=1}^{l} \\
&=&\left( \nu ^{k}\left( \mathrm{Id}_{X}\left( \cdot \right) \right) \right)
_{k=1}^{l} \\
&=&\overline{\nu }\text{.}
\end{eqnarray*}%
It remains to show that $\bar{\mu}$ cautiously believes each $E_{m}$, $%
m=1,...,n$. Fix any $E_{m}$. Since $\overline{\nu }$\ cautiously believes $%
\mathrm{Proj}_{X}\left( E_{m}\right) $, by definition there exists $\bar{k}%
\leq l$\ such that $\cup _{k\leq \bar{k}}\mathrm{Supp}\nu ^{k}=\mathrm{Proj}%
_{X}\left( E_{m}\right) $. Then, for all $k\leq \bar{k}$,%
\begin{equation*}
\mu ^{k}\left( E^{m}\right) =\nu ^{k}\left( \psi ^{-1}\left( E_{m}\right)
\right) \geq \nu ^{k}\left( \mathrm{Proj}_{X}\left( E_{m}\right) \right) =1%
\text{,}
\end{equation*}%
where the inequality follows from (\ref{set inclusion lemma extension LPS}).
Since $\overline{\mathrm{marg}}_{X}\bar{\mu}=\overline{\nu }$, we have%
\begin{equation*}
\tbigcup_{k\leq \bar{k}}\mathrm{Suppmarg}_{X}\mu ^{k}=\mathrm{Proj}%
_{X}\left( E_{m}\right) \text{.}
\end{equation*}%
Therefore, by Proposition 2.2, $E_{m}$\ is cautiously believed under $\bar{%
\mu}$\ at level $\bar{k}$.\hfill $\blacksquare $

\bigskip

Now we state and prove the main result of this section. To simplify the
proof, we find it convenient to define $R_{i}^{0}:=S_{i}\times T_{i}$ and $%
\hat{R}_{i}^{0}:=S_{i}\times T_{i}$\ for each $i\in I$ in a type structure $%
\mathcal{T}$.

\bigskip

\noindent \textbf{Theorem D.2} \textit{Fix a finite game }$G:=\left \langle
I,(S_{i},\pi _{i})_{i\in I}\right \rangle $\textit{\ and an associated
belief-complete type structure }$\mathcal{T}$. \textit{Then, for every }$%
n\in \mathbb{N}_{0}$\textit{,}

\textit{(i) }$\tprod_{i\in I}\hat{R}_{i}^{n}\subseteq \tprod_{i\in
I}R_{i}^{n}$\textit{, and}

\textit{(ii) }$\prod_{i\in I}\mathrm{Proj}_{S_{i}}R_{i}^{n}=\prod_{i\in I}%
\mathrm{Proj}_{S_{i}}\hat{R}_{i}^{n}$\textit{.}

\bigskip

\noindent \textbf{Proof}. The proof is by induction on $n\in \mathbb{N}_{0}$.

\textit{Induction Hypothesis} \textit{(}$n$\textit{)}: For each $i\in I$ and
for each $m\leq n$,\textbf{\ }$\mathrm{Proj}_{S_{i}}\left( R_{i}^{m}\right) =%
\mathrm{Proj}_{S_{i}}\left( \hat{R}_{i}^{m}\right) $ and $\hat{R}%
_{i}^{m}\subseteq R_{i}^{m}$.

The basis step ($n=0$) trivially holds by definition. Suppose that the
result is true for some $n\in \mathbb{N}_{0}$. We show that it is true for $%
n+1$. Fix any $\left( s_{i},t_{i}\right) \in \hat{R}_{i}^{n+1}$. Since $\hat{%
R}_{i}^{n+1}\subseteq \hat{R}_{i}^{n}$, it follows from the induction
hypothesis that $\left( s_{i},t_{i}\right) \in R_{i}^{n}$. We claim that $%
\left( s_{i},t_{i}\right) \in \mathbf{B}_{i}^{c}\left( R_{-i}^{n}\right) $.
To see this, note that, by the induction hypothesis, $\mathrm{Proj}%
_{S_{-i}}\left( R_{-i}^{n}\right) =\mathrm{Proj}_{S_{-i}}\left( \hat{R}%
_{-i}^{n}\right) $ and $\hat{R}_{-i}^{n}\subseteq R_{-i}^{n}$. Since type $%
t_{i}$ cautiously believes $\hat{R}_{-i}^{n}$, Remark 3 entails that $t_{i}$
cautiously believes $R_{-i}^{n}$. Since $\hat{R}_{i}^{n+1}:=\hat{R}%
_{i}^{n}\cap \mathbf{B}_{i}^{c}\left( \hat{R}_{-i}^{n}\right) $ and $%
R_{i}^{n+1}:=R_{i}^{n}\cap \mathbf{B}_{i}^{c}\left( R_{-i}^{n}\right) $, it
follows that $\hat{R}_{i}^{n+1}\subseteq R_{i}^{n+1}$. Thus, $\prod_{i\in I}%
\hat{R}_{i}^{n+1}\subseteq \prod_{i\in I}R_{i}^{n+1}$, which implies $%
\prod_{i\in I}\mathrm{Proj}_{S_{i}}\hat{R}_{i}^{n+1}\subseteq \prod_{i\in I}%
\mathrm{Proj}_{S_{i}}R_{i}^{n+1}$.

To prove the converse, fix any $s_{i}\in \mathrm{Proj}_{S_{i}}R_{i}^{n+1}$,
so that $\left( s_{i},t_{i}\right) \in R_{i}^{n+1}$ for some $t_{i}\in T_{i}$%
. Let $\overline{\nu }_{i}:=\left( \nu _{i}^{1},...,\nu _{i}^{n}\right) =%
\overline{\mathrm{marg}}_{S_{-i}}\beta _{i}\left( t_{i}\right) $ denote the
marginal first-order belief of $t_{i}$ about co-players. Note that%
\begin{equation*}
R_{i}^{n+1}=R_{i}\cap \left( \tbigcap \limits_{m=1}^{n}\mathbf{B}%
_{i}^{c}\left( R_{-i}^{m}\right) \right) =R_{i}\cap \left( \tbigcap
\limits_{m=0}^{n}\mathbf{B}_{i}^{c}\left( R_{-i}^{m}\right) \right) \text{,}
\end{equation*}%
where the first equality follows from Lemma C.2, and the second equality
follows from the fact that $R_{i}=R_{i}\cap \mathbf{B}_{i}^{c}\left(
R_{-i}^{0}\right) $. By the induction hypothesis, $\mathrm{Proj}%
_{S_{-i}}R_{-i}^{m}=\mathrm{Proj}_{S_{-i}}\hat{R}_{-i}^{m}$ for every $m\in
\left \{ 0,...,n\right \} $. Proposition 2.2 yields that $\overline{\nu }%
_{i} $ cautiously believes $\left( \mathrm{Proj}_{S_{-i}}\hat{R}%
_{-i}^{m}\right) _{m=0}^{n}$. By Lemma D.1, there exists $\bar{\mu}_{i}\in 
\mathcal{N}(S_{-i}\times T_{-i})$ such that

(1) $\overline{\nu }_{i}=\overline{\mathrm{marg}}_{S_{-i}}\bar{\mu}_{i}$, and

(2) for each $m\in \left \{ 0,...,n\right \} $, $\hat{R}_{-i}^{m}$\ is
cautiously believed under $\bar{\mu}_{i}$.

By belief-completeness of $\mathcal{T}$, there is a type $t_{i}^{\ast }\in
T_{i}$ such that $\beta _{i}\left( t_{i}^{\ast }\right) =\bar{\mu}_{i}$.
With this,%
\begin{equation*}
\left( s_{i},t_{i}^{\ast }\right) \in \hat{R}_{i}\cap \left( \tbigcap
\limits_{m=0}^{n}\mathbf{B}_{i}^{c}\left( \hat{R}_{-i}^{m}\right) \right) =%
\hat{R}_{i}\cap \left( \tbigcap \limits_{m=1}^{n}\mathbf{B}_{i}^{c}\left( 
\hat{R}_{-i}^{m}\right) \right) =\hat{R}_{i}^{n+1}\text{,}
\end{equation*}%
where the inclusion holds by construction, the first equality holds because $%
\hat{R}_{i}=\hat{R}_{i}\cap \mathbf{B}_{i}^{c}\left( \hat{R}_{-i}^{0}\right) 
$, and the second equality follows from an analogue of Lemma C.2 concerning
the sets $\left( \hat{R}_{i}^{m}\right) _{m\in \mathbb{N}}$. It follows that 
$s_{i}\in \mathrm{Proj}_{S_{i}}\hat{R}_{i}^{n+1}$. This shows that $%
\prod_{i\in I}\mathrm{Proj}_{S_{i}}R_{i}^{n+1}\subseteq \prod_{i\in I}%
\mathrm{Proj}_{S_{i}}\hat{R}_{i}^{n+1}$.\hfill $\blacksquare $

\bigskip

For completeness, we record the following result, which is an immediate
consequence of Theorems 1 and 3 in the main text.

\bigskip

\noindent \textbf{Theorem D.3} \textit{Fix a finite game }$G:=\left \langle
I,(S_{i},\pi _{i})_{i\in I}\right \rangle $\textit{\ and an associated type
structure }$\mathcal{T}:=\langle S_{i},T_{i},\beta _{i}\rangle _{i\in I}$ 
\textit{which is terminal with respect to the class of all finite type
structures. Then, for every }$n\in \mathbb{N}_{0}$\textit{,}%
\begin{equation*}
\tprod_{i\in I}\hat{R}_{i}^{n}\subseteq \tprod_{i\in I}R_{i}^{n}\text{%
\textit{.}}
\end{equation*}

\bigskip

\noindent \textbf{Proof}. The proof is by induction on $n\in \mathbb{N}_{0}$%
. The basis step ($n=0$) trivially holds by definition. Suppose that the
result is true for $n\in \mathbb{N}_{0}$. We show that it is true for $n+1$.
Fix any $\left( s_{i},t_{i}\right) \in \hat{R}_{i}^{n+1}$. Since $\hat{R}%
_{i}^{n+1}\subseteq \hat{R}_{i}^{n}$, it follows from the induction
hypothesis that $\left( s_{i},t_{i}\right) \in R_{i}^{n}$. We claim that $%
\left( s_{i},t_{i}\right) \in \mathbf{B}_{i}^{c}\left( R_{-i}^{n}\right) $.
To see this, note that, by Theorem 1 and Theorem 3, $\mathrm{Proj}%
_{S_{-i}}\left( R_{-i}^{n}\right) =\mathrm{Proj}_{S_{-i}}\left( \hat{R}%
_{-i}^{n}\right) $ and, by the induction hypothesis, $\hat{R}%
_{-i}^{n}\subseteq R_{-i}^{n}$. Since type $t_{i}$ cautiously believes $\hat{%
R}_{-i}^{n}$, Remark 3 in the main text entails that $t_{i}$ cautiously
believes $R_{-i}^{n}$. Since $\hat{R}_{i}^{n+1}=\hat{R}_{i}^{n}\cap \mathbf{B%
}_{i}^{c}\left( \hat{R}_{-i}^{n}\right) $ and $R_{i}^{n+1}=R_{i}^{n}\cap 
\mathbf{B}_{i}^{c}\left( R_{-i}^{n}\right) $, it follows that $\hat{R}%
_{i}^{n+1}\subseteq R_{i}^{n+1}$. Thus, $\prod_{i\in I}\hat{R}%
_{i}^{n+1}\subseteq \prod_{i\in I}R_{i}^{n+1}$.\hfill $\blacksquare $